\documentclass[fleqn,usenatbib]{mnras}

\def\WB#1{\textcolor{black}{#1}}

\def\CB#1{\textcolor{black}{#1}}

\usepackage{newtxtext,newtxmath}

\usepackage[T1]{fontenc}

\DeclareRobustCommand{\VAN}[3]{#2}
\let\VANthebibliography\thebibliography
\def\thebibliography{\DeclareRobustCommand{\VAN}[3]{##3}\VANthebibliography}

\usepackage{graphicx}	
\usepackage{amsmath}
\usepackage{xcolor}
\usepackage{subcaption}
\usepackage{float}

\usepackage{orcidlink}

\title[The frame-dragging on galaxy scales]{The frame-dragging vector potential on galaxy scales from Dark-Matter-only Newtonian $N$-body simulations}

\author[W. Beordo et al.]{
William Beordo$^{\orcidlink{0000-0002-5094-1306}}$,$^{1}$\thanks{E-mail: william.beordo@inaf.it}
Marco Bruni$^{\orcidlink{0000-0003-2305-5699}}$,$^{1,2,3}$
Cristian Barrera-Hinojosa$^{\orcidlink{0000-0003-4976-3028}}$$^{4}$
and Mariateresa Crosta$^{\orcidlink{0000-0003-4369-3786}}$$^{1}$
\\
$^{1}$INAF - Osservatorio Astrofisico di Torino, Via Osservatorio 20, 10025, Pino Torinese, Italy\\
$^{2}$Institute of Cosmology and Gravitation, University of Portsmouth, Dennis Sciama Building, Burnaby Road, Portsmouth PO1 3FX, UK \\
${{^{3}}}$INFN Sezione di Trieste, Via Valerio 2, 34127 Trieste, Italy\\
$^{4}$Instituto de F\'{i}sica y Astronom\'ia, Universidad de Valpara\'iso, Gran Breta\~na 1111, Valpara\'iso, Chile
}

\date{Accepted XXX. Received YYY; in original form ZZZ}

\pubyear{2025}

\begin{document}
\label{firstpage}
\pagerange{\pageref{firstpage}--\pageref{lastpage}}
\maketitle

\begin{abstract}
Effects of General Relativity are usually neglected in the non-linear evolution of structures, where Newtonian $N$-body simulations are traditionally employed. In the post-Friedmann expansion framework, a weak-field relativistic approximation purpose-built for cosmology, a frame-dragging gravito-magnetic vector potential arises at leading order, sourced by momentum currents. \WB{At this order, the vector potential contributes to the metric while leaving the dynamics of the matter fields unaffected, as it does not appear in the Euler equation. It can therefore be extracted a posteriori from standard N-body simulations, where the dynamics is purely Newtonian.} Using the Delaunay Tessellation Field Estimator code on the IllustrisTNG simulations, here we extend previous work in order to compute the power spectrum of this vector potential down to galactic scales. The magnitude of the vector potential is two orders of magnitude larger than predicted by perturbation theory, and is a $1\% \sim 0.1\%$ effect compared to the non-linear Newtonian scalar gravitational potential. In the redshift range considered here, the gravito-magnetic effect remains subdominant, without showing any enhancement during a particular phase in the evolution of structures, aside from the continuous growth of non-linearity at low redshift. Although this seems to suggest that, within the $\Lambda$CDM model, no significant gravito-magnetic effects contribute to the non-linear evolution of cosmic structures, i.e. to the dynamics of massive particles, possible observational consequences, e.g. in lensing, deserve further exploration.

\end{abstract}




\section{Introduction}

A central challenge in modern cosmology lies in understanding the interplay between the non-linear evolution of structures and general-relativistic (GR) effects. On large scales, the universe is typically described using the Friedmann-Lema\^itre-Robertson-Walker (FLRW) metric, with small perturbations treated within the framework of cosmological perturbation theory (PT). On smaller, highly non-linear scales, the standard approach involves Newtonian $N$-body simulations, which provide a robust description of gravitational clustering and accurately capture many aspects of non-linear structure formation. However, these simulations are limited by their Newtonian framework, which neglects certain GR-specific quantities, such as the difference between the two scalar potentials, gravitational waves, and the vector potential in the spacetime metric. As GR is fundamentally a non-linear theory that couples different scales, this approximation introduces uncertainties that must be addressed, as observations enter the era of precision cosmology. 

To compute these small GR effects is important for two reasons: {\it i)} to make sure that the theoretical predictions are not only precise as required by observations, but also accurate; {\it ii)} to predict effects possibly observable in the future.

This challenge becomes even more pressing when studying the evolution of galaxies, the Milky Way in particular.
Reconstructing its history with microarcsecond accuracy -- a feat now achieved by missions such as Gaia \citep{prustiGaiaMission2016} -- demands the application of similarly precise, fully GR models for highly accurate astrometry \citep[see e.g.][]{Klioner_2003, crostaRayTracingAnalytical2015d}.

Traditionally, investigations into the formation and dynamics of galaxies rely on hydrodynamical simulations, which are high-resolution zoom-in refinements of larger-scale cosmological $N$-body simulations. To accurately reconstruct their evolutionary history -- marked by multiple mergers and interactions -- one must consider the possible contributions of purely non-linear GR phenomena. These effects may in principle introduce non-trivial corrections that may influence the inferred dynamical properties of galaxies, hence their evolution. The leading order beyond-Newtonian effect that appears in GR in post-Newtonian type approximations is gravito-magnetism, a vector potential in the metric that is responsible for the dragging of inertial frames. Far from being limited to strong fields, i.e.\ rotating neutron stars and black holes, this effect is ubiquitous, and relevant also in weak fields.

Frame-dragging caused by the Earth's rotation, which is linked to the gravito-magnetic vector potential, has been successfully measured by \citet[]{everittGravityProbeB}. More recently, the presence of a gravitational dragging has been suggested by \citet{crostaTestingCDMGeometrydriven2020} to drive the rotation of the spacetime, effectively reproducing the flat rotation curve of the Milky Way measured with Gaia DR2 \citep{gaiacollaborationGaiaDataRelease2018}. This result has been confirmed with Gaia DR3 \citep{gaiacollaborationGaiaDataRelease2023} and poses serious challenges to standard models of galactic dynamics \citep{beordoGeometrydrivenDarkmattersustainedMilky2024e, beordoExploringMilkyWay2024}, since it may potentially signal the breakdown of Newtonian weak-field approximations when applied to extended objects like galaxies.



In any case, even if these effects turn out to be small, as is typically expected in perturbative approaches, they may still be observable and should no longer be neglected, given the increasing precision of current and forthcoming surveys.
\CB{Moreover, the presence of a gravito-magnetic field affects not only the dynamics of matter but also photon trajectories, through which it may leave imprints on weak-lensing observables~\citep{schneiderGravitationalLenses1992}. In this context, its contribution to the lensing convergence field has been found to be strongly suppressed relative to the dominant (standard) weak-lensing signal~\citep{Bartelmann:1999yn,Cuesta-Lazaro:2018uyv}. Similarly,~\citet{sagaWeakLensingInduced2015} showed that B-modes induced by this field remain below the expected noise level of galaxy surveys such as Vera Rubin LSST~\citep{LSST2019ApJ...873..111I}. However, more recently,~\citet{barrera-hinojosaLookingTwistProbing2022} demonstrated that a cross-correlation method combining a weak-lensing survey with Cosmic Microwave Background (CMB) maps -- particularly those containing the kinetic Sunyaev-Zel'dovich (kSZ) effect signal -- can significantly enhance the signal-to-noise ratio compared to auto-correlation methods. This approach may therefore enable a successful detection with data from forthcoming CMB experiments such as the Simons Observatory~\citep{SimonsObservatory}, in combination with galaxy surveys such as LSST and Euclid~\citep{euclid}. }
\CB{Hence, as observational sensitivity improves, relativistic effects such as frame dragging may provide novel constraints on models of dark matter, dark energy, and modified gravity theories, since these scenarios predict distinct signals~\citep{thomasGravityNonlinearScales2015}.}

Post-processing methods, such as those developed by \citet{bruniComputingGeneralrelativisticEffects2014a, thomasFullyNonlinearPostFriedmann2015}, enabled for the first time the extraction of the gravito-magnetic vector potential from cosmological Newtonian $N$-body simulations on fully non-linear scales. This approach is justified within the post-Friedmann formalism \citep{mililloMissingLinkNonlinear2015}, where, at the leading order in the Newtonian dynamical regime, the purely relativistic vector potential is sourced by the transverse component of the momentum density field.
\WB{Such vector potential, at the leading order, does not enter the Euler equation, so it does not affect matter dynamics. Therefore, even if the simulation is Newtonian, as it only accounts for the Newtonian scalar potential, the vector potential can be evaluated a posteriori to make a sensible estimate.}
While being about two or three orders of magnitude smaller than the scalar Newtonian potential, the vector potential computed from second-order perturbation theory \citep{luCosmologicalBackgroundVector2009} is found to be underestimated down to two orders of magnitude at fully non-linear scales.
Although the Newtonian simulations neglect feedback from the relativistic fields into the dynamics, these results confirm that at a sufficiently small scale, in the highly non-linear regime, the relativistic corrections to Newtonian dynamics are subdominant.

With some simplifying approximations, GR N-body codes, such as gevolution \citep{adamekNbodyMethodsRelativistic2014} and GRAMSES \citep{barrera-hinojosaGRAMSESNewRoute2020} have emerged in recent years. Although these codes take into account the back reaction of the vector potential into the simulation dynamics, they lead to very similar findings \citep{adamekGeneralRelativityCosmic2016, barrera-hinojosaVectorModesLCDM2021}. Thanks to the Adaptive Mesh Refinement (AMR) technique, which enables higher spatial resolution to be achieved in high-density regions, \citet{barrera-hinojosaVectorModesLCDM2021} also quantified the gravito-magnetic acceleration inside dark matter halos, where the ratio with respect to the standard Newtonian acceleration remains typically of the order of $10^{-3}$ for redshift $z < 3$ and halo mass in the range $10^{12.5}\text{--}10^{15} \; h^{-1}\, \mathrm{M}_{\odot}$.

Based on a fluid-description but otherwise fully relativistic, codes such as the Einstein Toolkit \citep{EinsteinToolkit} have also been used in cosmology in recent years \citep{BentivegnaBruni, MunozBruni} to study possible GR effects. In particular, using idealised initial conditions based on a simple spatial distribution in the growing mode of a curvature perturbation, it was shown in \citet{MunozBruni} that gravito-magnetic effects are always generated around halos and filaments. Current work based on the N-body relativistic code GRAMSES~\citep{barrera-hinojosaGRAMSESNewRoute2020} and assuming similar initial conditions confirms these effects \citep[]{Barrerainprep}.

However, these studies focused on large cosmological simulations with box size $L > 160 \; h^{-1}$ Mpc and mass resolutions greater than $10^{8} \; h^{-1}\, \mathrm{M}_{\odot}$. While these configurations are well-suited for studying large-scale structure formation, they lack the necessary spatial and mass resolution to accurately capture the internal dynamics of galaxies. \WB{On the other hand, widely employed simulations suites like IllustrisTNG \citep{nelsonIllustrisTNGSimulationsPublic2019} and EAGLE \citep{theeagleteamEAGLESimulationsGalaxy2017} are optimal for} resolving small-scale structures and are specifically designed for galactic studies. The use of such simulations \WB{can} also straightforwardly facilitate the extension to the hydrodynamical counterpart of the simulations in order to investigate the role of baryons in this context, which has yet to be evaluated in the literature.

\WB{The present work makes use of dark-matter-only simulations from the IllustrisTNG suite, following the post-processing method developed by \citet{bruniComputingGeneralrelativisticEffects2014a} and \citet{thomasFullyNonlinearPostFriedmann2015} to evaluate the gravito-magnetic potential in the post-Friedmann framework and its time evolution. This involves the use of the Delaunay Tessellation Field Estimator \citep[DTFE,][]{schaapContinuousFieldsDiscrete2000} to extract the continuous fields, and the spectral analysis for the solution of the Einstein's equations. In the present paper, in addition, the solution over the entire simulation volume is also reconstructed from spectral analysis.} 

\WB{It is important to recall again that the vector potential is in all respects a relativistic quantity that has no counterpart in the conventional Newtonian dynamics, but only appearing at the leading order of the expansion in the post-Friedmann approach. 
Since it does not affect the dynamics at this order, the vector potential can be consistently estimated a posteriori from classical Newtonian N-body simulations, where the dynamics is only determined by the Newtonian scalar potential.}

The remainder of this paper is organised as follows.
In Section \ref{sec:post-friedmann}, we introduce the post-Friedmann approach. Section \ref{sec:IllustrisTNG} describes the IllustrisTNG simulations adopted in this work.
In Section \ref{sec:B_extraction}, we outline the methodology used to extract the gravito-magnetic potential from the simulation data.
Section \ref{sec:results} presents our results for the source fields, the gravitational potentials, and their time evolution.
Finally, we summarise our conclusions in Section \ref{sec:conclusions}.

\section{the non-linear Post-Friedmann framework} \label{sec:post-friedmann}

Understanding cosmic structure formation across all relevant scales requires a framework that seamlessly transitions from the fully non-linear Newtonian regime to the relativistic domain on cosmological scales. Traditional approaches, such as post-Newtonian (PN) and post-Minkowskian approximations, are well-suited for weak-field scenarios (and slow-motion, in the case of PN) and have been applied to cosmology \citep{Tomita1988, Tomita1991, Futamase1988, Futamase1996, Kofman1995, Takada1999, Matarrese1996, Hwang2008, Szekeres2000}, but are typically limited to sub-Hubble scales. 

To address this limitation, the post-Friedmann (PF) approach to cosmology has been developed by \cite{mililloMissingLinkNonlinear2015} as a new approximation scheme that provides a unified framework for all scales, from the fully non-linear Newtonian regime to the largest scales where relativistic effects become important \citep{Bruni2012, Bruni_gaussianity2014}.
This approach generalises to a FLRW background spacetime the weak-field approximation on a Minkowski background, expanding Einstein field equations in inverse powers of the speed of light, following the pioneering work by \cite{PN_chandrasekhar} \citep[see][for a comprehensive modern treatment]{PoissonWill}. However, while in the PN formalism velocities are treated as small, in the PF approach  only {\it peculiar} velocities are small.

These expansions need to be performed differently in cosmology compared to those of the post-Newtonian formalism for several reasons. In the PN approach, the approximation is derived by assuming small velocities ($v/c \ll 1$) and neglecting time derivatives compared to spatial ones. However, in a cosmological context, 
the time evolution of physical distances between observers cannot be ignored.
In other words, if a Newtonian approach were adopted, imposing the condition $|\dot{\Vec{r}}| \ll c$, the resulting approximation would only be valid on very small scales compared to the Hubble horizon: $|\Vec{r}| \ll cH^{-1}$. Instead, assuming comoving coordinates $\Vec{x}$, in cosmology $\Vec{r}=a(t)\Vec{x}$, where $a(t)$ is the FLRW background scale factor, so that $\dot{\Vec{r}}=H\Vec{r}+a\dot{x}$, and in the PF approach it is only the peculiar velocity $\Vec{v}=\dot{\Vec{x}}$, i.e.\ the deviation from the Hubble flow $H\Vec{r}$ (where $H(t)$ is the Hubble function), that it is assumed to be small: $|\Vec{v}|/c\ll 1$.

Moreover, the PN formalism has traditionally been developed to capture relativistic corrections in the dynamics of isolated systems, concentrating on their equations of motion rather than providing a systematic approximation to the Einstein equations.
To overcome these limitations, the PF formalism adopts an active approach, treating the space-space and time-time components of the metric on an equal footing. 
This ensures a consistent treatment of all of Einstein’s equations order by order in the $c^{-n}$ expansion. 

When the set of equations is linearised, the formalism correctly reproduces first-order cosmological relativistic perturbation theory. On the other hand, at the leading order \citep[which we refer to as 0PF, following][]{mililloMissingLinkNonlinear2015} the Newtonian regime is recovered. 
Therefore, the scheme is capable of describing the evolution of structures on all scales of interest in a unified framework \citep{mililloMissingLinkNonlinear2015}. In the weak-field limit, however, besides the standard Newtonian scalar gravitational potential, a new vector gravitational potential appears as sourced by energy-momentum currents. This vector potential physically represents the gravito-magnetic field that manifests as the GR effect of frame-dragging.

\subsection{The formalism}\label{sec:PF_formalism}

The post-Friedmann approach assumes the standard $\Lambda$CDM cosmology, where the FLRW metric describes a homogeneous and isotropic universe. The perturbed flat background is expanded in the Poisson (or conformal-Newtonian) gauge \citep{Bertschinger:1996, Malik:2008im, Matarrese:1997ay, Ma:1995ey}, up to order $c^{-5}$, so that the components of the metric tensor can be written as
\begin{subequations}\label{eq:metricPF}
\begin{align}
    g_{00} &= - \left[ 1 - \frac{2U_\mathrm{N}}{c^2} + \frac{1}{c^4} \left( 2U_\mathrm{N}^2 - 4U_\mathrm{P} \right) \right] + O\left(\frac{1}{c^6}\right) \;, \label{eq:g00} \\
    g_{0i} &= - \frac{a}{c^3} B^\mathrm{N}_i - \frac{a}{c^5} B^P_i + O\left(\frac{1}{c^7}\right) \;, \label{eq:g0i} \\
    g_{ij} &= a^2 \left[ \left( 1 + \frac{2V_\mathrm{N}}{c^2} + \frac{1}{c^4} \left( 2V_\mathrm{N}^2 + 4V_\mathrm{P} \right) \right) \delta_{ij} + \frac{1}{c^4} h_{ij} \right] + O\left(\frac{1}{c^6}\right)\; , 
\end{align}
\end{subequations}
where $a$ and $\delta_{ij}$ are respectively the scale factor and flat spatial metric of the FLRW background\footnote{Latin indices refer to the spatial coordinates and can take the values $1$,$2$,$3$.}. 
The metric perturbations are characterized by the scalar potentials $U$ and $V$, the vector potential $\mathbf{B}$\footnote{Note that the vector potential $\mathbf{B}$ defined in the metric tensor~(\ref{eq:metricPF}) of the PF formalism (with $B_i$ components) is defined with a different convention compared to \citet[][with $\beta^i$ components]{barrera-hinojosaVectorModesLCDM2021}, so that $B_i \equiv a\beta^i$.}, and the tensor potential $h_{ij}$. 
These perturbations are split into Newtonian (N) and post-Friedmann (P) quantities, each corresponding to a different level of approximation.
Newtonian terms appear at orders $c^{-2}$ and $c^{-3}$ and are the only relevant ones at $0$PF order, where \WB{the Einstein's Equations (EEs)} reduce to the exact non-linear equations of Newtonian cosmology for a pressure-less matter fluid. Post-Friedmann terms, appearing with powers $c^{-4}$ and $c^{-5}$ at $1$PF order, contain the GR corrections modifying the Newtonian equations. The tensor perturbation $h_{ij}$ only enters the equations at $1$PF order and represents pure tensor modes.
The vector potential $\mathbf{B}$ and the tensor $h_{ij}$ only have two degrees of freedom each, since in the Poisson gauge the vector potential is divergence-less, namely $\nabla \cdot \mathbf{B} = 0$, while $h_{ij}$ is transverse and trace-free.

In a universe filled by collision-less cold dark matter (CDM) fluid, the energy-momentum tensor can be written as $T^\mu_{\;\ \nu} = c^2 \rho u^\mu u^\nu$, where $\rho$ is the mass density and $u^\mu$ the time-like dimensionless 4-velocity. 
The spatial component of the 4-velocity can be expressed in terms of the physical peculiar velocity $v^{i} =a(  dx^i/dt)$, with $x^i$ the spatial coordinate with respect to the comoving grid, such that
\begin{align}
    u^i &= \frac{dx^i}{c d\tau} = \frac{dx^i}{c dt} \frac{dt}{d\tau} = \frac{v^i}{c a} u^0 \;,
\end{align}
where, keeping terms up to order \( c^{-4} \), the time component is
\begin{align}
    u^0 = 1 &+ \frac{1}{c^2} \left( U_\mathrm{N} + \frac{1}{2} v^2 \right) \nonumber \\
    &+ \frac{1}{c^4} \left[ \frac{1}{2} U_\mathrm{N}^2 + 2 U_\mathrm{P} + v^2 V_\mathrm{N} + \frac{3}{2} v^2 U_\mathrm{N} + \frac{3}{8} v^4 - B^\mathrm{N}_i v^i \right] \; ,
\end{align}
with $v^2 = \delta_{ij}v^iv^j$. From this equation, the stress-energy tensor can be expanded accordingly.

The Einstein field equations are then computed for the metric~(\ref{eq:metricPF}), including the cosmological constant.
The equations for the conservation of energy and momentum are also expanded in powers of $1/c$ and calculated from the contracted Bianchi identities.
We refer the reader to \citet{mililloMissingLinkNonlinear2015} for the full system of expanded equations obtained from the Einstein and hydrodynamic equations.

\subsection{Leading order: frame-dragging vector potential in the Newtonian regime}
By retaining only the leading-order terms in the expansion, i.e. the $0$PF order, the Newtonian continuity and Euler equation result from the hydrodynamic equations 
\begin{align}
    \frac{d\delta}{dt} + \frac{v^{i}{}_{,i}}{a} (1 + \delta) &= 0 \label{eq:PF_continuity}\\
    \frac{dv_i}{dt} + \frac{\dot{a}}{a} v_i &= \frac{1}{a} U_{\mathrm{N},i} \label{eq:PF_euler} \;,
\end{align}
while Einstein's equations, once the FLRW background has been subtracted, give 
\begin{align}
    \frac{1}{c^2 a^2} \nabla^2 V_\mathrm{N} &= -\frac{4\pi G}{c^2} \bar\rho \delta \label{eq:PF_poisson}\\
    \frac{2}{c^2 a^2} \nabla^2 (V_\mathrm{N} - U_\mathrm{N}) &= 0 \label{eq:PF_scalar}\\
    \frac{1}{c^3} \left[ \frac{2\dot{a}}{a^2} U_{\mathrm{N},i} + \frac{2}{a} \dot{V}_{\mathrm{N},i} - \frac{1}{2a^2} \nabla^2 B^\mathrm{N}_i \right] 
    &= \frac{8\pi G \bar\rho}{c^3} (1 + \delta) v_i \; . \label{eq:PF_vector}
\end{align}
where $\bar\rho$ denotes the background density and $\delta$ the density contrast, defined as $\delta \equiv (\rho - \bar\rho)/\bar\rho$. Note that in this Newtonian regime, the density contrast is not required to be small.
\WB{In the above equations, the lower script "$,i$" correspond to the partial derivative with respect to the spatial component $i$.}

Equation~(\ref{eq:PF_poisson}) coincides with the Poisson equation, where $V_\mathrm{N}$, usually identified as a curvature perturbation in relativistic perturbation theory \citep[see e.g.][]{Bruni_gaussianity2014}, can be identified with the Newtonian gravitational potential generated by the matter field, $V_\mathrm{N} = -\phi_\mathrm{N}$. The scalar potential $U_\mathrm{N}$ enters in the Euler fluid equation~(\ref{eq:PF_euler}), driving the dynamical flows of matter, and therefore can also be identified with the Newtonian gravitational potential, $U_\mathrm{N} = -\phi_\mathrm{N}$.
The Einstein equations yield two additional constraints. Equation~(\ref{eq:PF_scalar}) forces the scalar potentials $V_\mathrm{N}$ and $U_\mathrm{N}$ to be equal, consistently with the fact that there is only one scalar potential in Newtonian theory, namely $V_\mathrm{N} = U_\mathrm{N} = -\phi_\mathrm{N}$. At higher orders in the $c^{-n}$ expansion, the difference between the two scalar potential is small but can be measured from relativistic $N$-body simulations \citep{adamekGeneralRelativityCosmic2016}.

The second constraint equation~(\ref{eq:PF_vector}) relates the leading-order vector gravitational potential $B_i^\mathrm{N}$ to the momentum-density field of the matter. The latter is completely determined from the standard Newtonian equations \eqref{eq:PF_continuity} and \eqref{eq:PF_euler}, therefore the vector potential is not dynamical at the leading order, as it does not enter the Euler equation, meaning that matter flow is not affected at this order of approximation.
However, even in the regime where the matter dynamics is correctly described by Newtonian theory, the frame-dragging potential $B^\mathrm{N}_i$ should not be set to zero. If this were the case, an extra constraint on the Newtonian dynamics would be set. 
In fact, taking the the curl of Equation~(\ref{eq:PF_vector}), the gravito-magnetic potential is sourced by the transverse part of the Newtonian momentum density $\mathbf{P}^\mathrm{N}$
\begin{align}\label{eq:vector_pot}
    \nabla \times \nabla^2 \mathbf{B}^\mathrm{N} &= - 16 \pi G a^2 \nabla \times \mathbf{P}^\mathrm{N} \; ,
   \\
   \mathbf{P}^\mathrm{N}&\equiv\bar\rho \left[(1+\delta)\mathbf{v}\right] \;.
\end{align}
Thus, imposing the condition $B^\mathrm{N}_i=0$ does imply a purely longitudinal energy-momentum current, which is clearly an artificial constraint on the Newtonian dynamics, since there is no reason why this should occur.

Even if the vector potential is not dynamical at this leading order, it appears in the metric and affects null geodesics and observables. For example, \citet{ThomasLensing} found that the power spectra of the B-mode of shear for the weak gravitational lensing, which is generated by the time derivative of the vector potential, is $\approx 10^{-5}$ times the E-mode, which is the main effect generated by the scalar potential. \citet{barrera-hinojosaLookingTwistProbing2022} also showed that the possibility of detecting the gravito-magnetic effect in the lensing-kSZ cross-correlation is realistically possible with future CMB experiments.

The density and velocity fields sourcing the vector potential via Eq.~\eqref{eq:vector_pot} are purely Newtonian quantities and can be extracted from standard Newtonian $N$-body simulations \citep{bruniComputingGeneralrelativisticEffects2014a, thomasFullyNonlinearPostFriedmann2015}.
Indeed, it is well known that transverse vector modes are generated by non-linearity, and in particular, vorticity is generated after shell crossing \citep{pueblasGenerationVorticityVelocity2009, jelic-cizmekGenerationVorticityCosmological2018, hahnPropertiesCosmicVelocity2015}. The extraction of the vector potential is illustrated in Section~\ref{sec:B_extraction}.

\WB{At the next-to-leading order, 1PF, the correction $B^\mathrm{P}_i$ to the frame-dragging potential is of the order of $c^{-5}$ (see Equation~\ref{eq:g0i}), meaning that its magnitude is $c^2$ times smaller than the leading order frame-dragging effect. In addition, at the 1PF expansion order, scalar and tensorial perturbations appear as $c^{-4}$ corrections to the metric terms  and these, together with the frame-dragging, have a non-trivial effect on the matter dynamics, which is not investigated in this paper.} 

\section{The Illustris-TNG simulations} \label{sec:IllustrisTNG}
As already mentioned in the introduction, the behaviour of the frame-dragging potential is poorly known at very small scales, where non-linearities are expected to generate rotational momentum flows of matter.
Although \citet{barrera-hinojosaVectorModesLCDM2021} probed the vector potential inside dark matter halos with masses $10^{12.5}\text{--}10^{15} \; h^{-1}\, \mathrm{M}_{\odot}$, thanks to the great resolution capabilities of the AMR technique, they focused on a large cosmological simulation of sole dark matter with a box size of $256\;h^{-1}$ Mpc and a particle mass resolution of $1.33 \times 10^{9} \; h^{-1}\, \mathrm{M}_{\odot}$ \citep[the same mass resolution was used by][for larger simulations]{adamekGeneralRelativityCosmic2016}.
On the other hand, \citet{thomasFullyNonlinearPostFriedmann2015} analysed smaller simulations ranging from $80$ to $320\;h^{-1}$ Mpc, but with mass resolution greater than $3.97 \times 10^{8} \; h^{-1}\, \mathrm{M}_{\odot}$ in all cases.

This paper employs high-resolution simulations from the IllustrisTNG suite, which allow for a detailed characterisation of the gravito-magnetic vector potential down to galaxy scales.
The IllustrisTNG project comprises a set of state-of-the-art Newtonian cosmological simulations designed to model the formation and evolution of galaxies in the universe within the framework of the $\Lambda$CDM cosmology \citep{nelsonIllustrisTNGSimulationsPublic2019,TNG_Pillepich,TNG_Springel, PillepichTNG, Pillepich50}. These simulations build upon the original Illustris project \citep{IllustrisP} and introduce significant improvements in the physical models for galaxy formation, including refined treatments of feedback from supernovae and active galactic nuclei, as well as enhanced resolution capabilities.

The suite consists of multiple simulation volumes, ranging from the small-scale TNG50 to the large-scale TNG300, with each box simulated both in a full-physics and a dark-matter-only configuration. 
This multi-scale approach allows for a comprehensive exploration of the evolution of cosmic structures from the largest galaxy clusters to the detailed internal properties of individual galaxies.
The full-physics runs employ a sophisticated hydrodynamical solver, based on the moving-mesh code AREPO \citep{Springel2010}, to self-consistently model baryonic processes such as gas cooling, star formation, black hole growth, and various feedback mechanisms \citep{Weinberger2018, Weinberger2017, Nelson50}.
These simulations provide realistic galaxy morphologies, scaling relations, and detailed predictions of the interplay between baryonic and dark matter components, making them a powerful tool for exploring the physics of galaxy formation and evolution \citep{GenelTNG, Torrey, Rodriguez-Gomez, Lovell}.

On the other hand, the dark-matter-only versions of these simulations are particularly useful for studying the non-linear evolution of dark matter halos and their substructures, gravitational potentials, and cosmic velocity fields, without the complicating effects of baryonic physics. \WB{In this case, the dark matter particles are more massive than in the case with baryons, precisely to compensate for the lack of baryons and preserve the matter density of the universe.}
Table~(\ref{tab:simulations}) lists the main parameters of the \WB{dark-matter-only} simulations analysed in this paper.
\begin{table}
	\centering
	\begin{tabular}{lccc} 
		\hline
		Name & Box size & $N_{\mathrm{DM}}$ & $m_{\mathrm{DM}}$ \\
            & [$h^{-1}$ Mpc] & & [$h^{-1}$ M$_\odot$] \\
		\hline
        \textbf{TNG50-2-Dark} & $35$ & $1080^3$ & $2.9 \times 10^6$ \\
        TNG50-3-Dark & $35$ & $540^3$ & $2.3 \times 10^7$ \\
        TNG50-4-Dark & $35$ & $270^3$ & $1.9 \times 10^8$ \\
        \textbf{TNG100-2-Dark} & $75$ & $910^3$ & $4.8 \times 10^7$ \\
        \textbf{TNG300-2-Dark} & $205$ & $1250^3$ & $3.8 \times 10^8$ \\
        TNG300-3-Dark & $205$ & $625^3$ & $3.0 \times 10^9$ \\
		\hline
	\end{tabular}
    \caption{Details of the IllustrisTNG simulations used in this paper: the box size, the number of dark matter particles $N_\mathrm{DM}$, and the mass of each dark matter particle $m_\mathrm{DM}$. \WB{Simulations used for the main results are highlighted in bold; those in plain text were only used for the convergence tests in Appendix.}}\label{tab:simulations}
\end{table}
The results presented in this paper are based on the simulations TNG50-2-Dark, TNG100-2-Dark, TNG300-2-Dark, which have box size of $35$, $75$, and $205\;h^{-1}$ Mpc, respectively.
Other simulations like TNG50-3-Dark, TNG50-4-Dark, and TNG300-3-Dark have fewer particles, hence poorer mass resolution, but are employed to perform convergence tests as shown in Appendix~\ref{app:main_gravito}.

In the smallest simulation, i.e. TNG50-2-Dark, dark matter particles have a mass of $2.9 \times 10^{6} \; h^{-1}\, \mathrm{M}_{\odot}$, meaning that the mass resolution is at least two orders of magnitude better than the studies mentioned above \citep{bruniComputingGeneralrelativisticEffects2014a,thomasFullyNonlinearPostFriedmann2015,thomasGravityNonlinearScales2015,barrera-hinojosaVectorModesLCDM2021, adamekGeneralRelativityCosmic2016}. On the other hand, the largest simulation is used to ensure consistency with these works.
Higher-resolution simulations, such as TNG50-1-Dark, provide finer details but will not be employed due to the prohibitive computational cost of extracting the relevant fields.

The \WB{IllustrisTNG} simulations \WB{were} evolved from redshift $z=127$ and assuming the Planck$2015$ cosmology \citep{planckcollaborationPlanck2015Results2016}\footnote{\WB{Note that the Planck2018 cosmology was not yet available when the IllustrisTNG project started.}}, with a Hubble parameter $h =0.6774$, matter density parameter $\Omega_\mathrm{m} = 0.3089$, and dark energy density parameter $\Omega_\Lambda = 0.6911$. Multiple snapshots from $z=20$ to $z=0$ are available \WB{on request, and we used them} in order to investigate the time evolution of the vector potential.
In future works, the use of the IllustrisTNG suite will also straightforwardly facilitate the extension to the hydrodynamical counterpart of the simulations in order to investigate the role of baryons in this context, which has yet to be evaluated in the literature.

\section{Extracting the gravito-magnetic potential}\label{sec:B_extraction}
In order to derive the vector potential from Equation~(\ref{eq:vector_pot}), one needs to extract the transverse component of the momentum density field from the simulation data.
However, the raw particle distribution in $N$-body simulations consists of a finite number of tracers that sample an underlying continuous field. To recover physical quantities such as density, velocity, and their spatial derivatives, we need to employ an interpolation scheme that reconstructs smooth fields while retaining the non-linear structure of the simulated system.

A common approach in cosmological simulations is grid-based interpolation, such as the Cloud-in-Cell (CIC) and Triangular-Shaped Cloud (TSC) methods~\citep{Hockney-Eastwood:1988}. These techniques work by assigning particle properties to a fixed spatial grid and averaging over nearby points to generate a continuous field representation. While efficient and straightforward to implement, such methods fail to accurately capture small-scale structures, as they can lead to over-smoothing.

\subsection{The Delaunay Tessellation Field Estimator}\label{sec:DTFE}

\WB{As already implemented by \citet{bruniComputingGeneralrelativisticEffects2014a} and \citet{thomasFullyNonlinearPostFriedmann2015},} a more adaptive and accurate alternative is the \textit{Delaunay Tessellation Field Estimator} \citep[][]{schaapContinuousFieldsDiscrete2000}, which reconstructs continuous fields using a Delaunay triangulation of the particle positions. This technique partitions the simulation domain into a unique tessellation of tetrahedra, whose vertices are defined by the particle positions and that do not contain any particle, and then draws the field on a regular grid. 
In detail, the density field is first evaluated on each vertex as inversely proportional to the volume of the contiguous tetrahedra, while the velocity field is already known \WB{since the particle velocity is given as input}. Then, for each tetrahedron, the field is linearly interpolated on $n$ random sampling points extracted inside the tetrahedron.\footnote{In this paper the standard value of $n=100$ is adopted. \WB{We did not check the results for different values, as \citet{bruniComputingGeneralrelativisticEffects2014a} varied $n$ up to 1000 and found no difference to the results.}} Finally, the field at each grid cell is computed as the volume-weighted average of the field over the entire volume of the grid cell, where the contribution of each sample point is distributed to the grid cell in which it resides.

The DTFE is a widely used method in cosmological simulations for extracting continuous fields and their derivatives from discrete particle data. 
Unlike fixed-grid methods, it dynamically adapts to the local density of particles, providing higher resolution in dense regions such as halos and filaments, while maintaining coarser resolution in cosmic voids.
This reduces artificial smoothing effects, such as assigning artificial zero velocities in voids, which is the major limitation of methods like CIC \citep{pueblasGenerationVorticityVelocity2009}. Moreover, the volume averaging of the fields reduces the Poisson noise.
This adaptability makes the DTFE particularly suited for estimating both the density and velocity fields in highly non-linear environments.
The DTFE also enables direct computation of spatial derivatives, including velocity divergence and vorticity, without requiring additional smoothing or convolution steps. 

However, the method is not without limitations. The computational cost of constructing the tessellation increases significantly with the number of particles, whereas the lack of a straightforward deconvolution of the window function prevents the correction for interpolation effects at very small scales.
Additionally, the method's reliability can be affected in regions with caustics or orbit-crossing events, where discontinuities in the underlying fields may lead to inaccuracies in the computed gradients \citep{hahnPropertiesCosmicVelocity2015}. Here, in fact, the velocity divergence and vorticity estimated via the finite difference scheme of DTFE are not completely reliable, whereas phase-space interpolation methods are expected to perform better \citep{Abel2012MNRAS.427...61A}. 


In this paper, the publicly available DTFE implementation by \citet{cautunDTFEPublicSoftware2019a} is employed to extract the density and velocity fields and their gradients\WB{, following the same methodology of \citet{bruniComputingGeneralrelativisticEffects2014a} and \citet{thomasFullyNonlinearPostFriedmann2015}}.
All the fields are interpolated into a cubic grid with $N_\mathrm{grid} = 1024^3$ nodes. This number ensures a good convergence of the results, as detailed in Appendix~\ref{app:grid_res}. 

In the following, in order to compare with $\delta$ and the velocity $\mathbf{v}$, we find convenient to work with the ``momentum field" $\mathbf{p}$ defined as \citet{thomasFullyNonlinearPostFriedmann2015}
\begin{equation}
    \mathbf{p} \equiv \frac{\mathbf{P}^{\mathrm{N}}}{\bar{\rho}}=(1+\delta) \mathbf{v}\; .
\end{equation}
This momentum and its transverse part are computed a posteriori on the mesh.


\subsection{Fourier analysis}\label{sec:Fourier_analysis}

\WB{In order to solve Equation~(\ref{eq:vector_pot}), \citet{bruniComputingGeneralrelativisticEffects2014a} and \citet{thomasFullyNonlinearPostFriedmann2015} proposed to derive the properties of the extracted fields in Fourier space,} where power spectra provide a compact statistical description of how structures are distributed across scales.
Following the convention below, the power spectrum of a vector field $\mathbf{v}$ is defined as the sum of the power spectra of its components:
\begin{equation}\label{eq:power_spectrum}
    \langle \widetilde{\mathrm{v}}^i(\mathbf{k}) \cdot \widetilde{\mathrm{v}}_i^*(\mathbf{k'}) \rangle = \left( 2 \pi \right)^3 \delta(\mathbf{k} - \mathbf{k'}) P_{\mathbf{v}}(k) \; ,
\end{equation}
where $P_{\mathbf{v}}(k)$ is given by spherically averaging over the Fourier modes, as the field is assumed statistically isotropic. Here, the reader must note that a more rigorous approach would involve considering a power spectrum tensor with six degrees of freedom, which includes contributions from both transverse and longitudinal components. In fact, the definition~(\ref{eq:power_spectrum}) only represents the trace of this power spectrum tensor. Moreover, the assumption of statistical isotropy may not be valid for simulations with small boxes, like IllustrisTNG, where structures evolve in the highly non-linear regime.
However, the present paper does not explore these details, as the primary focus at this stage is on the overall magnitude of the vector potential, leaving a more in-depth analysis to future work.

The computation of the power spectra is performed through the python library \textsc{nbodykit} \citep{handNbodykitOpensourceMassively2018}, where the Fourier coefficients are averaged over spherical shells in $k$-space. The width of the bins is set to $4\pi / L$, with $L$ denoting the size of the simulation box. All power spectra are truncated at the Nyquist wave number, namely $k_\mathrm{max} = \pi N_{\mathrm{grid}}^{1/3}/L$, while the minimum wave number is set to the fundamental mode $k_\mathrm{box} = 2 \pi /L$.
For the gravitational potentials, the results are expressed in terms of the dimensionless power spectra, defined as
\begin{equation}
    \Delta(k) \equiv \dfrac{k^3}{2 \pi^2} P(k) \; ,
\end{equation}
where $P(k)$ is the dimensional power spectrum.

Since the goal is to analyse the spectral properties of the fields, gradients are computed directly in Fourier space using operators defined via Fast Fourier transform (FFT) methods. In Fourier space, the divergence and curl of a vector field are proportional to the parallel and transverse projections of its Fourier transform, respectively. As a consequence, the following relation holds:
\begin{equation}
    P_{\mathbf{v}} = k^2 (P_{\nabla \times \mathbf{v}} + P_{\nabla \cdot \mathbf{v}}) \; .
\end{equation}
This procedure is more robust than computing the divergence and curl in real space and then Fourier transforming them, particularly in the presence of discontinuities or caustics \citep{hahnPropertiesCosmicVelocity2015}.

\noindent From Equation~(\ref{eq:vector_pot}), the power spectrum of the vector potential becomes
\begin{equation}
	 	P_{ \mathbf{B^\mathrm{N}}}(k) = \left(\dfrac{16 \pi G a^2 \bar{\rho}}{k^3} \right)^2 P_{\nabla \times \mathbf{p}}(k) \;,
    \label{eq:vector_pot_PS}
\end{equation}
and can thus be inferred from the power spectrum of the momentum vorticity $P_{\nabla \times \mathbf{p}}$.
Similarly, the power spectrum of the scalar potential derived from the Poisson equation~(\ref{eq:PF_poisson}) is given by
\begin{equation}
	P_{ \phi_\mathrm{N}}(k) = \left( \dfrac{4 \pi G a^2 \bar{\rho}}{k^2} \right)^2 P_{\delta}(k) \; ,
\end{equation}
where $P_{\delta}$ is the matter power spectrum measured from the simulation.

To interpret the amplitude and scale dependence of the vector potential, it is useful to compare these power spectra with analytical expectations from perturbation theory.
Here, the vector potential appears as a second-order term as it is sourced by the product of first-order density and velocity divergence, in the case of the CDM fluid. The power spectrum of the vector potential at the second-order approximation is given by \citet{luCosmologicalBackgroundVector2009}, which is equivalent to the integral in the form provided by \citet[][ Equation 30]{barrera-hinojosaVectorModesLCDM2021}:
\begin{align}
    \Delta_{\mathbf{B}}(k) =& \dfrac{9 \Omega_{\mathrm m}^2 H_0^4}{2 k^2} \int_0^\infty \mathrm{d}\mathit{w} \int_{|1-\mathit{w}|}^{|1+\mathit{w}|} \mathrm{d} \mathit{u} \dfrac{4 \mathit{w}^2 - (1+\mathit{w}^2-\mathit{u}^2)^2}{\mathit{u}^2 \mathit{w}^4} \nonumber\\
    & \times \left[\Delta_\delta(k \mathit{u}) \Delta_{\mathit{v}}(k \mathit{w}) - \dfrac{\mathit{w}}{\mathit{u}} \Delta_{\delta \mathit{v}}(k \mathit{u}) \Delta_{\delta \mathit{v}}(k \mathit{w}) \right]
    \label{eq:2nd_order}
\end{align}
where $\Delta_\delta$ and $\Delta_\mathit{v}$ are the dimensionless power spectra of the density contrast and velocity potential $\mathit{v}$, while $\Delta_{\delta \mathit{v}}$ their cross-spectrum. The integration variables are defined as 
\begin{align}
    \mathit{w} &= k'/k \\
    \mathit{u} &=\sqrt{1+\mathit{w}^2-2\mathit{w}\cos{\theta}} \\
    \cos{\theta} &= \mathbf{k} \cdot \mathbf{k'}/(k k') \;.
\end{align}
The velocity potential $\mathit{v}$ is defined so that its Laplacian is equal to the velocity divergence, i.e. $\nabla^2 \mathit{v}=\nabla \cdot \mathbf{v}$. In this way, according to the continuity equation~(\ref{eq:continuity}) in linear theory, the power spectrum of the velocity potential is determined by the matter power spectrum, namely $\Delta_\mathit{v} = k^{-4} \Delta_\delta $.
The reader must note that Equation~(\ref{eq:2nd_order}) differs from Equation 30 in \citet{barrera-hinojosaVectorModesLCDM2021} by a factor $a^2$. This is due to the different definition of the vector potential $\mathbf{B}$, already mentioned in Section~\ref{sec:PF_formalism}.

In the plots, we will refer to the linear power spectrum of $\mathbf{B}$, i.e. $\Delta_{\mathbf{B}}^\mathrm{lin}$, as calculated from Equation~(\ref{eq:2nd_order}) with the linear matter power spectrum. 
Instead, for a first comparison with its non-linear analogue, the same integral is computed using the non-linear matter power spectrum from HaloFit \citep{smith2003halofit}, a commonly used semi-analytical model for non-linear growth of structures, and we will refer to it as the non-linear power spectrum of $\mathbf{B}$, namely $\Delta_{\mathbf{B}}^\mathrm{n.l.}$.

\subsubsection{Global solution}\label{sec:globalSol}
Alternatively \WB{to the power spectrum method presented by \citet{bruniComputingGeneralrelativisticEffects2014a} and \citet{thomasFullyNonlinearPostFriedmann2015}}, Equations~(\ref{eq:vector_pot}) and~(\ref{eq:PF_poisson}) can also be globally solved in the Fourier space. \WB{Therefore, contrary to previous analyses, this allows us to reconstruct the solution over the entire volume of the simulation.}
For the scalar potential, the Poisson equation in Fourier space becomes
\begin{equation}\label{eq:Poisson_Fourier}
    \widetilde{\phi}(\mathbf{k}) = - 4 \pi G a^2 \bar{\rho} \dfrac{\widetilde{\delta}(\mathbf{k})}{k^2} \;,
\end{equation}
where $\widetilde{\phi}(\mathbf{k})$ and $\widetilde{\delta}(\mathbf{k})$ are the Fourier transforms of the scalar potential and the density contrast, respectively.
On the other hand, the equation for the gravito-magnetic vector potential becomes
\begin{equation}\label{eq:vectorPot_Fourier}
    \mathbf{k} \times \mathbf{\widetilde{B}(k)} = 16 \pi G a^2 \bar{\rho} \dfrac{\mathbf{k} \times \mathbf{\widetilde{p}(k)}}{k^2} \; ,
\end{equation}
where $\mathbf{\widetilde{B}(k)}$ and $\mathbf{\widetilde{p}(k)}$ are the Fourier transforms of the vector potential and the momentum field, respectively. 
This equation constitutes a system of three equations, of which only two are linearly independent. In order to close the system, the additional constraint $\nabla \cdot \mathbf{B} = 0$ is imposed. This constraint ensures that the vector potential is divergence-free and directly follows from the choice of the Poisson gauge in the post-Friedmann approach (see Section~\ref{sec:PF_formalism}).
Therefore, the solution reads:
\begin{align}
    \widetilde{B_x}(\mathbf{k}) &= \dfrac{16 \pi G a^2 \bar{\rho}}{k^4} \left[ k_y^2 \widetilde{p_x} - k_x k_y \widetilde{p_y} + k_z (k_z \widetilde{p_x} - k_x \widetilde{p_z}) \right] \;, \\
    \widetilde{B_y}(\mathbf{k}) &= \dfrac{16 \pi G a^2 \bar{\rho}}{k^4} \left[ k_x^2 \widetilde{p_y} - k_x k_y \widetilde{p_x} + k_z (k_z \widetilde{p_y} - k_y \widetilde{p_z}) \right] \;, \\
    \widetilde{B_z}(\mathbf{k}) &= \dfrac{16 \pi G a^2 \bar{\rho}}{k^4} \left[ (k_x^2 + k_y^2) \widetilde{p_z} + k_z (k_x \widetilde{p_x} + k_y \widetilde{p_y}) \right] \;.
\end{align}
These equations allow the derivation of $\phi$ and $\mathbf{B}$ globally across the entire simulation volume, taking advantage of periodic boundary conditions. These are defined up to an adding constant, therefore the two potentials are conventionally normalized in the real space so that they have zero mean in the simulation box, i.e. $\langle \phi \rangle = 0$ and $\langle B_i \rangle =0$. This is consistent with the simulation box being a sample of an isotropic FLRW universe.
The Fourier transforms are numerically computed through the FFT algorithm implemented in \textsc{numpy} \citep{harris2020numpy}, where the periodic boundary conditions are automatically incorporated. However, the inverse Fourier transform introduces spurious quantities at the caustic locations, where the velocity field is discontinuous and the derivative is infinite \citep[\WB{see}][\WB{for an extended comparison between spectral and real space derivatives}]{hahnPropertiesCosmicVelocity2015}. \WB{Since such spurious quantities are typically localized at the edges of filaments, these are smoothed out by the $k^2$ factor of the elliptic equation~(\ref{eq:vectorPot_Fourier})}. Therefore, \WB{even though} one should not expect this method to give accurate quantitative information \WB{near caustics}, \WB{we expect a minimal impact on the overall solution of} the gravito-magnetic potential.  

\section{Results} \label{sec:results}

In this section, the main numerical results obtained from the IllustrisTNG dark-matter-only simulations are presented. These include the reconstruction of the source fields and the computation of the vector gravitational potential in the post-Friedmann framework, following the methodology outlined in the previous sections.

\subsection{The source fields}
Although the gravito-magnetic potential is sourced directly by the transverse component of the momentum density field, a careful reconstruction of all the contributing fields—namely the density, velocity, and their derivatives—is essential to properly characterize these quantities in the highly non-linear regime probed by the simulations.

\subsubsection{Matter density}\label{sec:matter_dens}
\begin{figure}
	\includegraphics[width=\columnwidth]{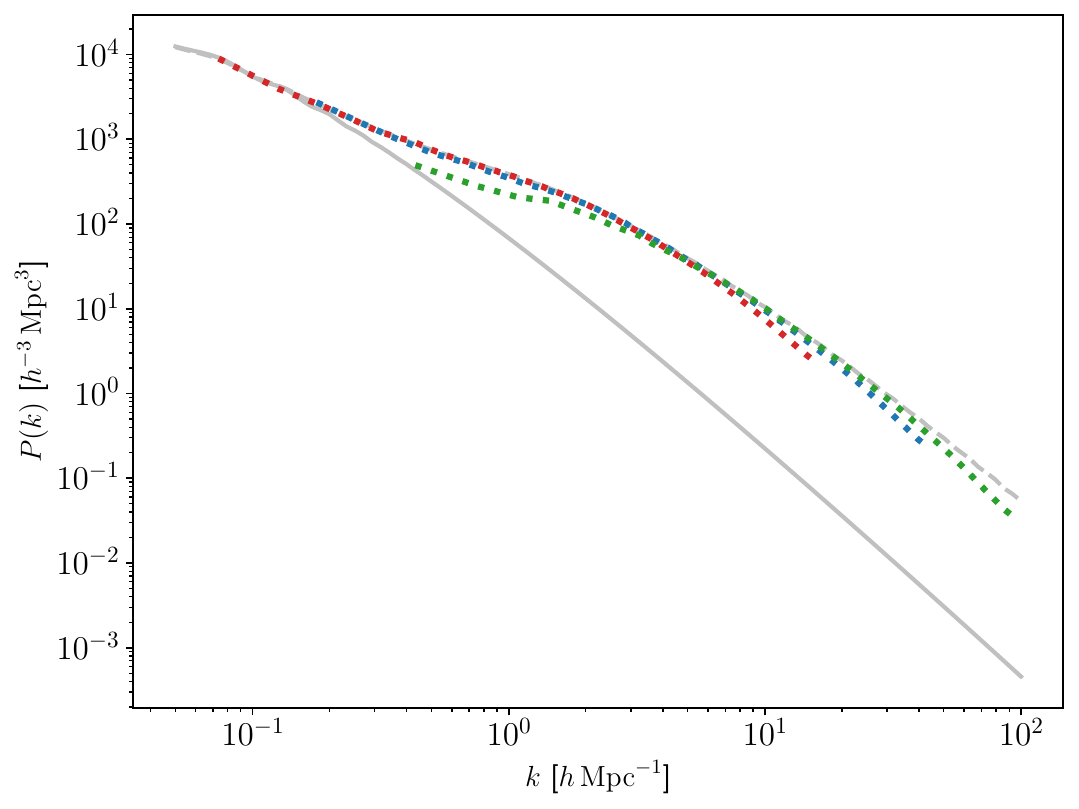}%
    \caption{Matter power spectra for the TNG300-2-Dark (red), TNG100-2-Dark (blue), and TNG50-2-Dark (green) simulations. The linear matter power spectrum is plotted as reference (grey solid line), together with the non-linear HaloFit prediction (grey dashed line).}
    \label{fig:matter_PS}
\end{figure}

The matter power spectra of the three simulations are shown in Figure~\ref{fig:matter_PS}, along with two reference spectra: the linear theory prediction and the non-linear approximation provided by HaloFit \citep{smith2003halofit}.
These are computed using the Code for Anisotropies in the Microwave Background \citep[CAMB,][]{camb}, where the HaloFit is computed from \citet{mead2020Halofit}.
Hereafter, in the plots, we will refer to TNG300-2-Dark, TNG100-2-Dark, and TNG50-2-Dark using the colours red, blue, and green, respectively.

While for the simulations with $75$ and $205 \, h^{-1}$ Mpc, the matter power spectra agree well with the non-linear prediction, for the smaller simulation, i.e. with $35 \, h^{-1}$, the power spectrum results suppressed at large scales close to the simulation box size. The small size of the simulation box, in fact, limits the perturbations on the largest scales which are already in the non-linear regime. 
In fact, the use of the periodic boundary conditions implies that the average density in the simulation box is the same as the average density in the universe; that is to say, perturbations at the scale of the simulation volume (and at larger scales) are ignored \citep{Bagla2009}.
In other words, being IllustriTNG evolved from cosmological initial conditions at $z=127$, the small size of the box is such that there is not enough time for structures with the size close to the simulations box to evolve and reach the full non-linearity.

This is a well-known limitation of cosmological $N$-body simulations performed on such small volumes. For instance, \citet{Bagla2009} suggests that the size of the simulation volume should be chosen so that the amplitude of fluctuations at the box scale (and at larger scales) is ignorable, or equivalently the starting redshift \citep[][]{Power2003}, while \citet{adamekGeneralRelativityCosmic2016} found that in order to get a good agreement between simulation results and perturbation theory, the box should be large enough to contain the matter–radiation equality scale, which is $ k_\mathrm{eq}\approx 0.010 \; h^{-1}$ Mpc \citep{Planck2019}.
In the next sections, we will see how this problem affects the calculation of the gravito-magnetic potential on these scales even more clearly and how to deal with it correctly.

At small scales, the matter power spectra deviate from the HaloFit prediction due to the limited spatial resolution of the DTFE and its window function.
It is important to ensure that our numerical results are robust and independent of the simulation parameters, like the mass resolution, the box size, as well as the grid resolution of the tessellation algorithm. Such convergence checks are presented in Appendix~\ref{app:main_gravito}, where a DTFE grid of $1024^3$ points is found to provide good convergence for the particle-mesh interpolation.
Although not shown here, it has been confirmed that the Cloud-in-Cell assignment performs much worse than DTFE for the same grid resolution, as expected \citep{pueblasGenerationVorticityVelocity2009, thomasFullyNonlinearPostFriedmann2015}.

\subsubsection{Velocity and momentum density}
\begin{figure}
    \centering
    \includegraphics[width=\columnwidth]{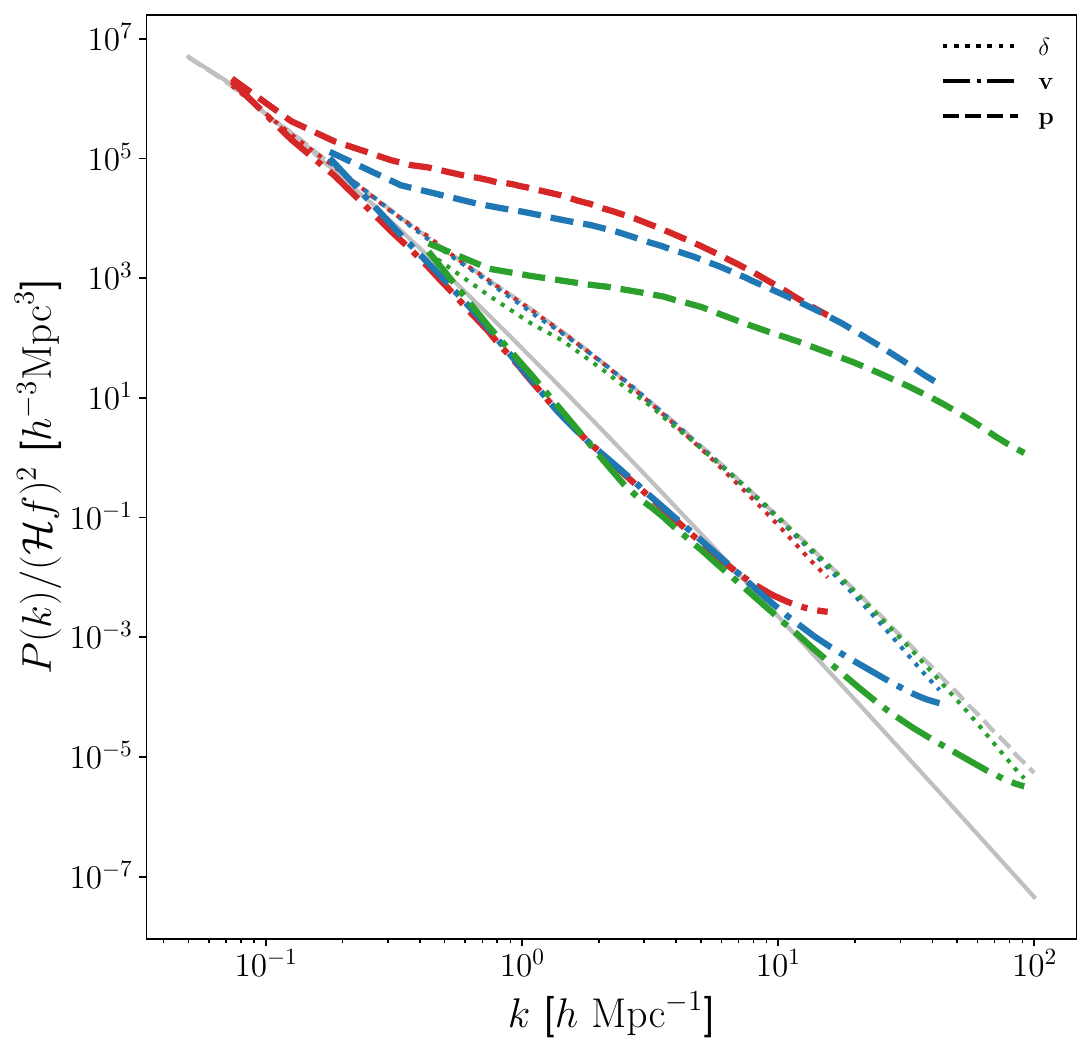}%
    \caption{Power spectra for the velocity and momentum fields, plotted with dot-dashed and dashed lines respectively, for the TNG300-2-Dark (red), TNG100-2-Dark (blue), and TNG50-2-Dark (green) simulations, both normalized by $(\mathcal{H} f)^2$ . The linear matter power spectrum is plotted as reference (grey solid line), together with the matter power spectra of the three simulations (coloured dotted lines). All matter power spectra are normalized by $k^2$.}
    \label{fig:velocityMomentum}
\end{figure}

Figure~\ref{fig:velocityMomentum} shows the power spectra of the velocity and momentum density fields, on top of Figure~\ref{fig:matter_PS}.
Here, the velocity and momentum power spectra are normalised by the factor $(\mathcal{H} f)^2$, where $\mathcal{H}$ is the conformal Hubble parameter, i.e. $\mathcal{H} = a H$, and $f$ is the logarithmic derivative of the linear growth factor, i.e. $f = d \ln{D} / d \ln{a}$. For the $\Lambda$CDM cosmology adopted here, following \citet{barrera-hinojosaVectorModesLCDM2021}, it is assumed to be \citep{Linder}
\begin{equation}
    f = \Omega_{m}(a)^{6/11} \; ,\text{ with } \; \Omega_{m}(a) = \dfrac{\Omega_{m}H_0^2}{a^3 H^2} \,.
\end{equation}
In this way, up to a factor $k^2$, the velocity and momentum power spectra share the same units of the matter power spectrum in the linear regime, where the continuity equation 
\begin{equation}\label{eq:continuity}
    \mathcal{H} f \delta = - \nabla \cdot \mathbf{v}
\end{equation}
is expected to hold.

At large scales, the power spectrum of the momentum field $\mathbf{p}$ \WB{matches} the power spectrum of the velocity field (Figure~\ref{fig:velocityMomentum}). 
This is expected in linear theory since the term $\delta \mathbf{v}$ behaves as a second-order perturbation, while at smaller scales, this term becomes relevant as a result of the non-linear evolution.

As already visible in the matter power spectra, the non-linear regime for the smaller simulation onsets at smaller scales than expected. 
The power spectrum of the momentum field, in fact, turns out to be 
systematically suppressed as the box size of the simulation decreases.
This feature is much more apparent than in the matter power spectrum, whereas it is completely absent in the velocity power spectrum. This indicates that the lack of power due to the finite size of the simulation box affects second-order non-linear terms to a greater extent, as a consequence of non-linearity being reached at smaller scales. 
This is the first time the size of the simulation box is clearly shown to have a direct impact on the momentum distribution in cosmological simulations.

\subsubsection{Vorticity and divergence}
\begin{figure}
    \centering
    \includegraphics[width=\linewidth]{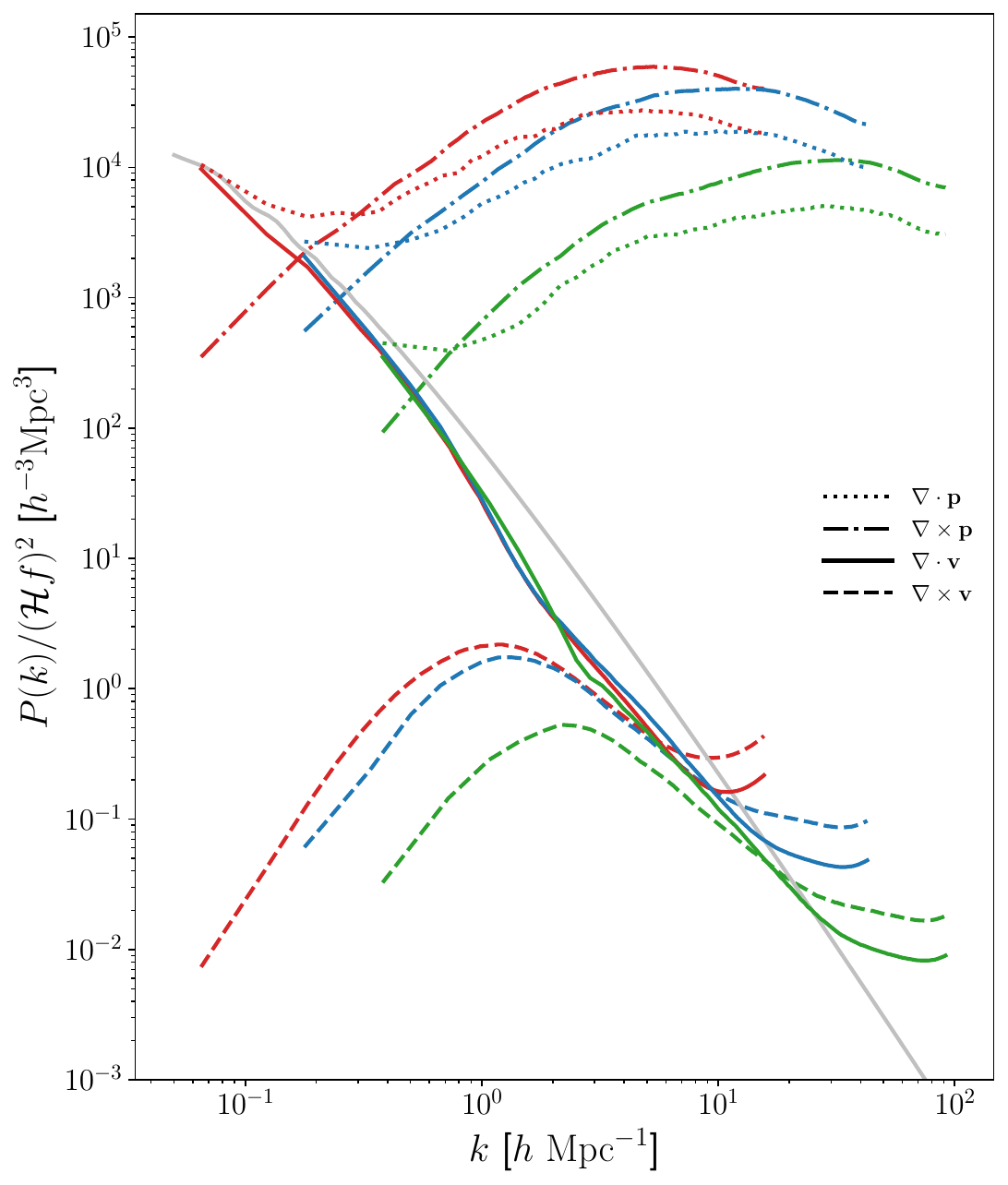}
    \caption{Power spectra of the gradients of the velocity and momentum density fields for the TNG300-2-Dark (red), TNG100-2-Dark (blue), and TNG50-2-Dark (green) simulations. The divergence and vorticity of the velocity are represented as solid and dashed lines, respectively, while the divergence and vorticity of the momentum are shown with dotted and dash-dotted lines. The linear matter power spectrum is plotted as a reference (grey solid line).}
    \label{fig:velMom_PS}
\end{figure}

In Figure~\ref{fig:velMom_PS}, the power spectra of the divergence and vorticity of both the velocity and momentum fields are shown.
As expected, the velocity divergence shows good agreement with linear theory at large scales, while exhibiting a dip at around $k\approx1.5 \; h$ Mpc$^{-1}$ for the 205 and 75 $ h^{-1}$ Mpc simulations, and at around $k\approx3 \; h$ Mpc$^{-1}$ for the 35 $ h^{-1}$ Mpc simulation \citep{barrera-hinojosaVectorModesLCDM2021}.

On the contrary, the momentum divergence deviates from linear theory except for the fundamental mode, underlying that the structure formation process has reached the deep non-linear regime almost at all scales. Again here, it is worth noting that, as the simulation box decreases in size, the scale at which agreement with linear theory occurs also diminishes, with an increase in the power suppression.

For the velocity vorticity, the power spectrum exhibits a peak. The peak is located at around the same scale the divergence shows the dip, i.e. $k\approx1.2 \; h$ Mpc$^{-1}$, consistently with \citet{barrera-hinojosaVectorModesLCDM2021} and \citet{jelic-cizmekGenerationVorticityCosmological2018}, with the smaller simulation showing a peak slightly shifted towards smaller scales, ie $k\approx2.5 \; h$ Mpc$^{-1}$. \WB{The dip and the peak in the velocity divergence and vorticity power spectra, respectively, being at the similar position, have been interpreted as the consequence of shell crossing occurring around that scales, where the angular momentum can be large enough to inhibit structure growth as particles are forced into orbits rather than falling inward \citep{barrera-hinojosaGRAMSESNewRoute2020, jelic-cizmekGenerationVorticityCosmological2018}.}

Beyond the peak, the power spectrum of the velocity vorticity becomes comparable to that of the divergence, dominating at very small scales. In contrast, in the case of the momentum, the power spectrum of the vorticity already exceeds the power spectrum of the divergence at scales close to the simulation box size.

This suggests that as non-linear structures form, a fraction of the power in the velocity divergence is transferred to vorticity, meaning that the conservation of angular momentum prevents further infall of matter and forces the particles to rotate around structures \citep{jelic-cizmekGenerationVorticityCosmological2018}.

In the context of cold dark matter, the pressure-less perfect fluid is described in the single-stream regime. In this approximation, at each point in space, the velocity field is uniquely defined, that is to say dark matter particles move coherently without crossing each other's trajectories. Under this assumption, vorticity is expected to vanish at all orders of perturbation theory \citep{luCosmologicalBackgroundVector2009}.

\begin{figure}
    \includegraphics[width=\columnwidth]{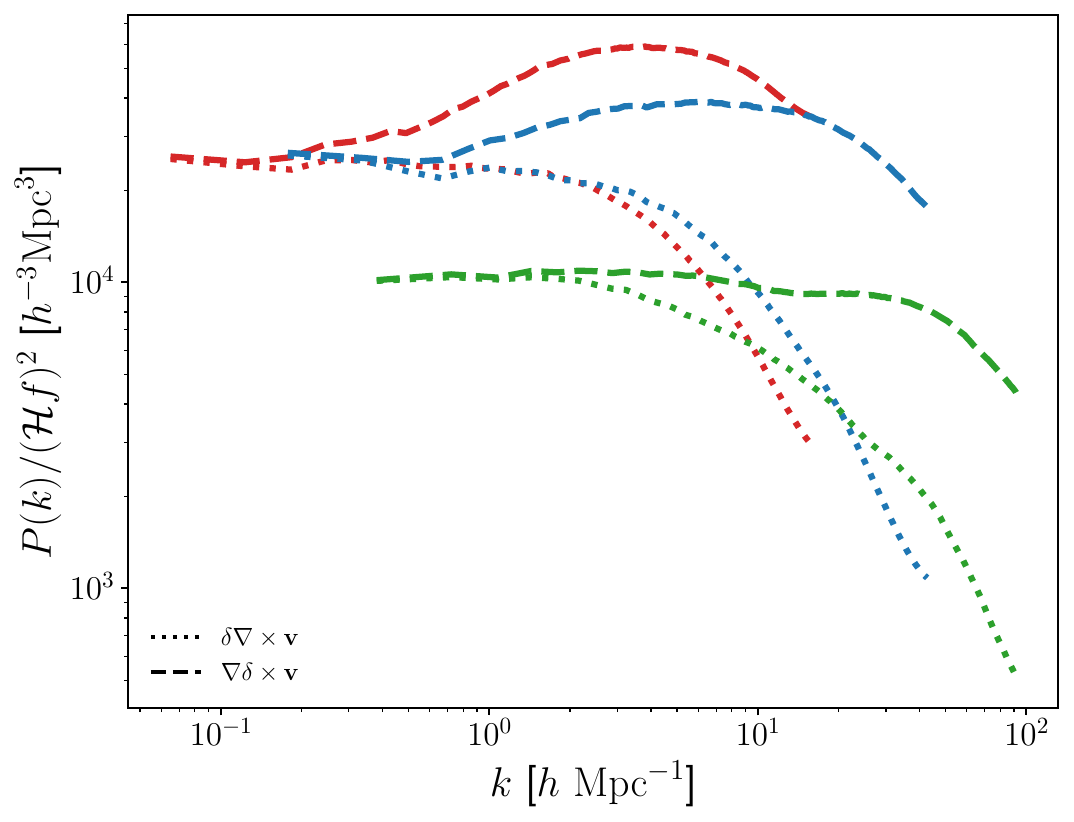}
    \caption{Power spectra of the main sources of momentum vorticity, i.e. $\delta \nabla \times \mathbf{v}$ (dotted lines) and $\nabla \delta \times \mathbf{v}$ (dashed lines), for the TNG300-2-Dark (red), TNG100-2-Dark (blue), and TNG50-2-Dark (green) simulations.}
    \label{fig:curlComponents_PS}
\end{figure}
However, in $N$-body simulations, as structures grow and gravitational collapse becomes non-linear, the single-stream approximation breaks down due to orbit crossing. In this multi-stream regime, particles with different velocities can coexist at the same spatial location, leading to the generation of vorticity \citep{pueblasGenerationVorticityVelocity2009}.

The emergence of vorticity in simulations is strongly dependent on the mass resolution \citep{jelic-cizmekGenerationVorticityCosmological2018}. If the resolution is too low, the fine details of orbit crossing may not be well captured, leading to an underestimation of vorticity. Conversely, a higher resolution allows for a more accurate representation of the small-scale velocity field, and a better resolution of the transition to the multi-stream regime.
In Appendix~\ref{app:mass_res}, three different mass resolutions for the simulation TNG50-Dark are tested. Although, at small scales, a coarser resolution attributes more spurious power, the shape and amplitude of the vorticity power spectrum are preserved. For the momentum vorticity, such dependence on the mass resolution is weaker, as well as for the divergence of both fields.

Additionally, the spatial resolution also plays a crucial role, as a finer grid enables a more detailed description of the multi-stream velocity field in collapsing regions.
Appendix~\ref{app:grid_res} shows that a grid resolution of $N_{\mathrm{grid}}=1024^3$ ensures a good convergence.

The momentum vorticity can also be split into three contributions:
\begin{equation}\label{eq:components}
    \nabla \times \mathbf{p} =  \nabla \times \mathbf{v} +  \delta \nabla \times \mathbf{v} + \nabla \delta \times \mathbf{v} \; .
\end{equation}
The power spectra of the last two terms are plotted in Figure~\ref{fig:curlComponents_PS}. In agreement with several previous works \citep{bruniComputingGeneralrelativisticEffects2014a, thomasFullyNonlinearPostFriedmann2015, barrera-hinojosaVectorModesLCDM2021}, the main component that sources the momentum vorticity is $\nabla \delta \times \mathbf{v}$, while the velocity vorticity alone is subdominant, as expected being $\delta\gg 1$.

\subsection{The gravito-magnetic potential}\label{sec:gravito-magneticPot}
\begin{figure}
    \centering
    \includegraphics[width=\linewidth]{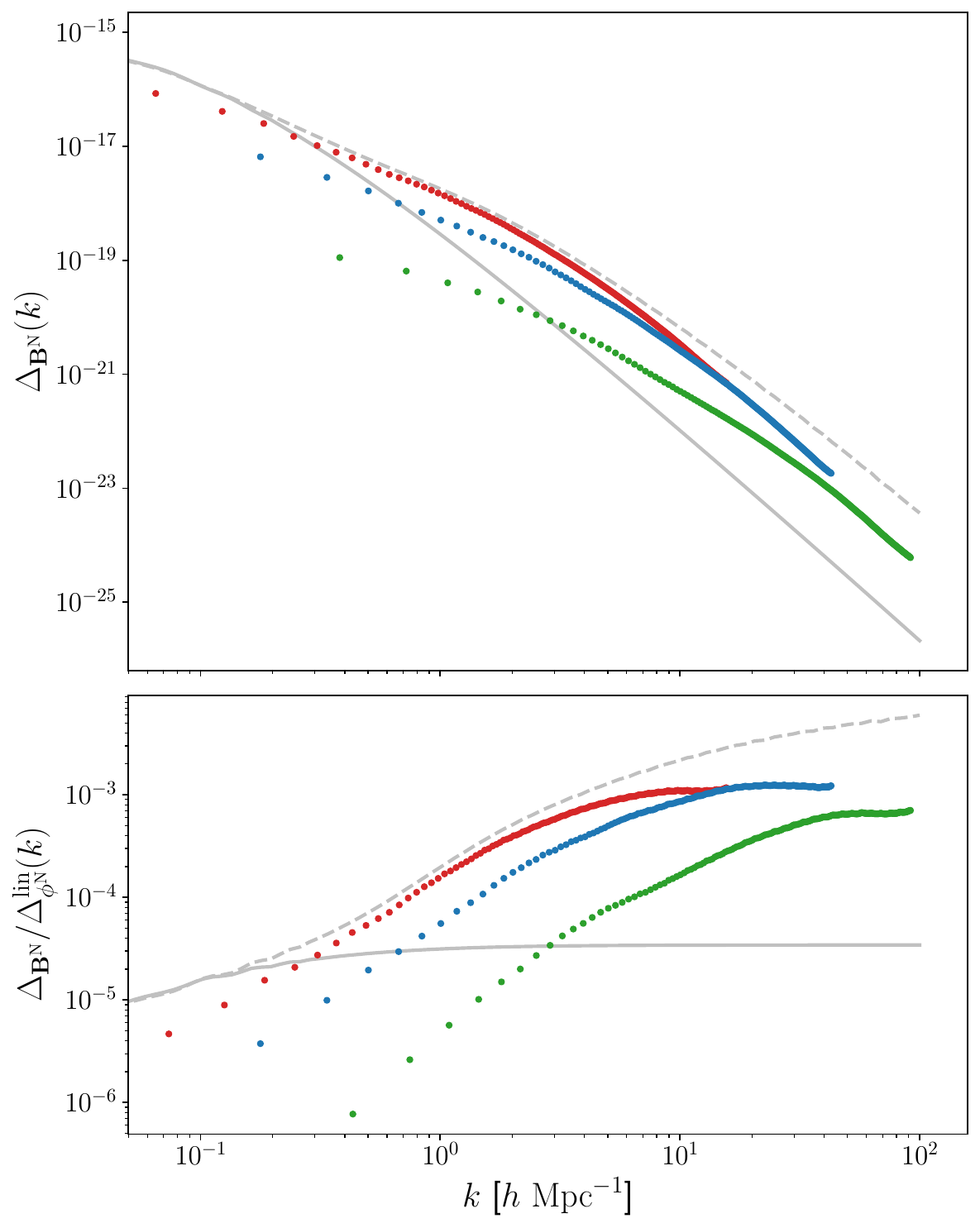}
    \caption{Top: power spectra of the vector potential for the TNG300-2-Dark (red), TNG100-2-Dark (blue), and TNG50-2-Dark (green) simulations. The power spectrum of the vector potential predicted with the second-order perturbation theory from Equation~(\ref{eq:2nd_order}) is shown as reference (grey solid line), along with the non-linear prediction computed from the same equation with the non-linear matter power spectrum of HaloFit (grey dashed line). Bottom: same plot but normalised for the power spectrum of the linear scalar potential $\Delta_{\phi^{\mathrm{N}}}^\mathrm{lin}$.}
    \label{fig:potentials_PS}
\end{figure}

The power spectra of the gravito-magnetic vector potential, computed from Equation~(\ref{eq:vector_pot_PS}) are plotted in Figure~\ref{fig:potentials_PS}. 
Two reference power spectra are plotted, namely the 2nd-order perturbation theory prediction $\Delta_{\mathbf{B}}^\mathrm{PT}$ and the non-linear analogue $\Delta_{\mathbf{B}}^\mathrm{n.l.}$, which are computed from Equation~(\ref{eq:2nd_order}) using linear and non-linear matter power spectra, respectively (see Section~\ref{sec:Fourier_analysis}). The reader must note that $\Delta_{\mathbf{B}}^\mathrm{n.l.}$ represent just a first approximation of the fully non-linear vector potential, since Equation~(\ref{eq:2nd_order}) is obtained from the 2nd-order perturbation theory and thus cannot account for the full non-linearity.

Figure~\ref{fig:potentials_PS} also shows the ratio of the vector potential power spectra with respect to the power spectrum of the linear scalar potential $\Delta_{\phi}^\mathrm{lin}$ as a function of the scale.
The power spectrum of the vector potential becomes more than an order of magnitude larger than that predicted by second-order PT towards smaller scales. On the contrary, it remains smaller than the non-linear expectation over the full range of scales, with the largest simulation, i.e. TNG300-2-Dark, providing the closest match. This means that the non-linear prediction $\Delta_{\mathbf{B}}^\mathrm{n.l.}$ actually gives an accurate estimate of the vector potential power spectrum in the non-linear regime.

Again here, the vector power spectrum shows even more clearly the power suppression due to the missing perturbation modes larger than the simulation box, as pointed out in the previous sections for the source fields.
Although \citet{thomasFullyNonlinearPostFriedmann2015} concluded that there is no evidence for a systematic dependence of the vector potential spectrum on box size for boxes smaller than $200\; h^{-1}$ Mpc (with the smallest considered being of $80\; h^{-1}$ Mpc), a clear box size effect seems already visible in all their plots \citep[][, see in particular Figure RB6 in Supplementary Material]{thomasFullyNonlinearPostFriedmann2015}. 

\begin{figure}
    \centering
    \includegraphics[width=\linewidth]{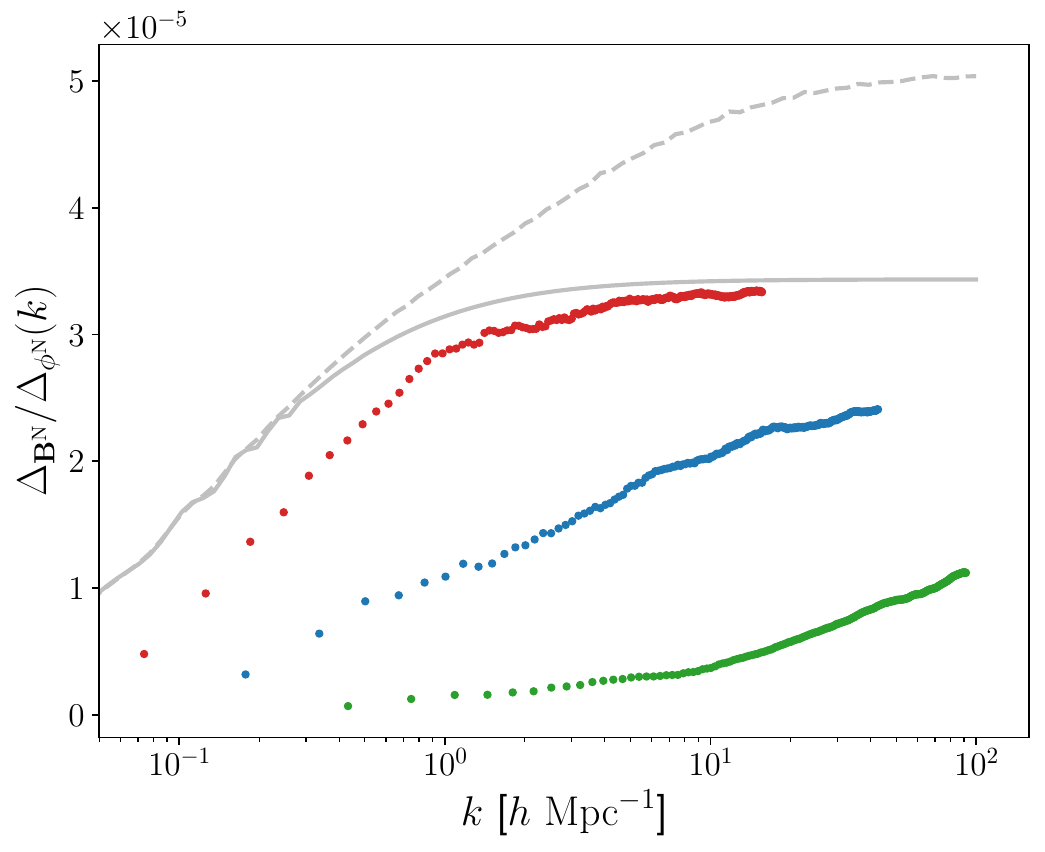}
    \caption{The ratio between the power spectra of the fully non-linear scalar and vector potentials, both derived from the simulation, for TNG300-2-Dark (red), TNG100-2-Dark (blue), and TNG50-2-Dark (green). The grey solid line represents the ratio between the second-order perturbation theory $\Delta_{\mathbf{B}}^\mathrm{PT}$ and the linear theory for the scalar potential $\Delta_{\phi}^\mathrm{lin}$, while the grey dashed line stands for the ratio between the non-linear analogues, namely $\Delta_{\mathbf{B}}^\mathrm{n.l.}$ and $\Delta_{\phi}^\mathrm{n.l.}$.}
    \label{fig:ratio_No_correction}
\end{figure}

In Figure~\ref{fig:ratio_No_correction}, the ratio between the fully non-linear scalar and vector potentials obtained from the simulations is shown.
For TNG300 and TNG100, this ratio turns out to be around $1\text{--}3 \times 10^{-5}$, in agreement with previous results for $z=0$ \citep{bruniComputingGeneralrelativisticEffects2014a, thomasGravityNonlinearScales2015, adamekGeneralRelativityCosmic2016, barrera-hinojosaVectorModesLCDM2021}, as well as with 2nd-order PT \citep{luCosmologicalBackgroundVector2009}. However, for the smallest simulation, TNG50, the ratio drops down of an order of magnitude due to the box size effect. 

The integral~(\ref{eq:2nd_order}) couples different modes of the matter density and velocity potential. This means that at a single scale the vector power spectrum receives contributions from different wave numbers.
Therefore, in order to compare the vector power spectra obtained from the simulations with the second-order perturbation theory in a consistent way, when performing the integration, one would, in principle, need to set both large-scale and small-scale cut-offs to account for the limited size of the simulation box and the finite sampling (the latter comprising both the limited mass and spatial resolutions of the simulation), respectively.
\citet{parkbaghyeonImpactBaryonicPhysics2018} developed a procedure to correct the momentum power spectrum for the missing power in the context of the \CB{kSZ} effect prediction, which, to the best of our knowledge, was the first attempt of this kind in the literature. In the present work, in order to retain the perturbation theory prediction as a reference, we apply this procedure to the power spectrum of the vector potential obtained from the simulations. The current paper is also the first one applying the \citet{parkbaghyeonImpactBaryonicPhysics2018} method to correct the missing power in the gravito-magnetic potential power spectrum.

To this end, the missing power is computed from Equation~(\ref{eq:2nd_order}), which is evaluated integrating for scales larger than the simulation box, i.e. for $k' < k_\mathrm{box} = 2 \pi / L$.
Here, the input matter power spectrum is assumed to be the non-linear estimation of HaloFit provided by CAMB, which is a good approximation for IllustrisTNG, as seen in Section~\ref{sec:matter_dens}. Note that the matter power spectrum derived from the simulation cannot be used here, as it is already suppressed at scales close to the box size, particularly for TNG50, and therefore it would underestimate the missing power.

The resulting power spectrum containing the missing power is then added to the measured power spectrum of the vector potential. In the linear regime, this corrected power spectrum should match the reference second-order perturbation theory prediction $\Delta_{\mathbf{B}}^\mathrm{PT}$. Therefore, we checked that the corrected power spectrum actually matches the full power spectrum at $z=20$ (the highest redshift for which the simulation snapshot is available), where the second-order perturbation theory 
is still expected to hold. Only for the $75\; h^{-1}$ Mpc simulation box, this correction overestimates the reference perturbation theory prediction by $5$\%, meaning that more power is added than necessary. 
The same result was found by \citet{parkbaghyeonImpactBaryonicPhysics2018} for the Illustris simulation of $75\; h^{-1}$ Mpc, where they calibrated the upper integration limit with a fudge factor $f_c$ to compensate for this overshooting, so that $k' < f_c k_\mathrm{box}$. 
Performing the same calibration, it is found that $f_c = 1$ for the $205$ and $35$ $h^{-1}$ Mpc boxes, indicating no overestimation, whereas a value of $f = 0.9$ is obtained for TNG100-2-Dark.

\begin{figure}
    \includegraphics[width=\columnwidth]{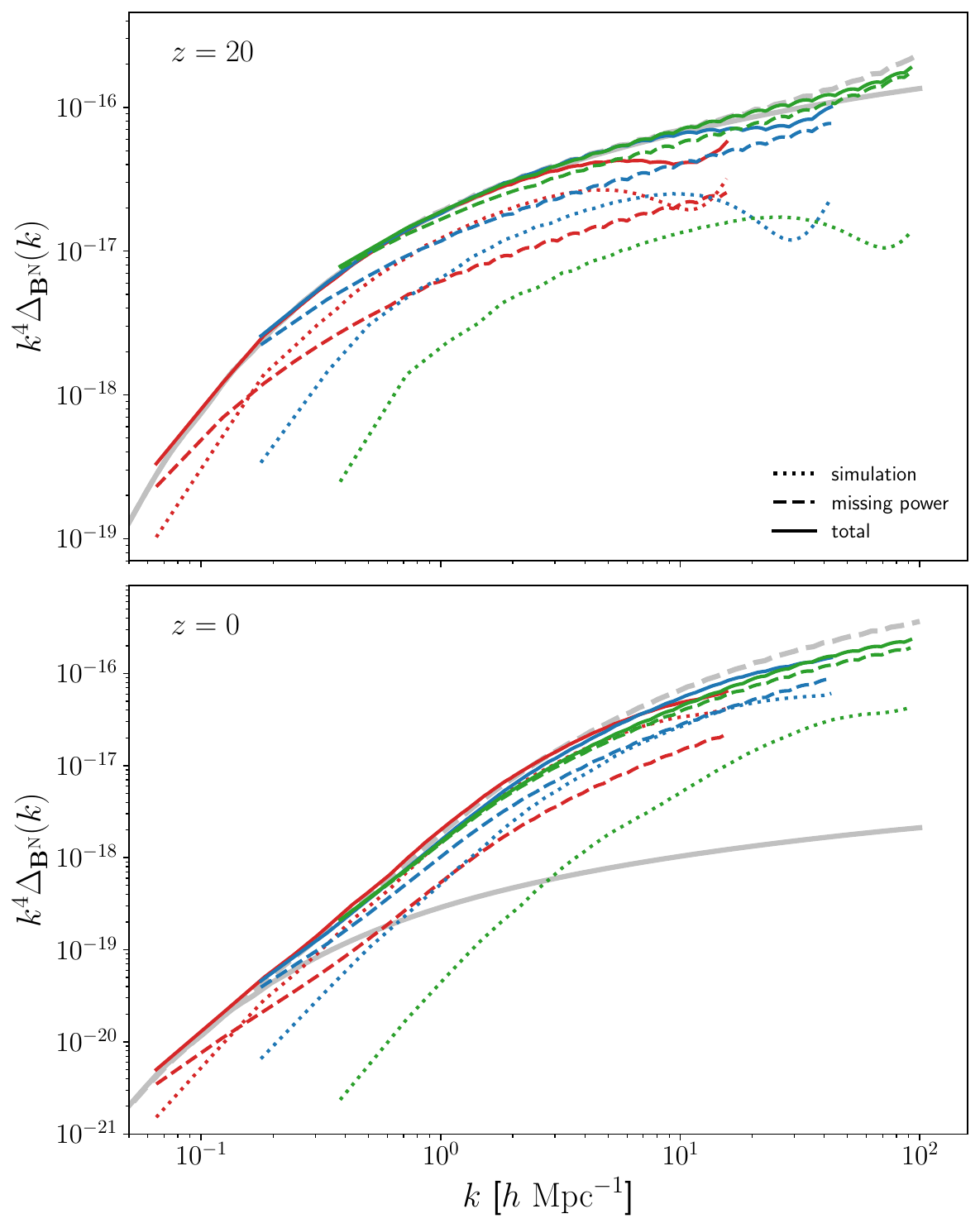}
    \caption{Power spectra of the gravito-magnetic potential measured from the simulations (dotted lines), missing power computed from Equation~(\ref{eq:2nd_order}) with the non-linear matter power spectrum (dashed lines) and their sum (solid lines) for the TNG300-2-Dark (red), TNG100-2-Dark (blue), and TNG50-2-Dark (green) simulations. The $y$-axis is scaled by $k^4$ to facilitate visualization. Top: relative to redshift $z=20$, adopted for calibrating the missing power. Bottom: relative to redshift $z=0$.}
    \label{fig:missing_power}
\end{figure}

Figure~\ref{fig:missing_power} demonstrates how the correction for the missing power operates.
The top panel shows the excellent agreement between the 2nd order PT and the result at $z=20$ for the three simulations, confirming the validity of the correction. 
For TNG100 and TNG50, the missing power exceeds the power spectrum derived from the simulation over the entire range of scales, while for TNG300 this is the case only at very large scales close to the box size.
By looking at $\Delta_{\mathbf{B}}^\mathrm{PT}$ and $\Delta_{\mathbf{B}}^\mathrm{n.l.}$, one can note that at this epoch the perturbations have been evolving linearly almost at all scales, with the non-linearity starting to onset at scales $k \gtrapprox 10\; h^{-1}$ Mpc. This is to say that most of the missing power comes from perturbations in the linear regime that the simulation box has cut off. Such perturbation modes are relevant for sourcing the non-linearity, and thus the vector potential, already at $z=20$.

The same correction for $z=0$ is shown in the bottom panel. As for $z=20$, the missing power is larger than the power measured from the simulation, but here, the latter has become larger than the second-order PT prediction at sufficiently small scales. 
In other words, the small-scale perturbations had enough time to evolve in the fully non-linear regime and consequently source the non-linear vector potential. 

\begin{figure}
    \centering
    \includegraphics[width=\linewidth]{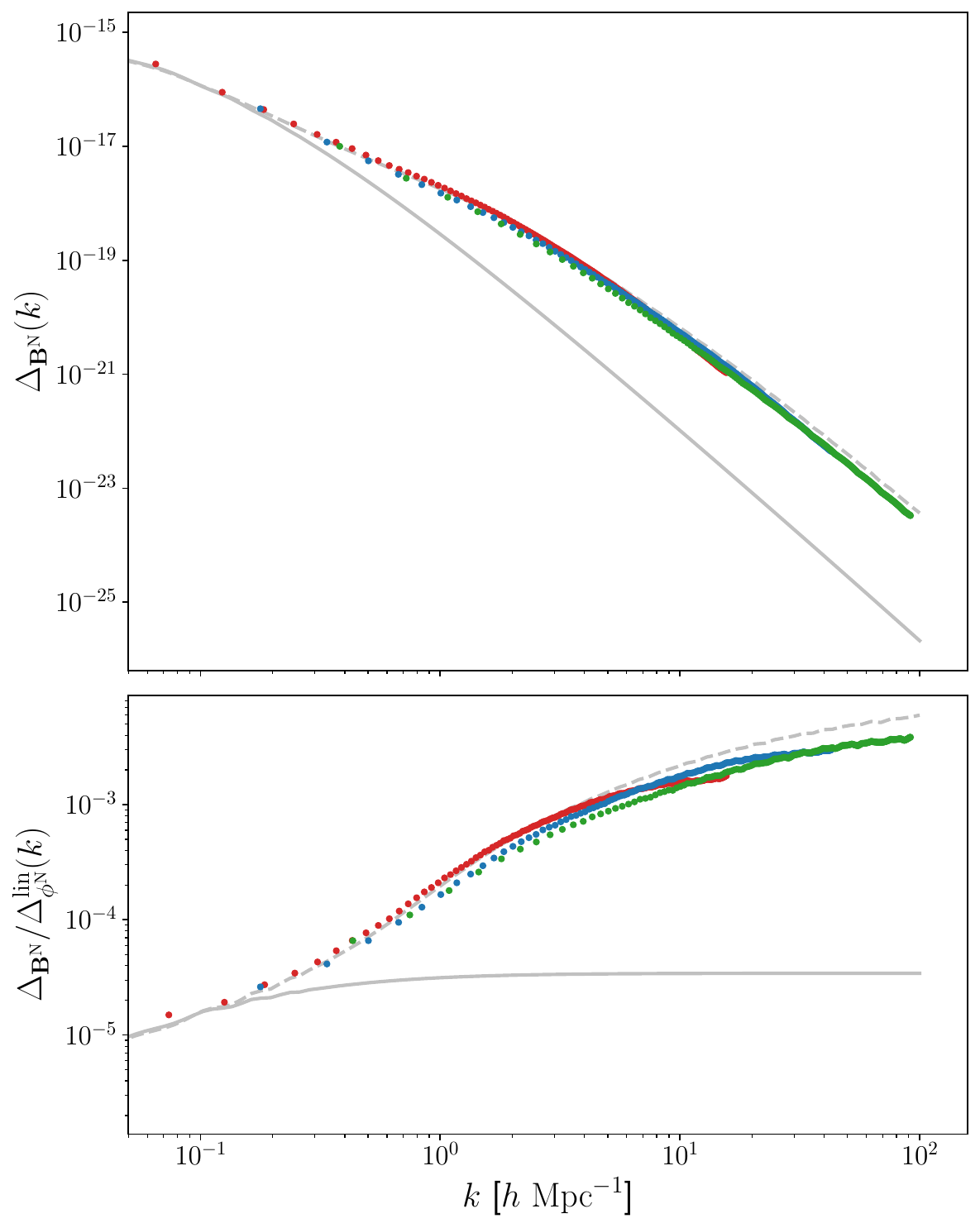}
    \caption{Same as Figure~\ref{fig:potentials_PS}, but the power spectra of the vector potential have been corrected for the missing power. Details about this procedure are given in the main text.}
    \label{fig:potentials_psCorrected}
\end{figure}

The power spectrum of the vector potential corrected for the missing power is shown in Figure~\ref{fig:potentials_psCorrected}.
\begin{figure}
    \centering
    \includegraphics[width=\linewidth]{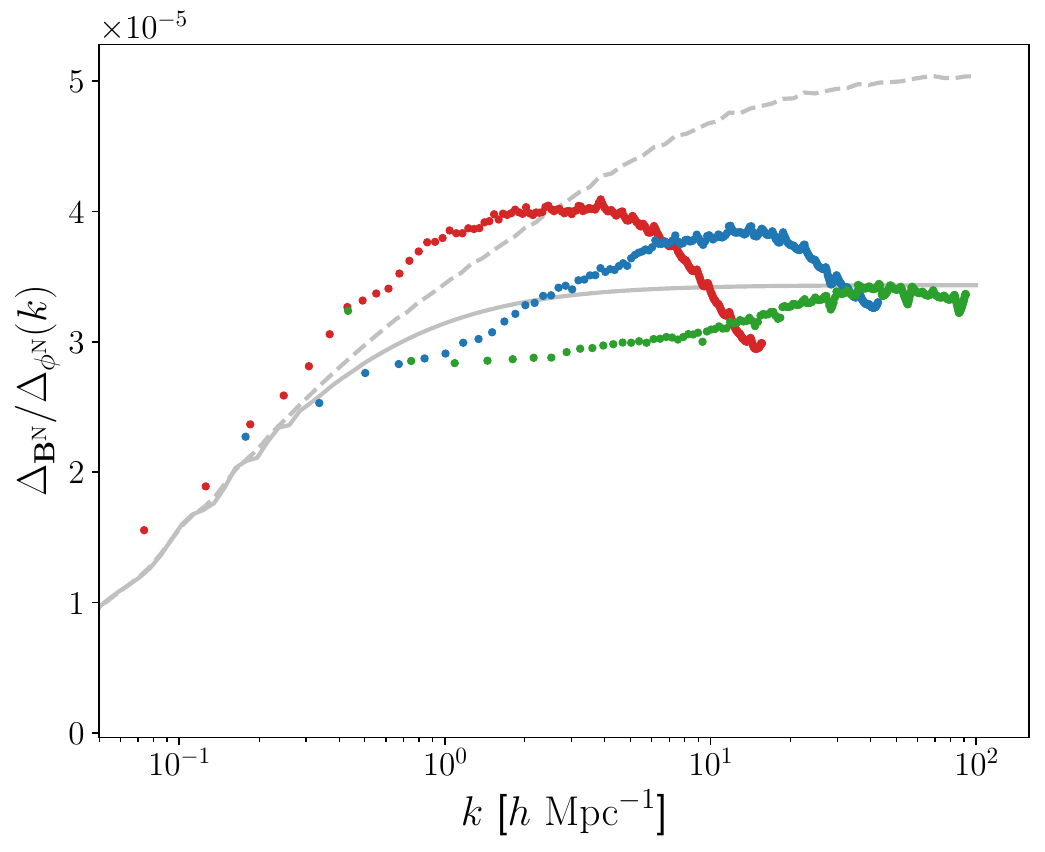}
    \caption{Same as Figure~\ref{fig:ratio_No_correction}, but the power spectra of the vector potential have been corrected for the missing power. Here, the corrected power spectrum of the vector potential is divided by $\Delta_{\phi}^{\mathrm{n.l.}}$, which is the non-linear prediction for the scalar potential computed using the HaloFit matter power spectrum, and not by the non-linear scalar potential measured from the simulation as in Figure~\ref{fig:ratio_No_correction}. This is because the first is used to compute the missing power, while the latter is already suppressed at scales close to the box size (see Section~\ref{sec:matter_dens}).}
    \label{fig:ratio_Corrected}
\end{figure}
The corrected power spectrum for TNG300 now matches the second-order perturbation theory prediction at very large scales, while following $\Delta_{\mathbf{B}}^\mathrm{n.l.}$ up to $k\approx4 \; h$ Mpc$^{-1}$. As already stated, this ensures the consistency of the procedure adopted and also indicates that the non-linear prediction for $\mathbf{B}$ is a good approximation for the power spectrum of the vector potential for simulation boxes of this size.
For the smaller boxes of 75 and 35 $ h^{-1}$ Mpc, the corrected power spectra now align with that of TNG300, although they still remain slightly lower. Overall, we must say that the agreement among the various simulations is quite good, considering the significant suppression caused by the box size effect, which affects around an order of magnitude the results for TNG50-2-Dark. 

The corrected ratio between the power spectra of the two gravitational potentials is shown in Figure~\ref{fig:ratio_Corrected}. For all three simulations, the ratio lies in the range $2$--$4 \times 10^{-5}$, in agreement with the previous studies. \citep{bruniComputingGeneralrelativisticEffects2014a, thomasFullyNonlinearPostFriedmann2015, barrera-hinojosaVectorModesLCDM2021}.
Note that, here, the corrected power spectrum of the vector potential is divided by $\Delta_{\phi}^{\mathrm{n.l.}}$, which is the non-linear prediction for the scalar potential computed using the HaloFit matter power spectrum, and not by the non-linear scalar potential measured from the simulation as in Figure~\ref{fig:ratio_No_correction}. In fact, as previously mentioned, the matter power spectrum obtained directly from the simulation cannot be adopted to evaluate the missing power since it is already suppressed at large scales (see Figure~\ref{fig:matter_PS}).
The reader must also note that the scale of the plot is linear, therefore, an exact match between the linear and non-linear predictions is not expected, given the variability of different realizations of the simulation as well as possible offsets stemming from slightly different configurations in simulations and models.

Moreover, the ratio in Figure~\ref{fig:ratio_Corrected} decreases at small scales. This occurs because the HaloFit matter power spectrum, used to compute $\Delta_{\phi}^{\mathrm{n.l.}}$, becomes larger at the smallest scales compared to the matter power spectrum obtained from the simulation via DTFE. As already mentioned, this discrepancy arises from the limited spatial resolution of the tessellation algorithm that interpolates the density field.
Therefore, a more refined approach to the missing mass correction may be required in the future to overcome this limitation and provide a more accurate estimate of the ratio between the scalar and vector potentials beyond these scales.

To conclude, the ratio of the vector and scalar power spectra remains of $\mathcal{O}(10^{-5})$ for the three simulations at $z=0$, confirming both the results found by \citep{bruniComputingGeneralrelativisticEffects2014a} and \citet{thomasFullyNonlinearPostFriedmann2015} from Newtonian $N$-body simulations, and those of \citep{barrera-hinojosaGRAMSESNewRoute2020} and \citep{adamekGeneralRelativityCosmic2016} from GR simulations.
This means that, up to the scales investigated in this study, i.e. $k\lessapprox 92 \; h\,$Mpc$^{-1}$, which at the present time correspond to a physical scale of $\approx 100$ kpc, the fully non-linear gravito-magnetic potential is around $10^{-2}\text{--}10^{-3}$ times the scalar potential, namely that the frame-dragging is potentially an order $1\text{--}0.1\%$ effect also on the scale of galaxies.

\subsection{Representation of the fields}
In this section, we provide a visual representation of the source fields extracted from the simulations through the DTFE algorithm and the gravitational potential reconstructed with Fast Fourier Transforms as described in Section~\ref{sec:globalSol}.

\begin{figure*}
    \centering
    \scalebox{.9}{ 
    \begin{minipage}{\textwidth}
    \includegraphics[width=0.49\textwidth]{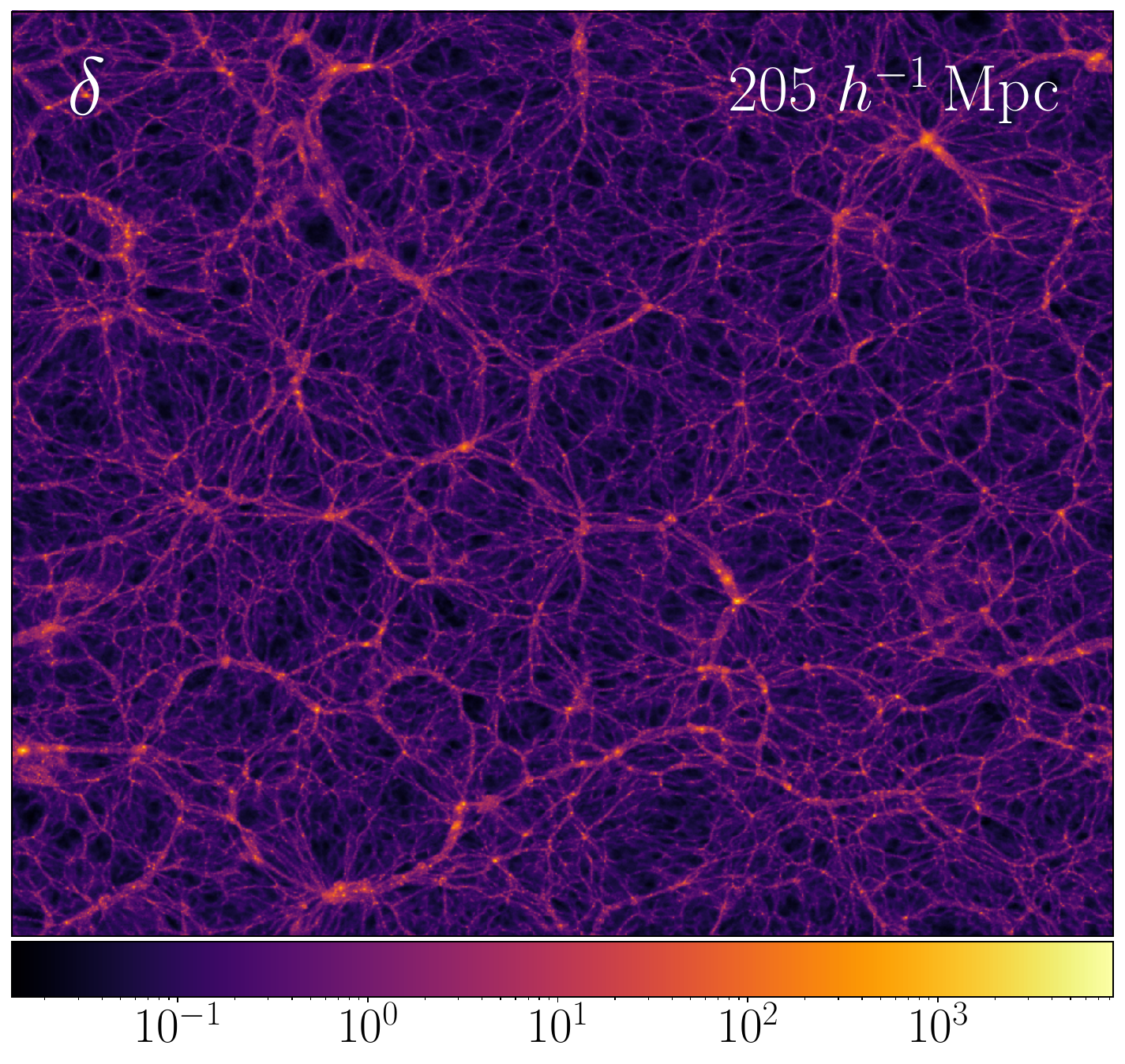}
    \hfill
    \includegraphics[width=0.49\textwidth]{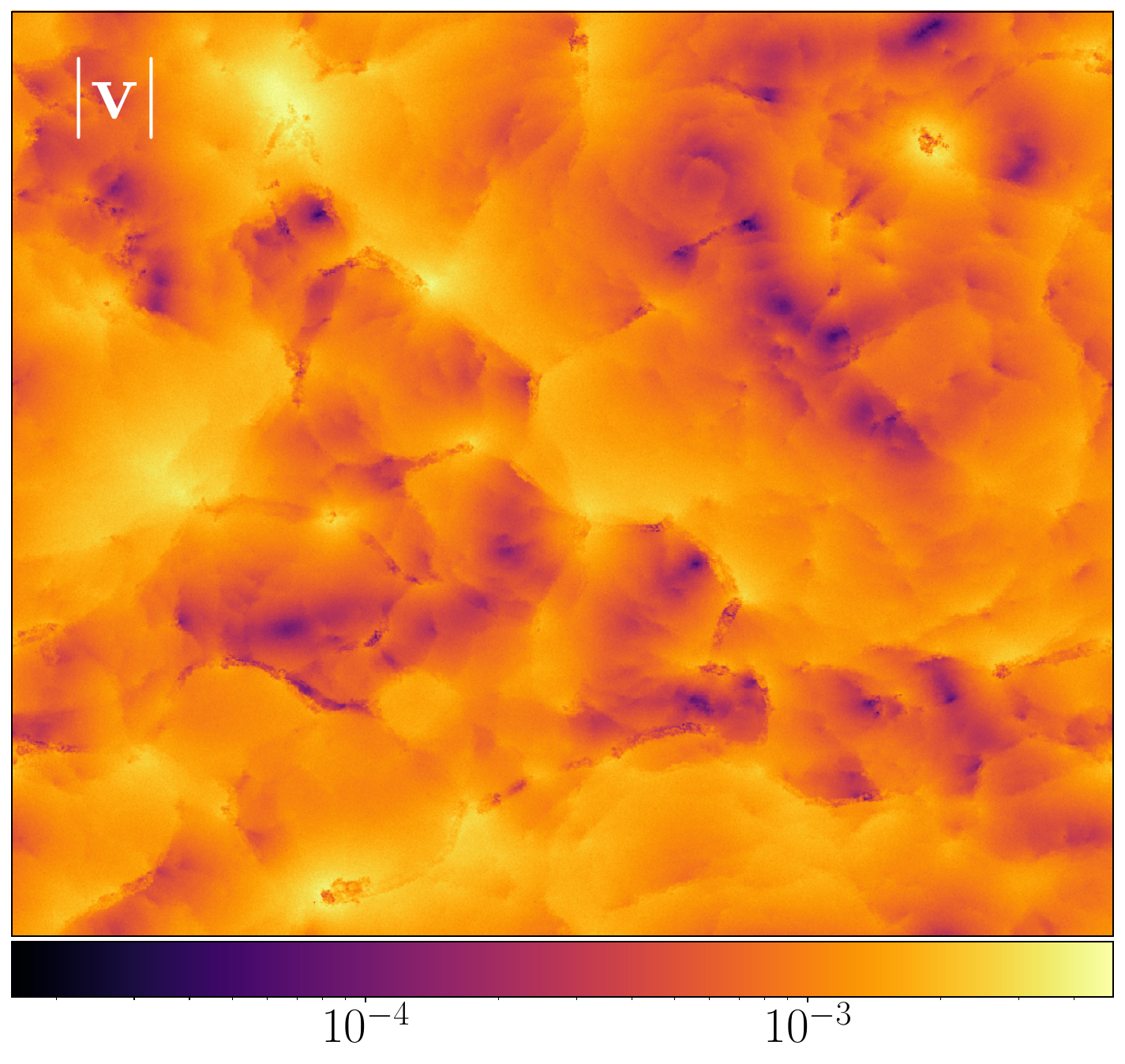}
    \vfill
    \includegraphics[width=0.49\textwidth]{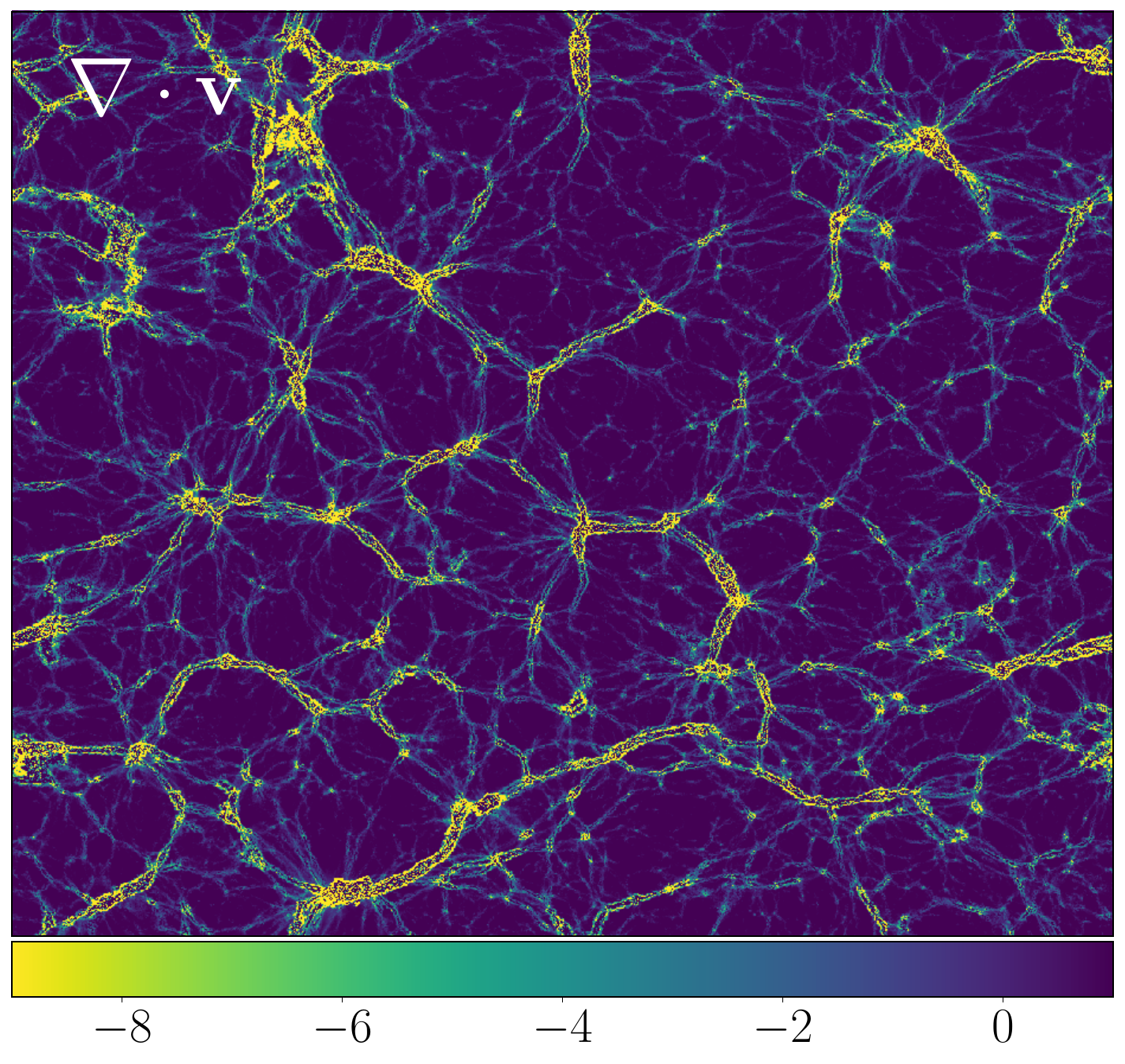}
    \hfill
    \includegraphics[width=0.49\textwidth]{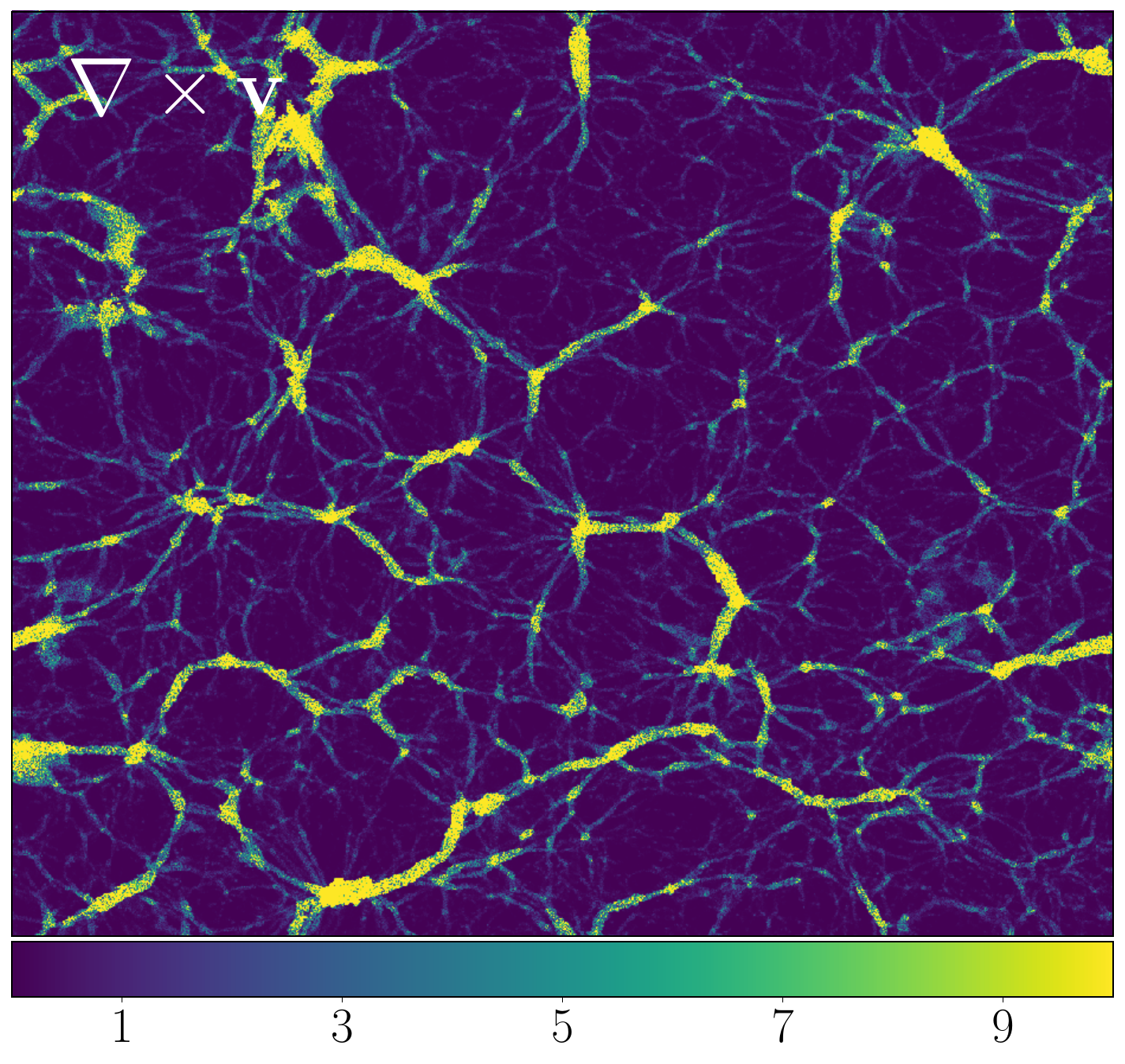}
    \vfill
    \includegraphics[width=0.49\textwidth]{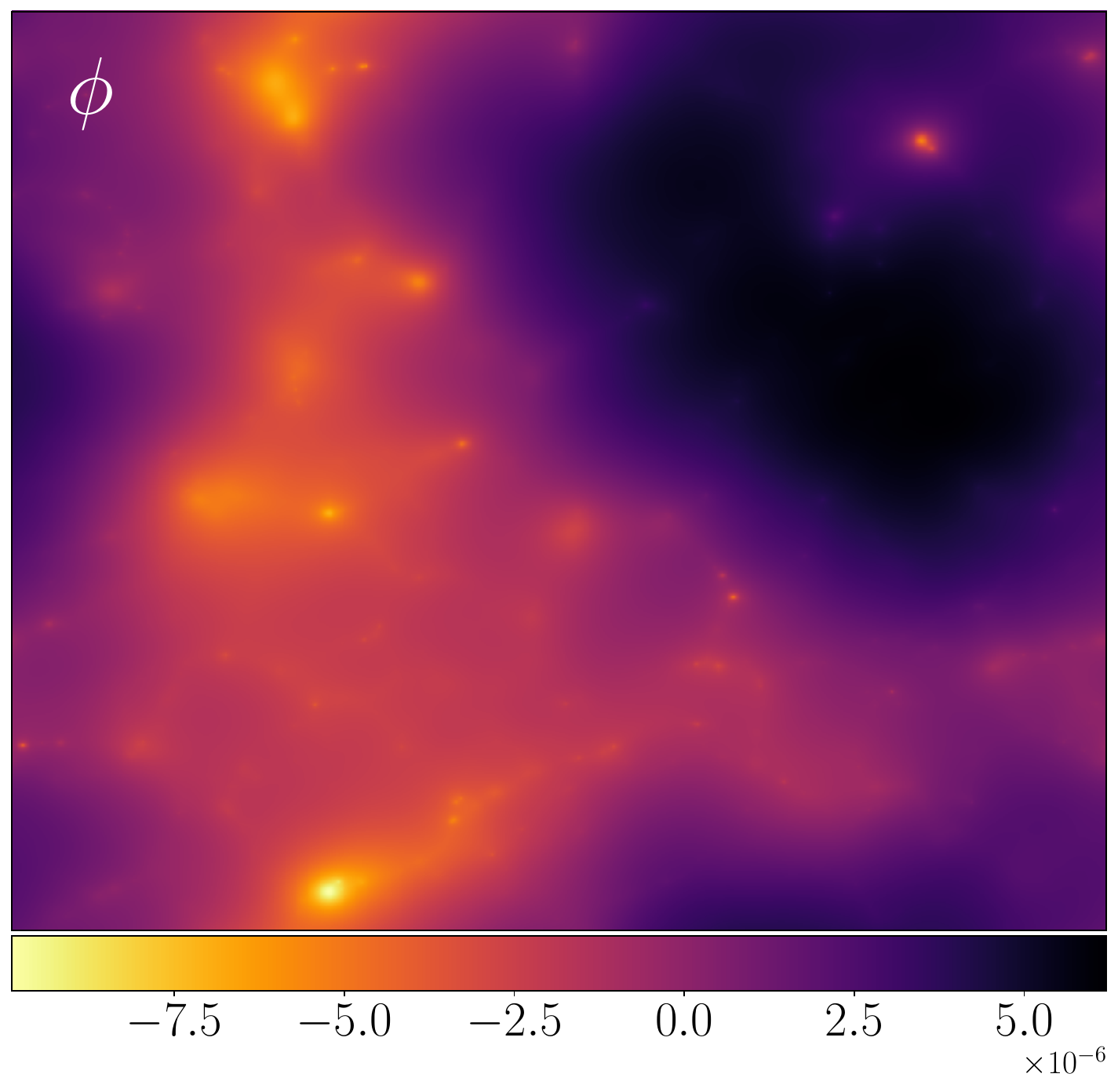}
    \hfill
    \includegraphics[width=0.49\textwidth]{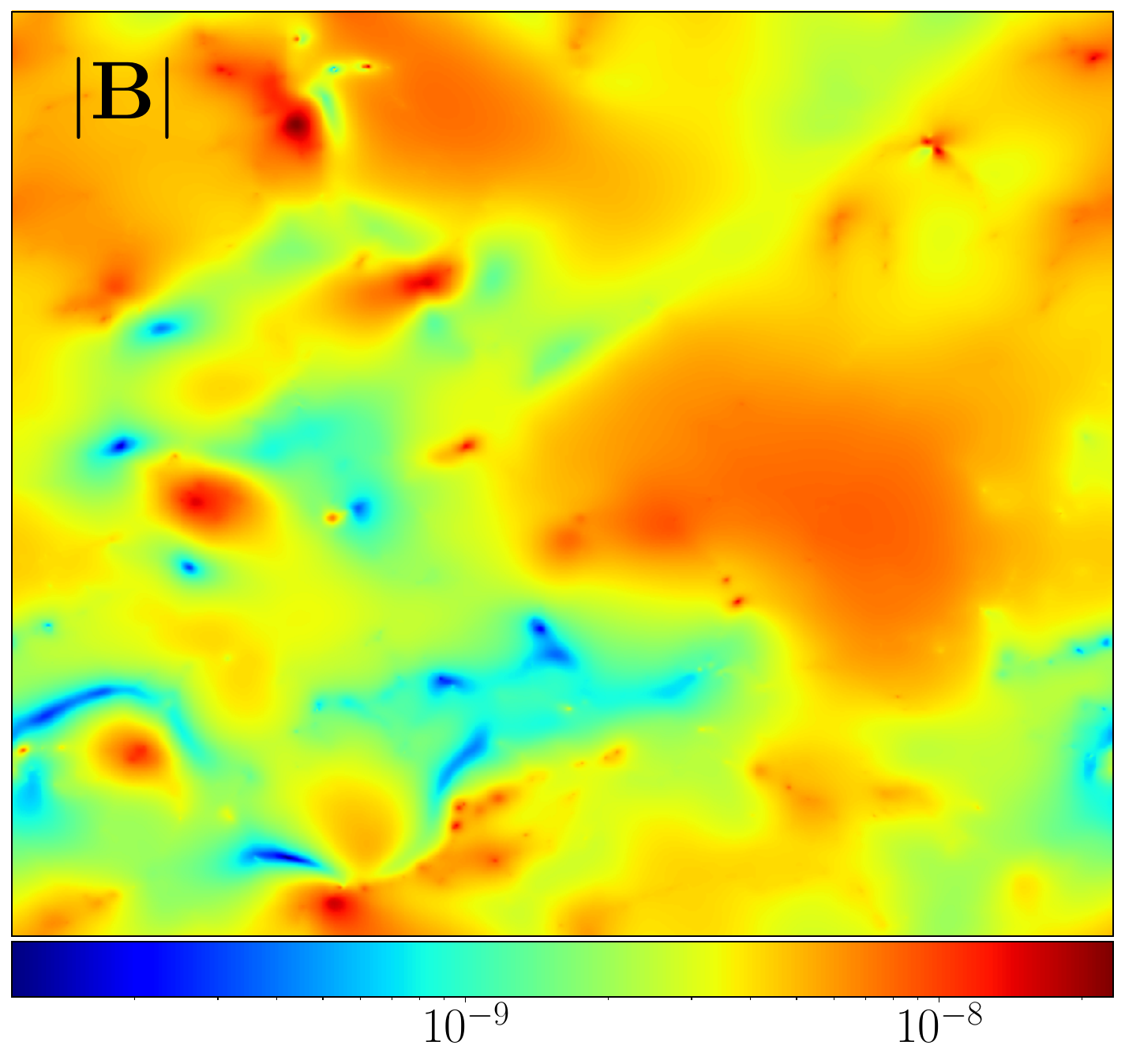}
    \end{minipage}
    }
    \caption{Representation of the fields evaluated across a section of the simulation box for the snapshot at redshift $z=0$ of the simulation TNG300-2-Dark. Respectively, the density contrast (top left), velocity (top right), velocity divergence (middle left), velocity vorticity (middle right), scalar potential (bottom left), and vector potential (bottom right). Divergence and vorticity are normalized by the factor $\mathcal{H}f$, therefore they are dimensionless. For the vector fields, the plots represent the magnitude.}
    \label{fig:2Dplots_TNG300}
\end{figure*}

\begin{figure*}
    \centering
    \scalebox{.9}{ 
    \begin{minipage}{\textwidth}
    \includegraphics[width=0.49\textwidth]{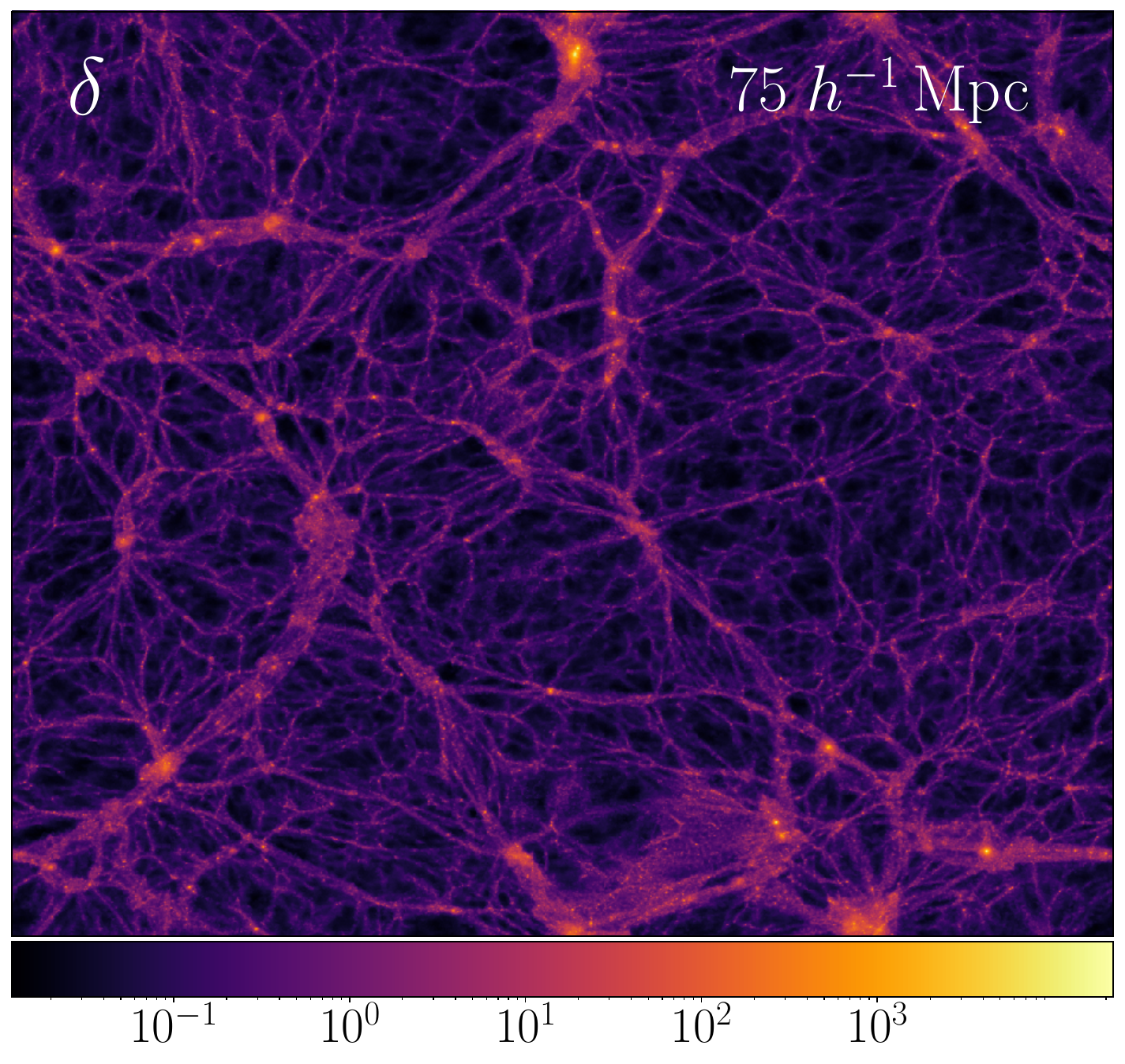}
    \hfill
    \includegraphics[width=0.49\textwidth]{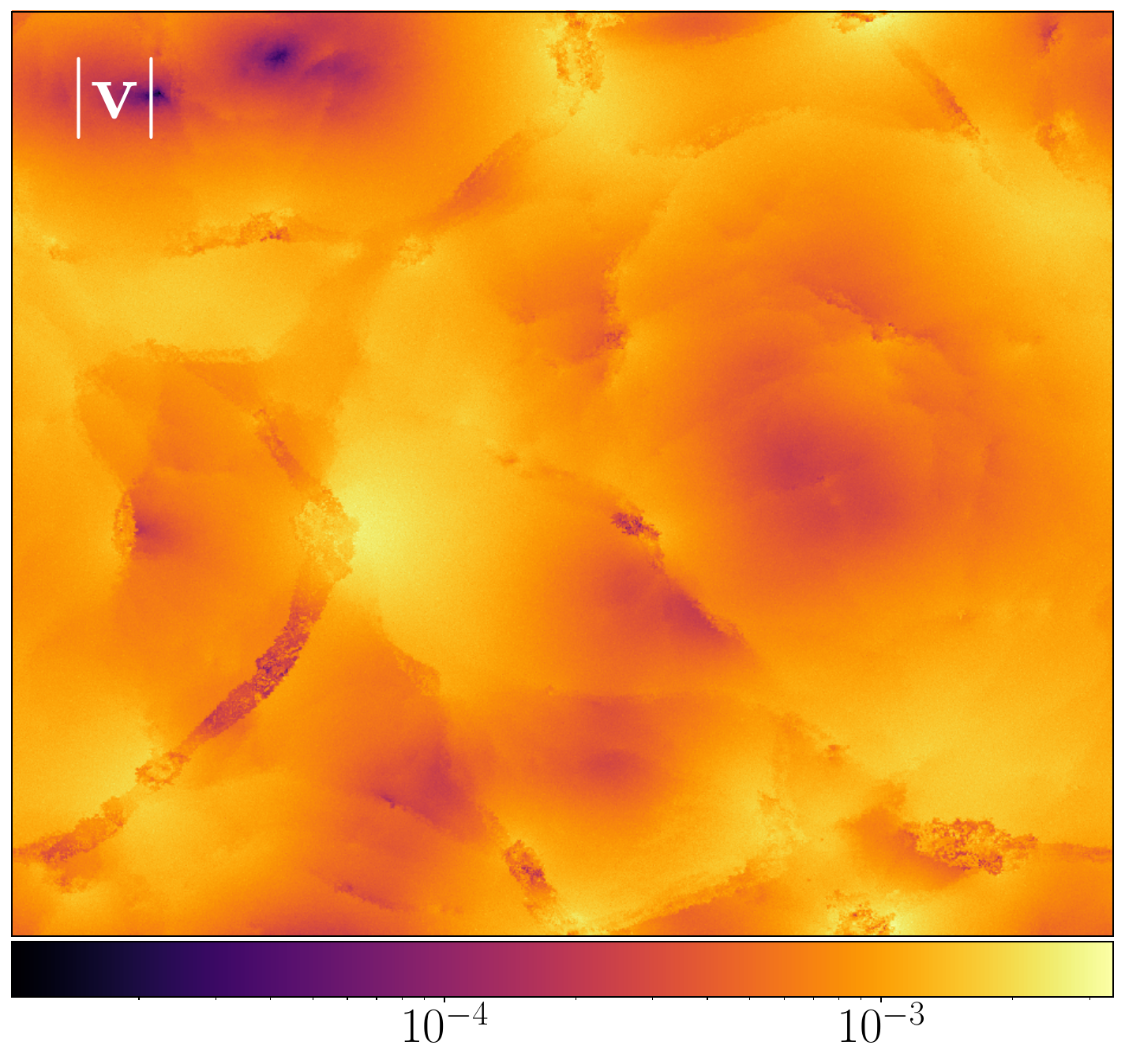}
    \vfill
    \includegraphics[width=0.49\textwidth]{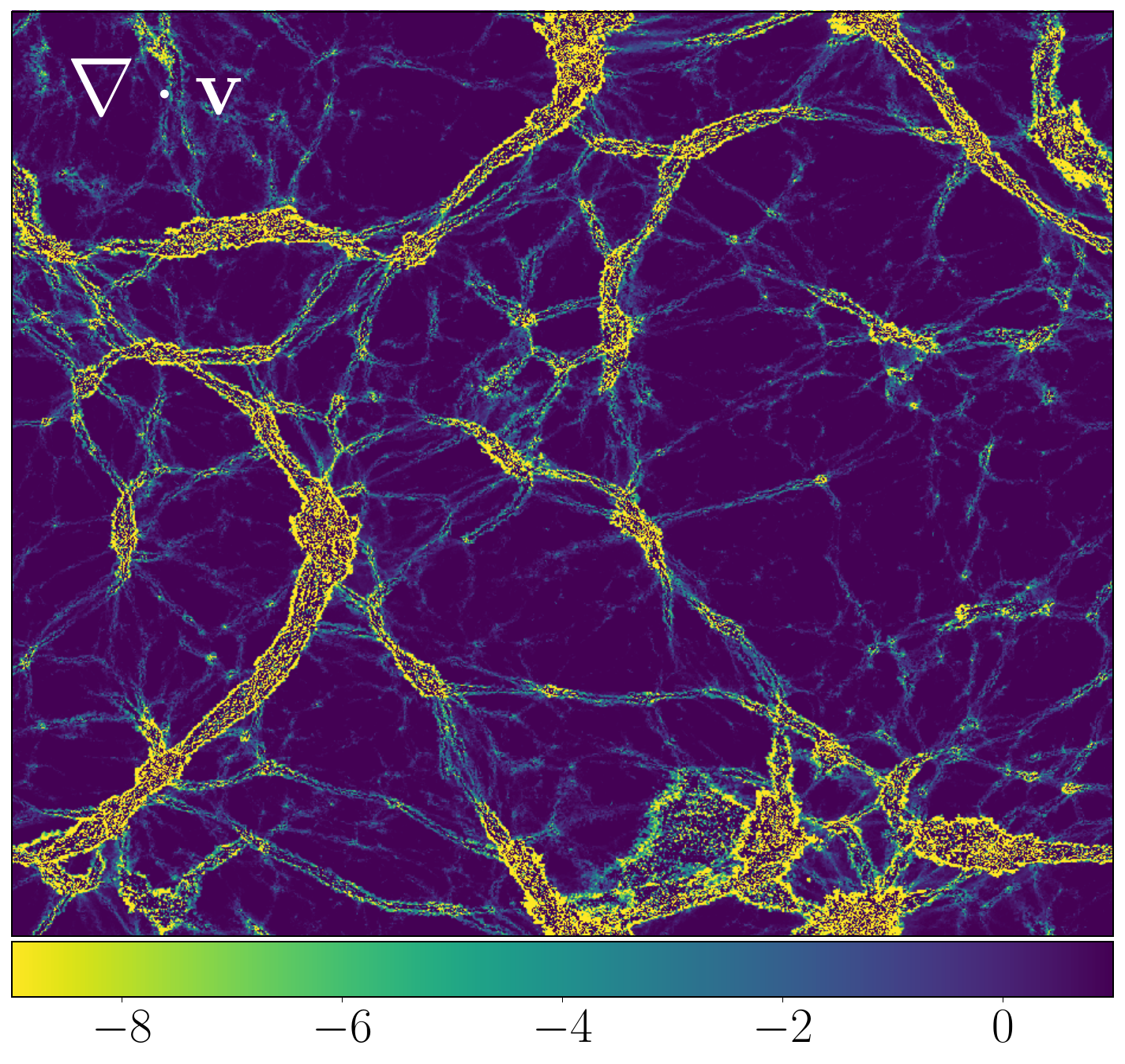}
    \hfill
    \includegraphics[width=0.49\textwidth]{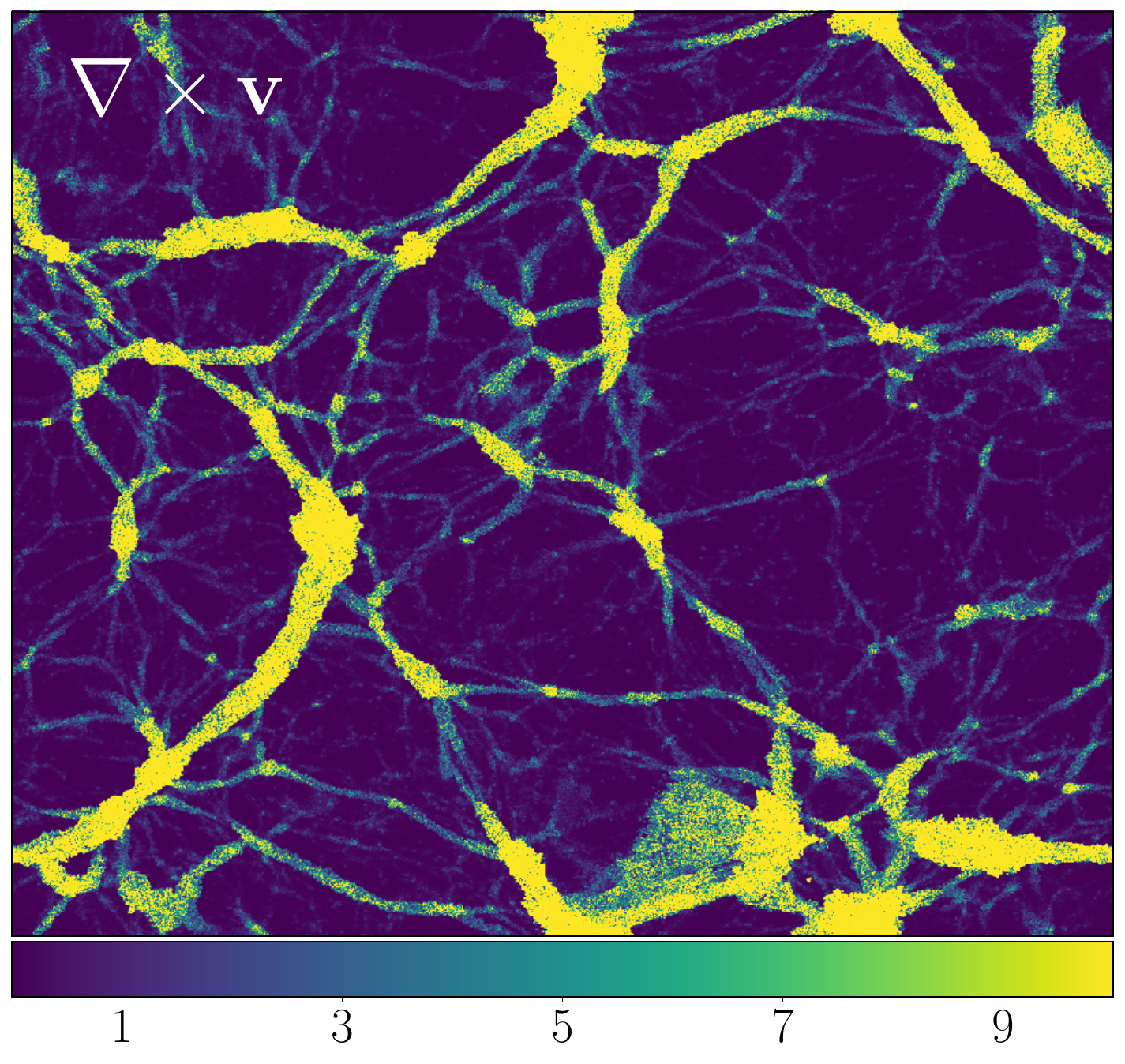}
    \vfill
    \includegraphics[width=0.49\textwidth]{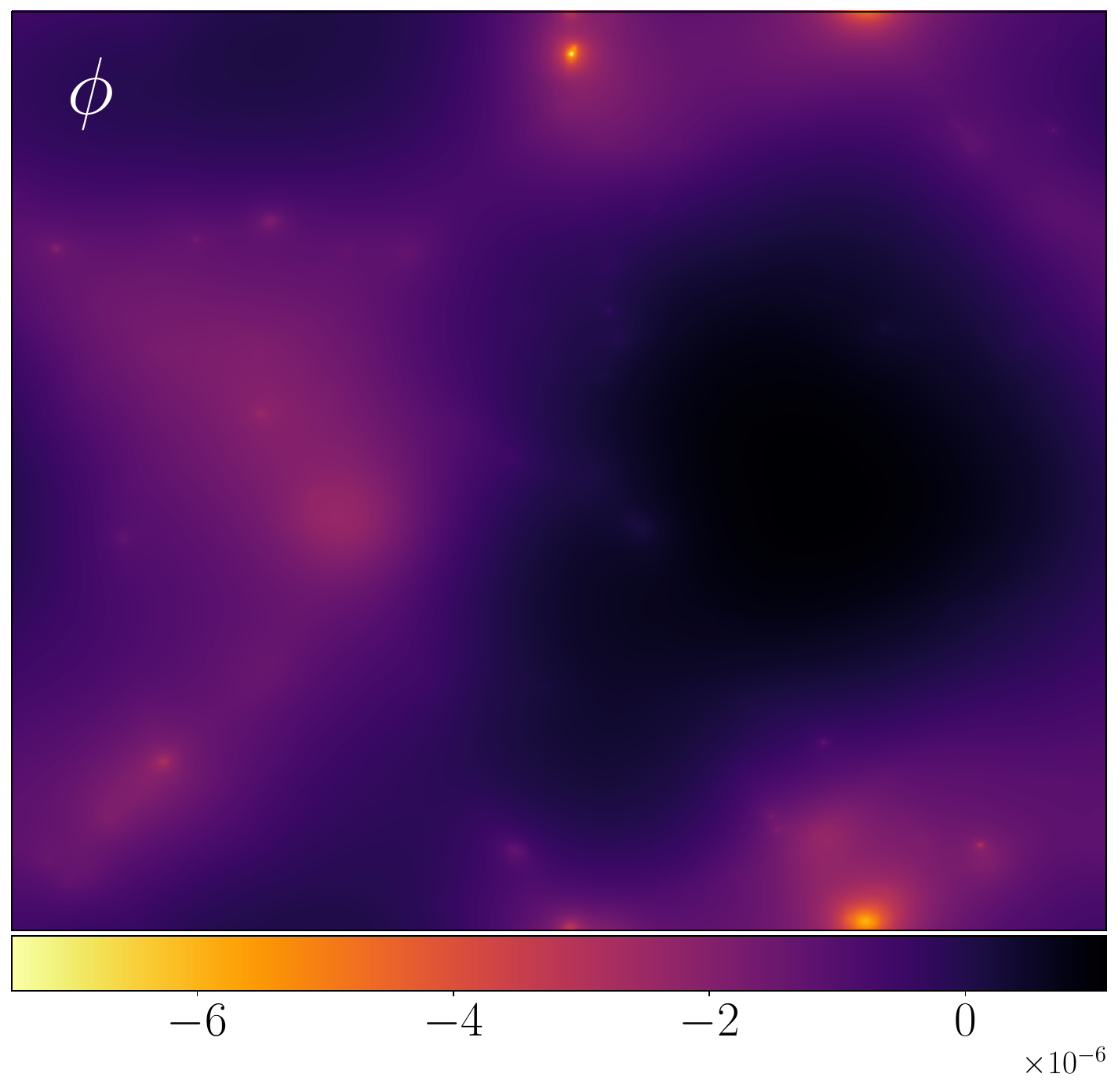}
    \hfill
    \includegraphics[width=0.49\textwidth]{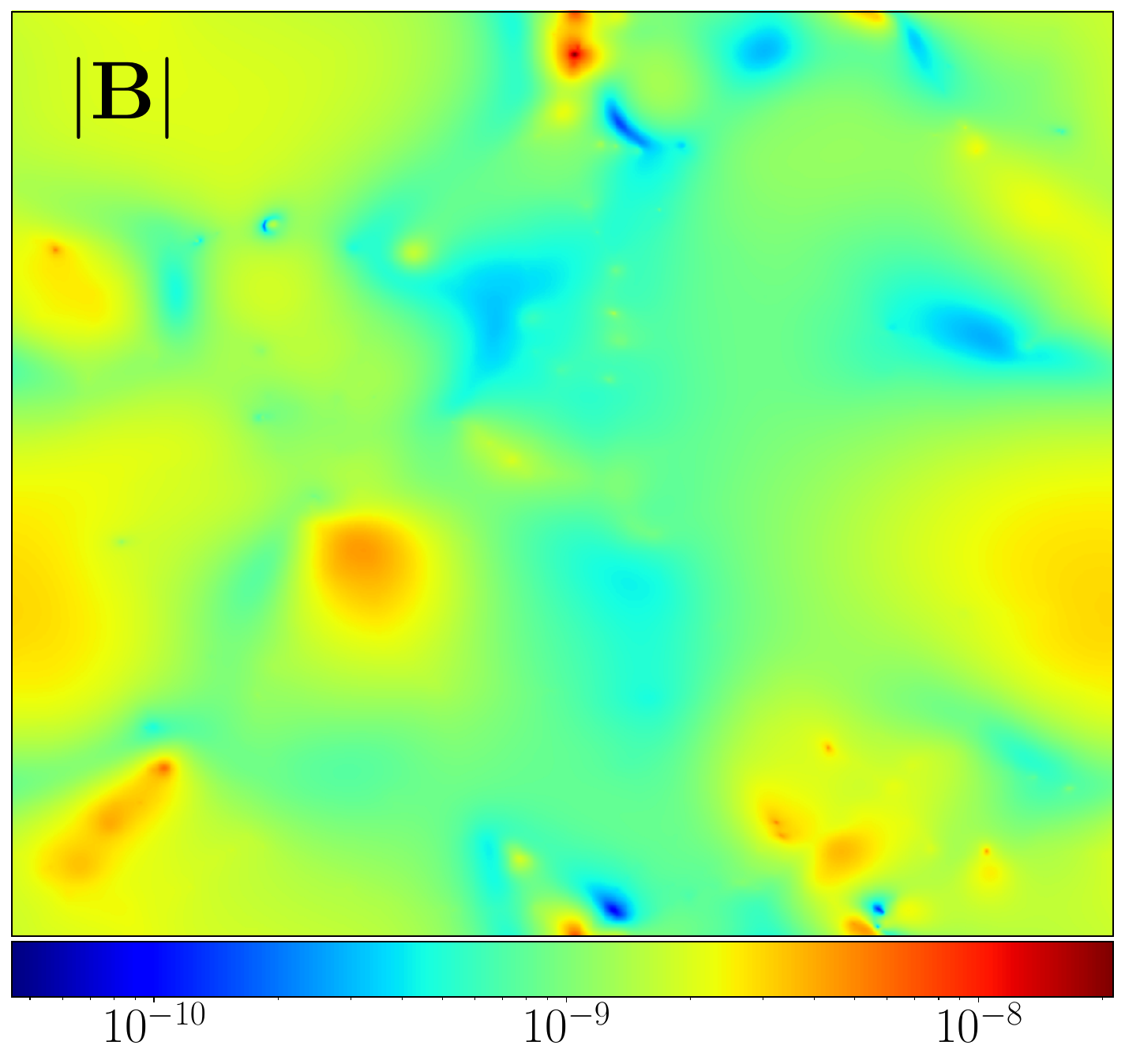}
    \end{minipage}
    }
    \caption{Same as Figure~\ref{fig:2Dplots_TNG300} but for the simulation TNG100-2-Dark.}
    \label{fig:2Dplots_TNG100}
\end{figure*}

\begin{figure*}
    \centering
    \scalebox{.9}{ 
    \begin{minipage}{\textwidth}
    \includegraphics[width=0.49\textwidth]{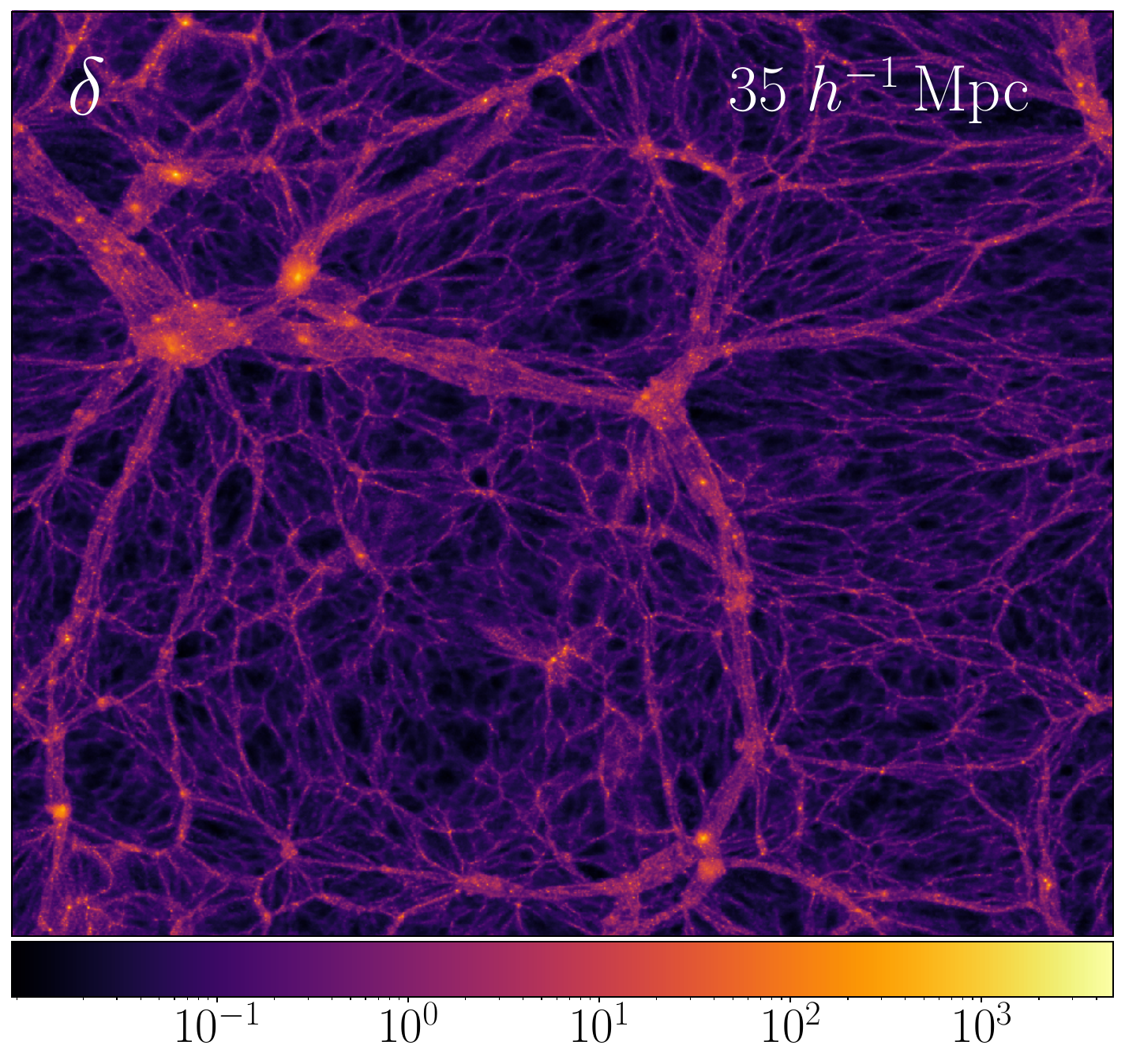}
    \hfill
    \includegraphics[width=0.49\textwidth]{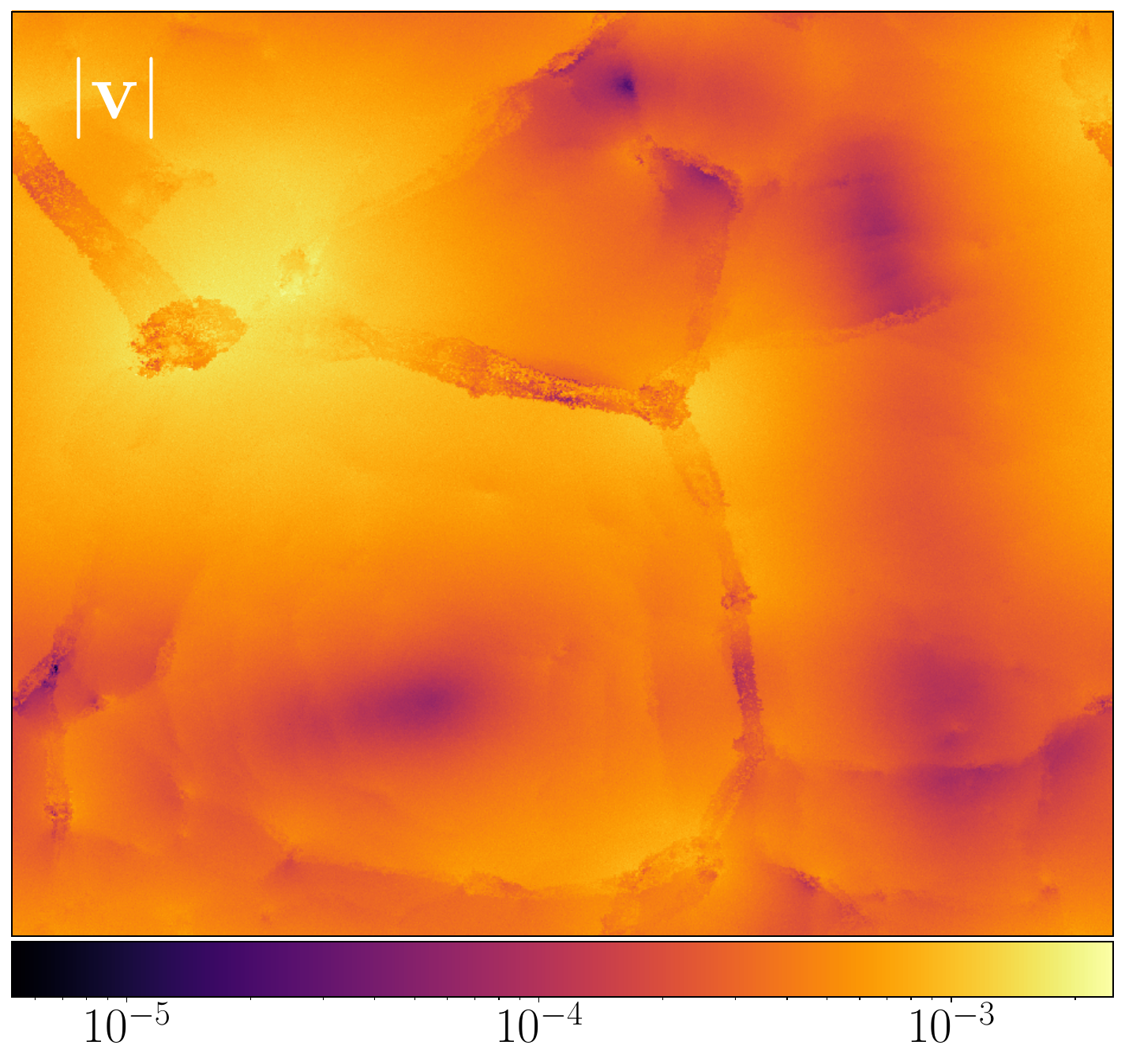}
    \vfill
    \includegraphics[width=0.49\textwidth]{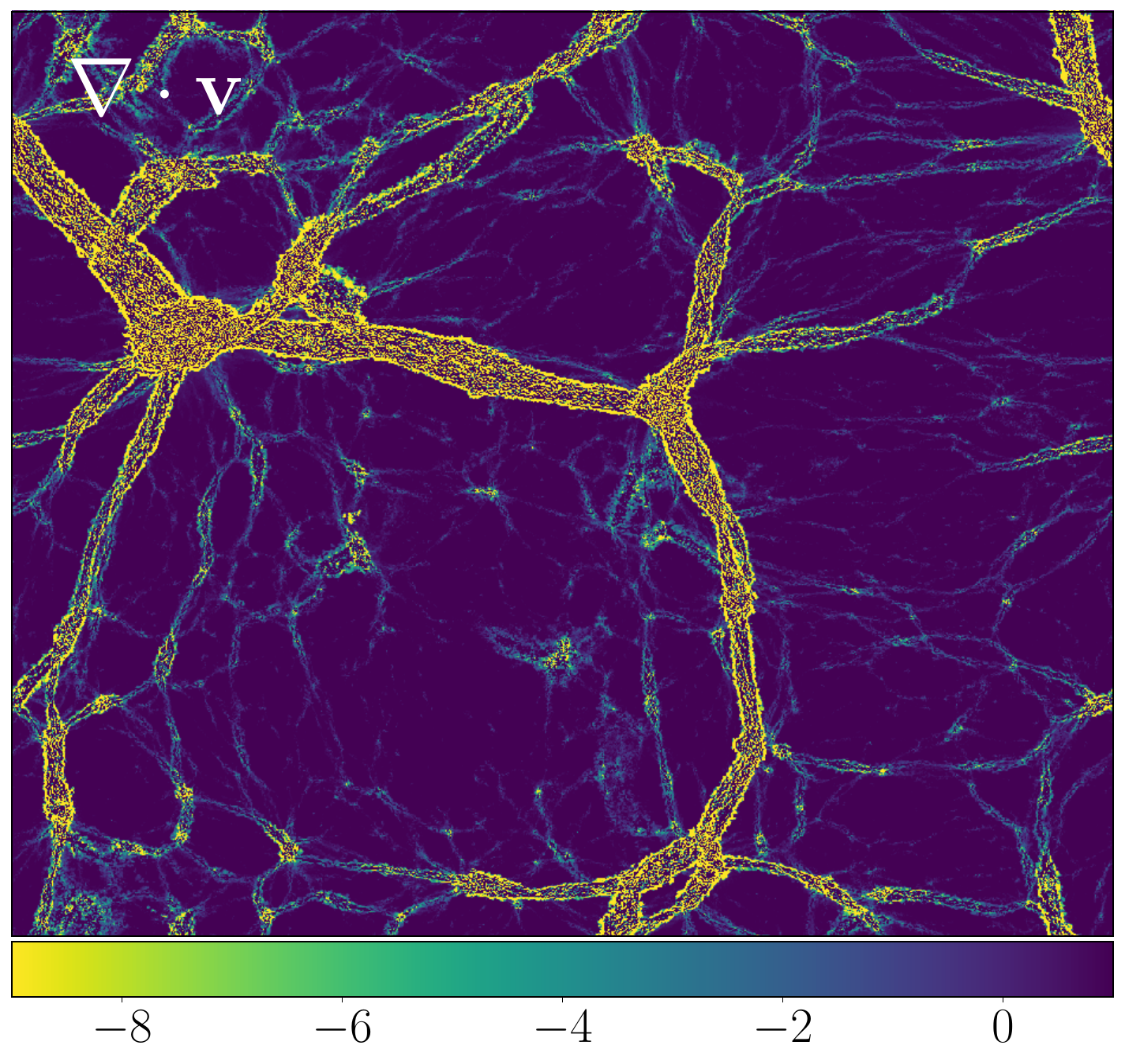}
    \hfill
    \includegraphics[width=0.49\textwidth]{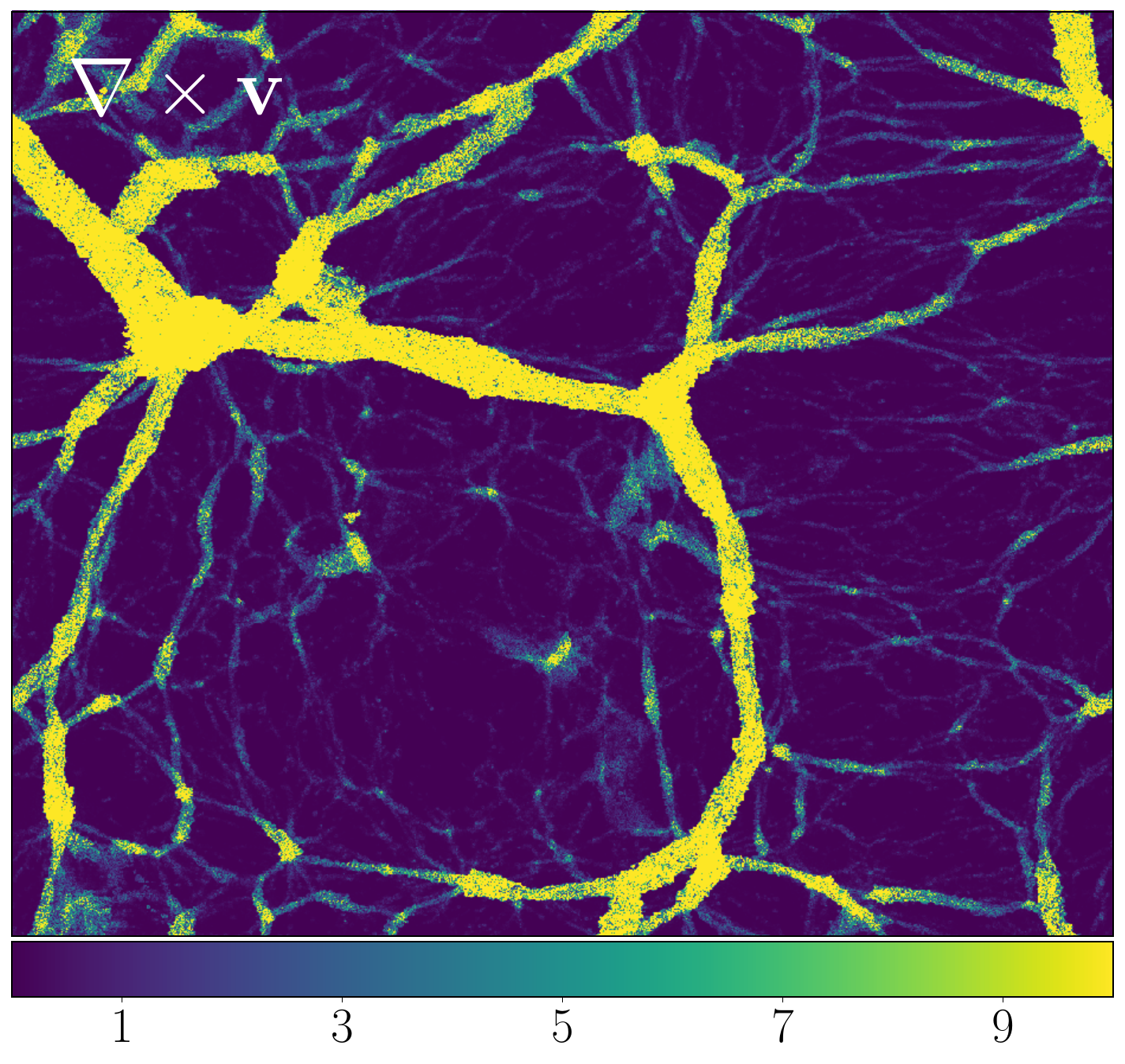}
    \vfill
    \includegraphics[width=0.49\textwidth]{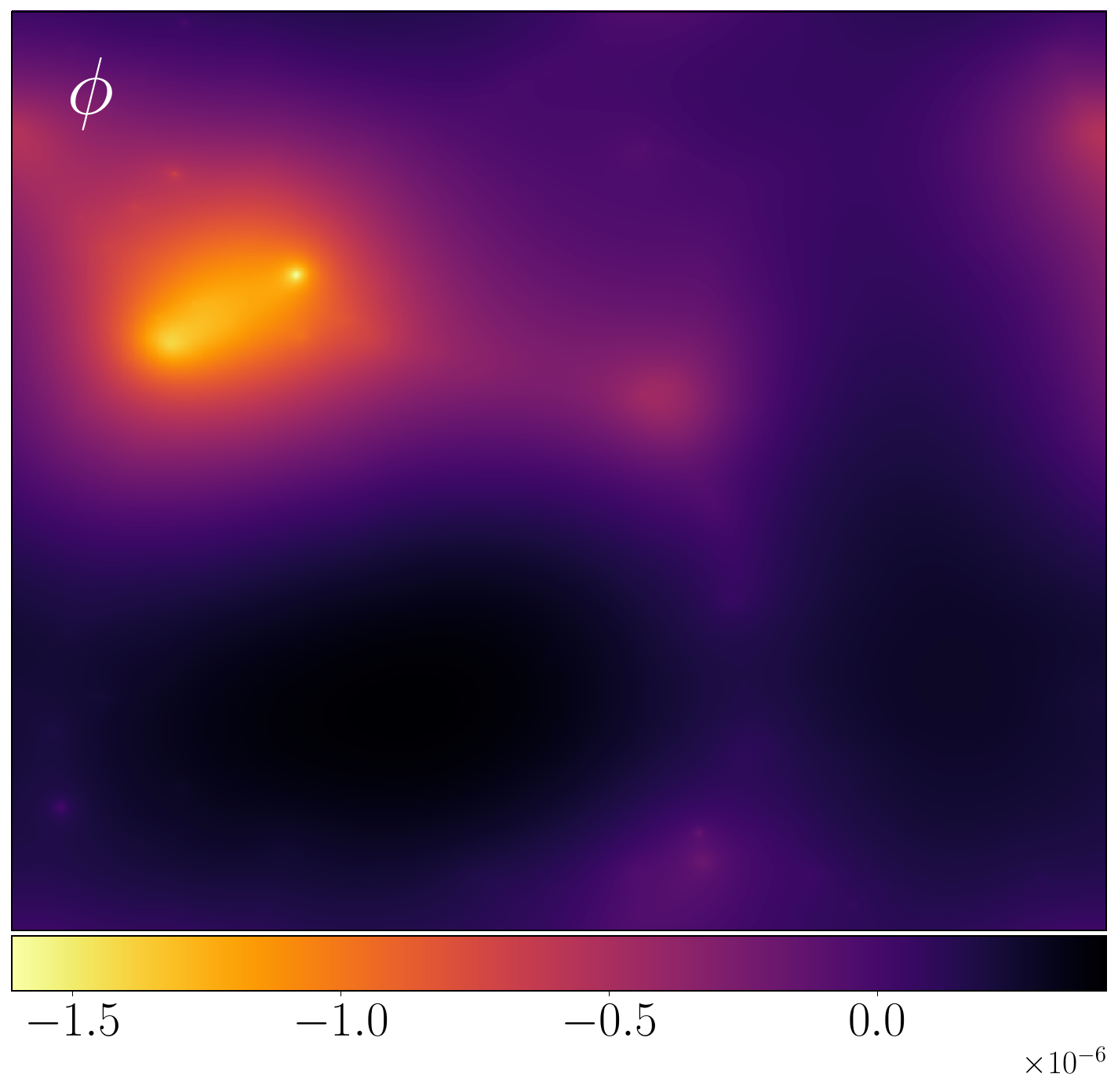}
    \hfill
    \includegraphics[width=0.49\textwidth]{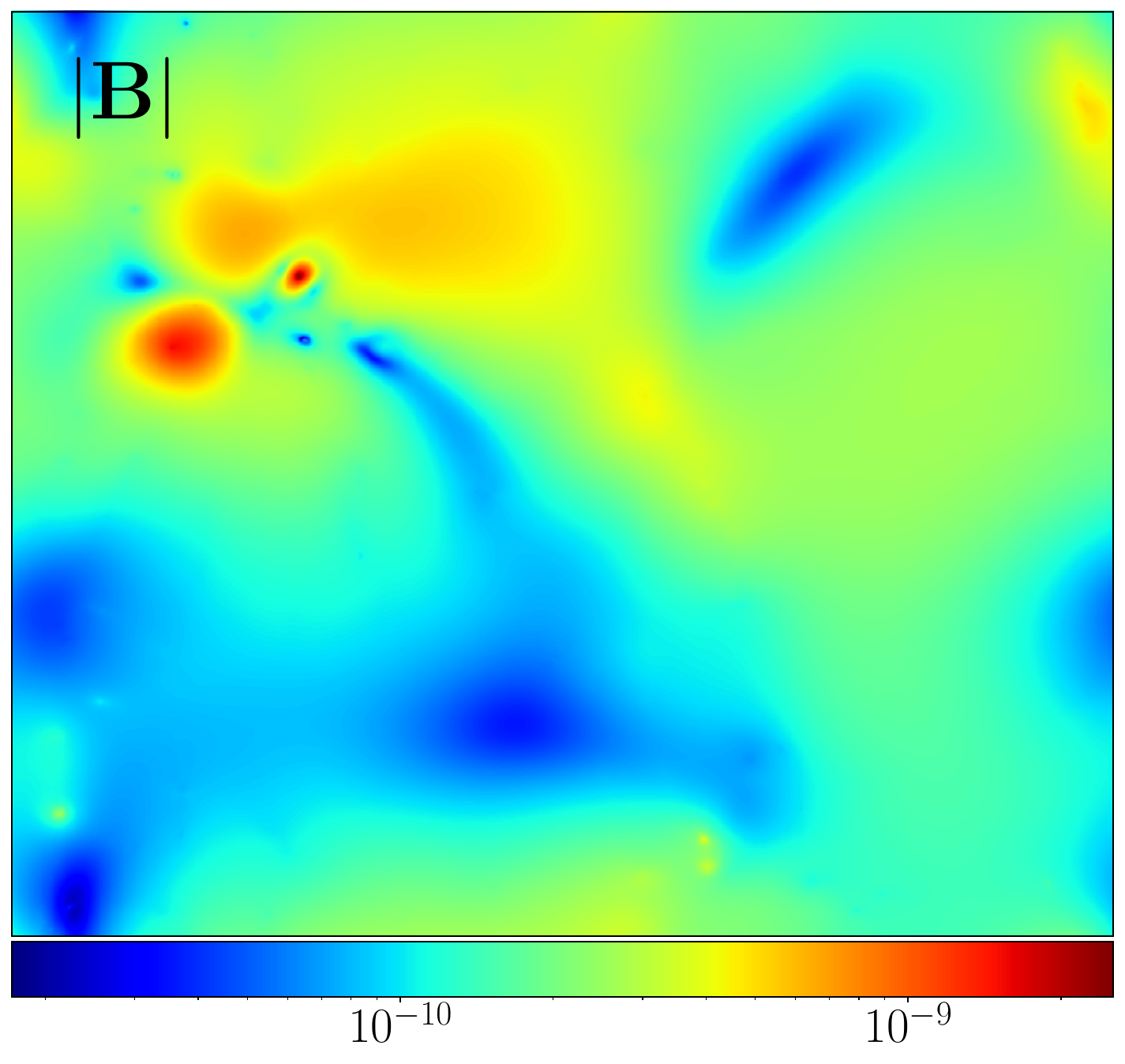}
    \end{minipage}
    }
    \caption{Same as Figure~\ref{fig:2Dplots_TNG300} but for the simulation TNG50-2-Dark.}
    \label{fig:2Dplots_TNG50}
\end{figure*}

Figures~\ref{fig:2Dplots_TNG300}--\ref{fig:2Dplots_TNG50} show 2D distributions across a slice of the simulations at $z=0$ for the following fields: the density contrast $\delta$, the magnitude\footnote{The magnitude here is defined with the flat metric $\delta_{ij}$, so that $|\mathbf{v}| = (\mathrm{v}_x^2 + \mathrm{v}_y^2 + \mathrm{v}_z^2)^{1/2}$.} of the velocity field $|\mathbf{v}|$, the velocity divergence $\nabla \cdot \mathbf{v}$, the magnitude of the velocity vorticity $|\nabla \cdot \mathbf{v}|$, the scalar Newtonian potential $\phi$, and the magnitude of the vector potential $|\mathbf{B}|$. 
This figure clearly shows that the density field shares a similar large-scale distribution with the velocity divergence, in line with the continuity equation~(\ref{eq:continuity}) of linear perturbation theory. High-density regions exhibit enhanced peculiar velocities, a direct consequence of gravitational infall as matter accumulates within deep potential wells. The divergence of the velocity field further reinforces this picture: negative values indicate regions of gravitational collapse, coinciding with overdensities, whereas positive values trace underdense voids where matter is being displaced outward.

The density field also shows a clear correlation with the velocity vorticity. The latter is peaked locally in the collapsing regions, where the orbit crossing of the particles occurs more frequently. 
However, as mentioned in Section~\ref{sec:DTFE}, the velocity divergence and vorticity estimated by DTFE are not completely reliable near caustics \citep{hahnPropertiesCosmicVelocity2015}, where velocity gradients are found to diverge. Therefore, such maps only provide qualitative information and an accurate picture on large scales.

The scalar potential is also very highly correlated with the velocity field, as expected from the Euler equation~(\ref{eq:PF_euler}). Compared to the matter field, its distribution is dominated by large-scale modes, while small-scale variations are smoothed out by the Laplacian operator in the Poisson equation~(\ref{eq:PF_poisson}), which introduces a $k^2$ factor in the Fourier space (see Equation~\ref{eq:Poisson_Fourier}).  

A similar behaviour is expected for the magnitude of the gravito-magnetic vector potential (see Equation~\ref{eq:vectorPot_Fourier}), which indeed shows a smoother distribution.
Although not sourced directly by the density fields and velocity gradients (but by the rotational part of the momentum density field), some degree of correlation between the vector potential and these quantities is clearly visible, particularly in very high-density regions.
Interestingly, the magnitude of the vector fields also seems to show non-local features: for instance, in the proximity of regions with large scalar potential, where low-value areas are paired with high-value ones, as well as near voids where $|\mathbf{B}|$ does not vanish. 

\begin{figure}
    \centering
    \includegraphics[width=\linewidth]{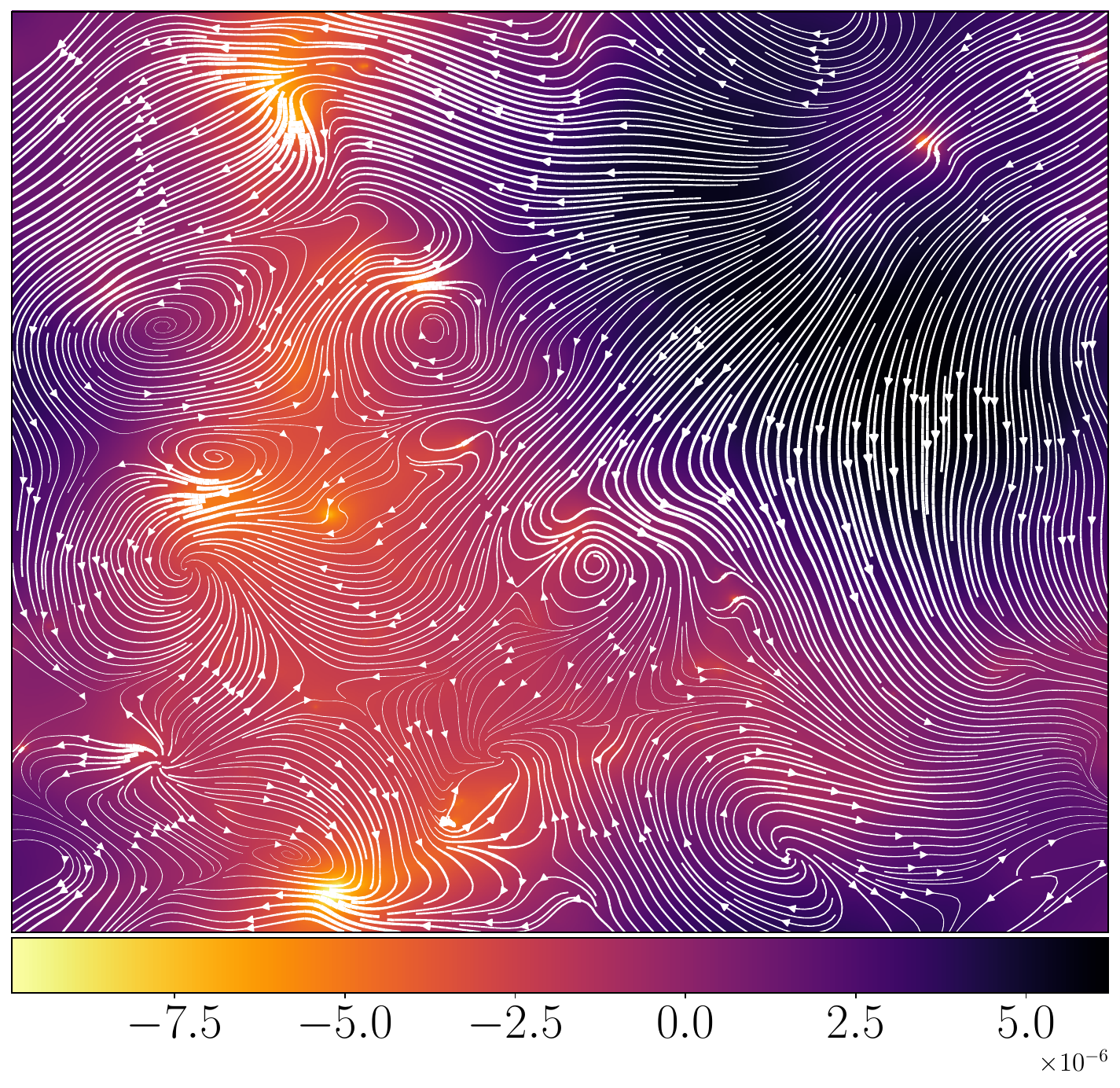}
    \caption{Representation of the stream lines of the vector potential, on top of the scalar potential distribution (bottom left panel of Figure~\ref{fig:2Dplots_TNG300}) for TNG300-2-Dark; streamlines with higher thickness represent regions where the magnitude of the vector potential is larger.}
    \label{fig:streamplots300}
\end{figure}

\begin{figure}
    \centering
    \includegraphics[width=\linewidth]{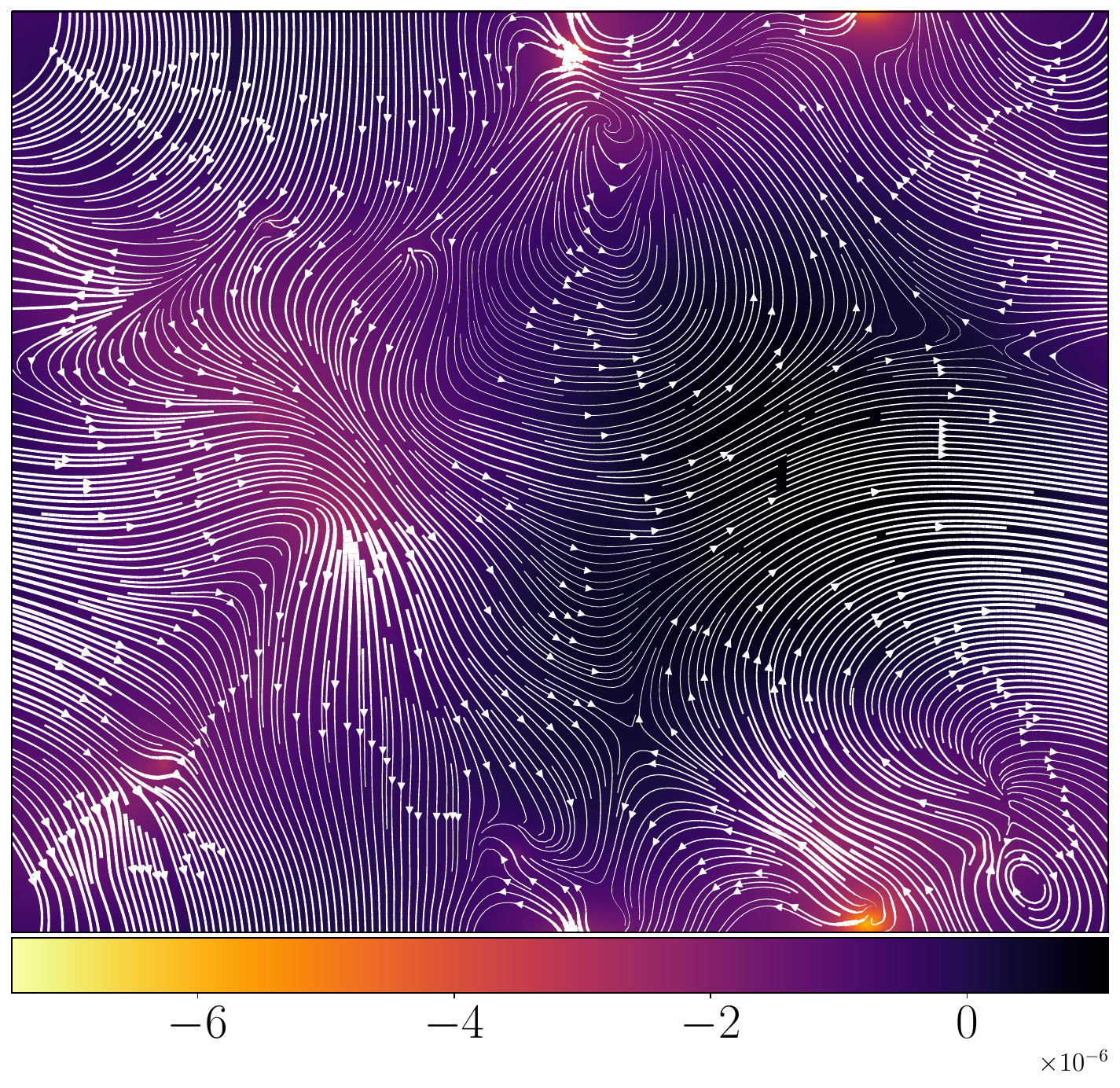}
    \caption{Same as Figure~\ref{fig:streamplots300} but for TNG100-2-Dark.}
    \label{fig:streamplots100}
\end{figure}

\begin{figure}
    \centering
    \includegraphics[width=\linewidth]{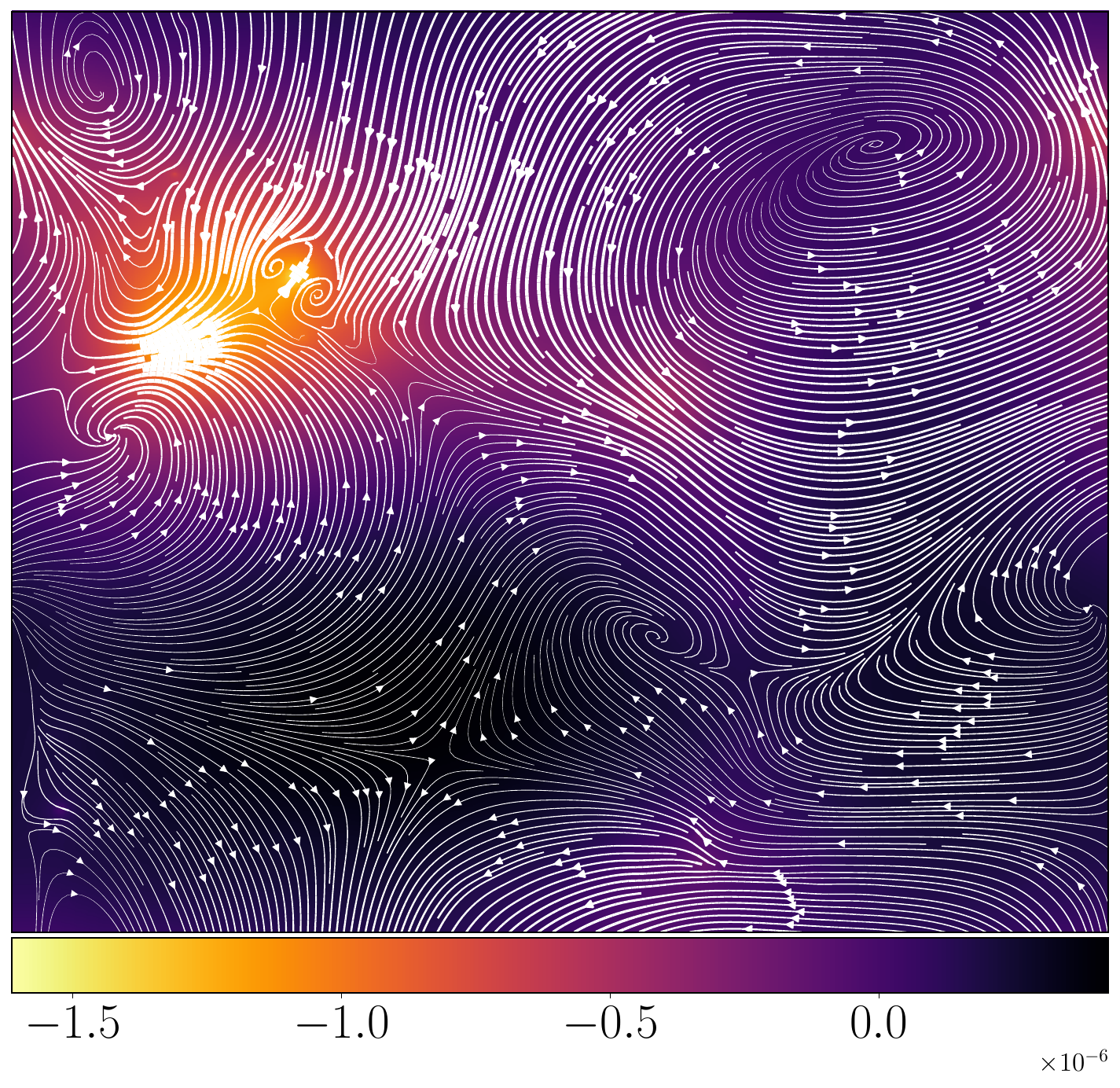}
    \caption{Same as Figure~\ref{fig:streamplots300} but for TNG50-2-Dark.}
    \label{fig:streamplots50}
\end{figure}

Figures~\ref{fig:streamplots300}--\ref{fig:streamplots50} show the streamlines of the gravito-magnetic vector field across the same slice, drawn on top of the scalar potential.
A striking feature of the visualization is the emergence of vortical structures in the vector potential, particularly around deeper scalar potential wells.
These organised vortical patterns in the vector potential first confirm that the reconstruction of the gravito-magnetic potential, outlined in Section~\ref{sec:globalSol}, produced consistent results, as they underline the solenoidal behaviour of the potential. Secondly, they suggest that, in line with the main subject of this paper, while traditional treatments of dark matter assume a predominantly irrotational velocity field, also the rotational dark matter flows may influence the angular momentum transport and thus the evolving cosmic structures in a non-trivial way.
In the words of \citet{adamekGeneralRelativityCosmic2016}, this indicates how \textit{spacetime is dragged around} by vortical matter flows.





Although in Section~\ref{sec:gravito-magneticPot} this effect has been quantified as $1$--$0.1$\% compared to that of the Newtonian potential, it will be possibly measured in the future also on the scale of galaxies. 
Moreover, while frame-dragging remains small even in highly non-linear regimes—making it unlikely to play a significant role in galaxy rotation within a $\Lambda$CDM framework—the possible contribution of baryons to the gravito-magnetic potential in cosmological hydrodynamical simulations such as IllustrisTNG remains an open question, which is left for future investigations.



\subsection{Time evolution}
\begin{figure*}
    \centering
    \includegraphics[width=0.49\textwidth]{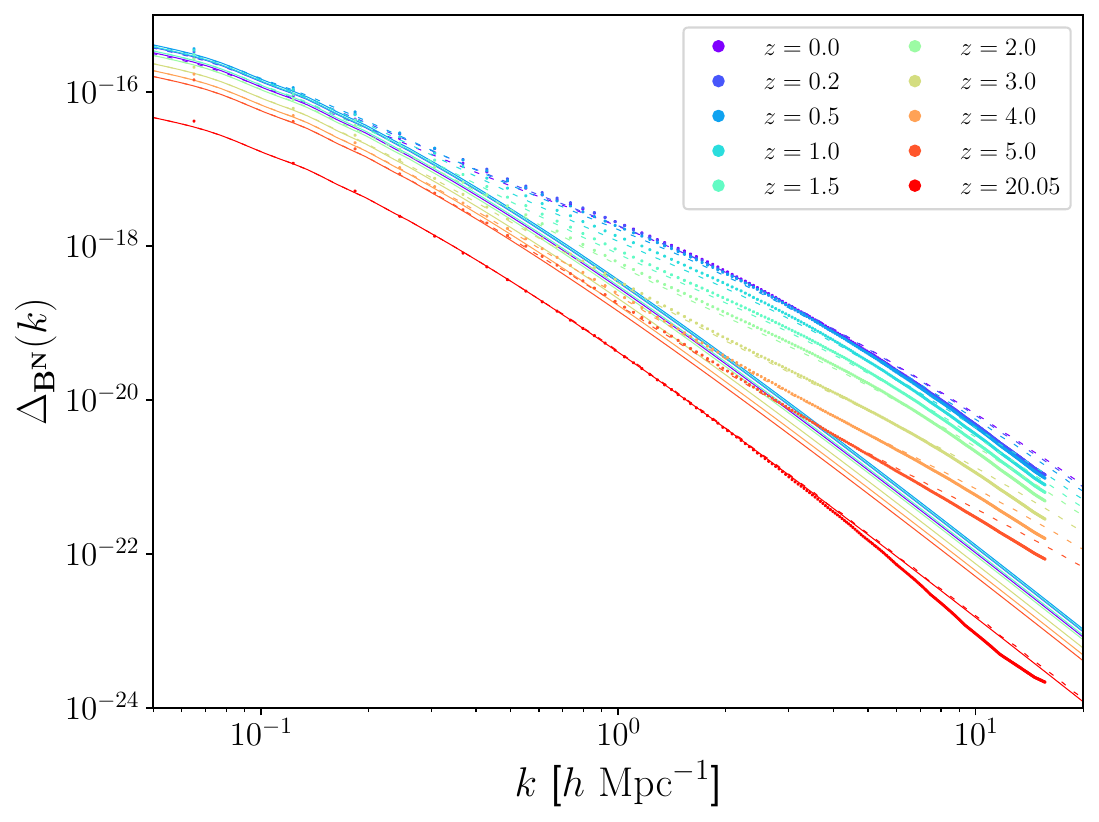}
    \hfill
    \includegraphics[width=0.49\textwidth]{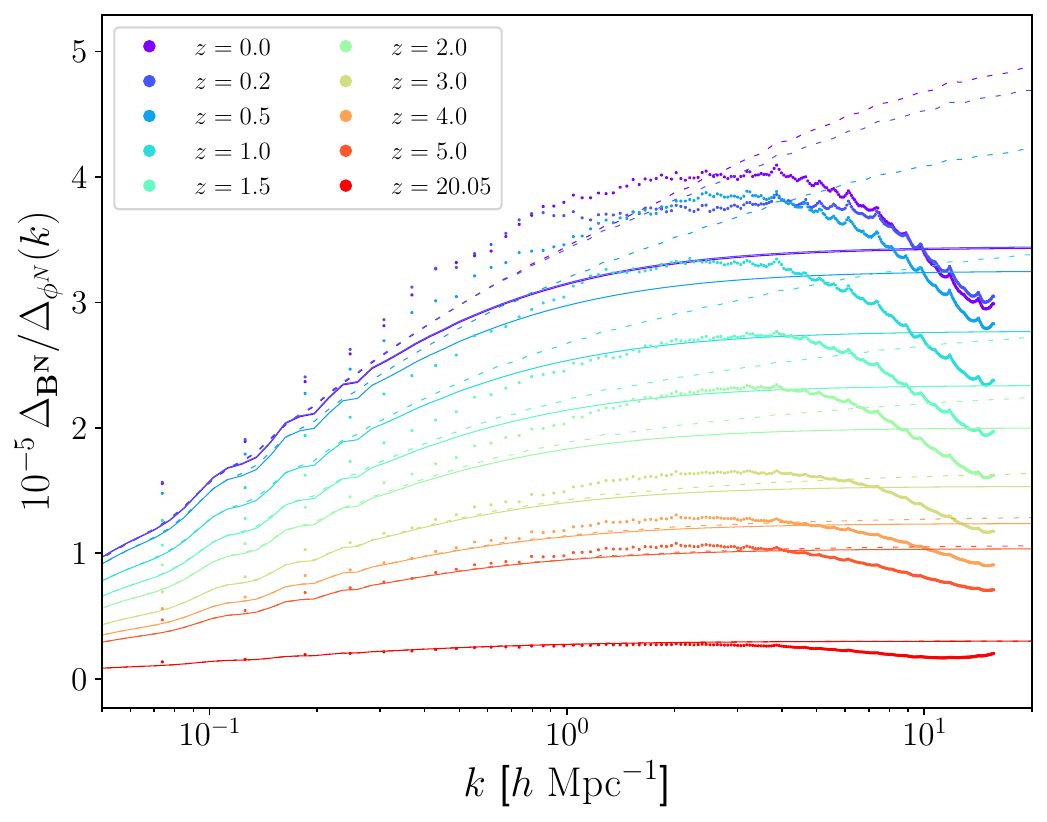}
    \caption{Time evolution of the vector potential power spectrum for TNG300-2-Dark, plotted for different redshifts. Left: same as the top panel of Figure~\ref{fig:potentials_psCorrected}. Right: same as Figure~\ref{fig:ratio_Corrected}.} 
    \label{fig:timeEv_PS_TNG300}
\end{figure*}

\begin{figure*}
    \centering
    \includegraphics[width=0.49\textwidth]{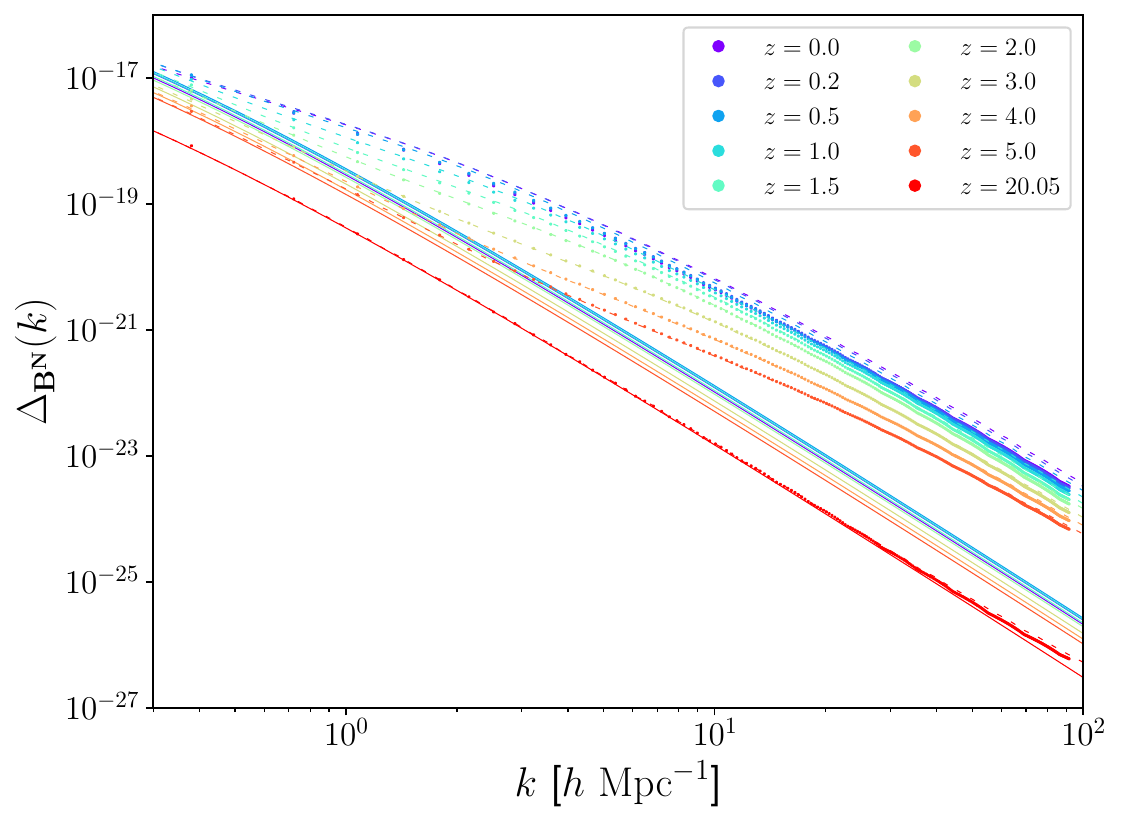}
    \hfill
    \includegraphics[width=0.49\textwidth]{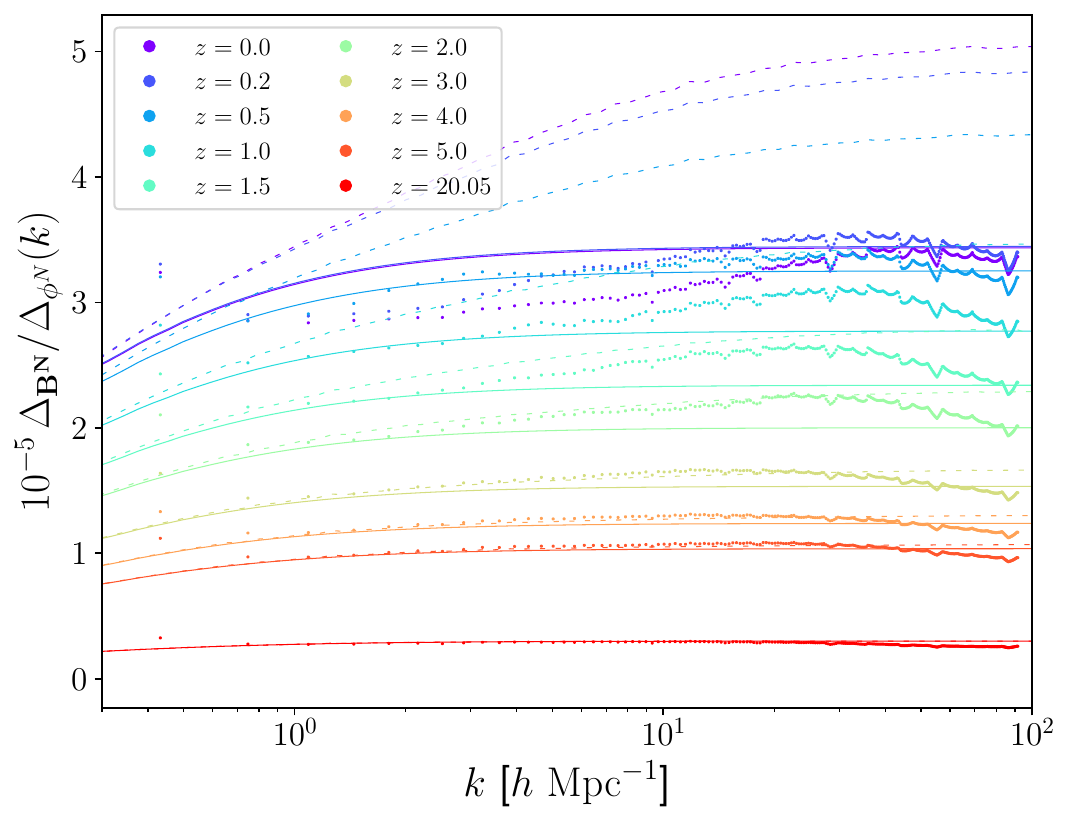}
    \caption{Same as Figure~\ref{fig:timeEv_PS_TNG300} but for TNG50-2-Dark.}
    \label{fig:timeEv_PS_TNG50}
\end{figure*}

The power spectra at different redshifts, from $z=0$ to $z=5$ and $z=20.05$, are shown in Figures~\ref{fig:timeEv_PS_TNG300} and~\ref{fig:timeEv_PS_TNG50} for the TNG300-2-Dark and TNG50-2-Dark simulations, respectively.
From these plots we can see how the power spectrum of the vector potential increases with time, as expected, with the ratio with respect to the scalar potential growing from $10^{-5}$ at $z=5$ to $3\text{--}4\times10^{-5}$ at $z=0$. As already mentioned in Section~\ref{sec:gravito-magneticPot} for $z=0$, the non-linear prediction $\Delta_{\mathbf{B}}^\mathrm{n.l.}$ slightly overestimates the actual power spectra measured from the simulations. In fact, given the correction for the missing power and the limited spatial resolution of the simulation and the tessellation algorithm, one cannot expect to get an exact match.

Compared to the right panel of Figure 5 in \citet{barrera-hinojosaVectorModesLCDM2021}, where the power spectrum of the vector potential is found to decrease over time, from $z=1.5$ to $z=0$, the reader should keep in mind that there is a factor $a^2$ difference between the adopted conventions for the power spectrum of $\mathbf{B}$. Moreover, in \citet{barrera-hinojosaVectorModesLCDM2021}, the ratio is found to increase towards the smallest scales, in contrast with our results. This is most likely due to the Adaptive Mesh Refinement technique implemented in GRAMSES, which enables a much higher spatial resolution than DTFE. It is still unclear, however, whether this increase manifests as a result of spurious power introduced by the AMR or rather as a pure GR indication.

\begin{figure*}
    \includegraphics[width=0.49\textwidth]{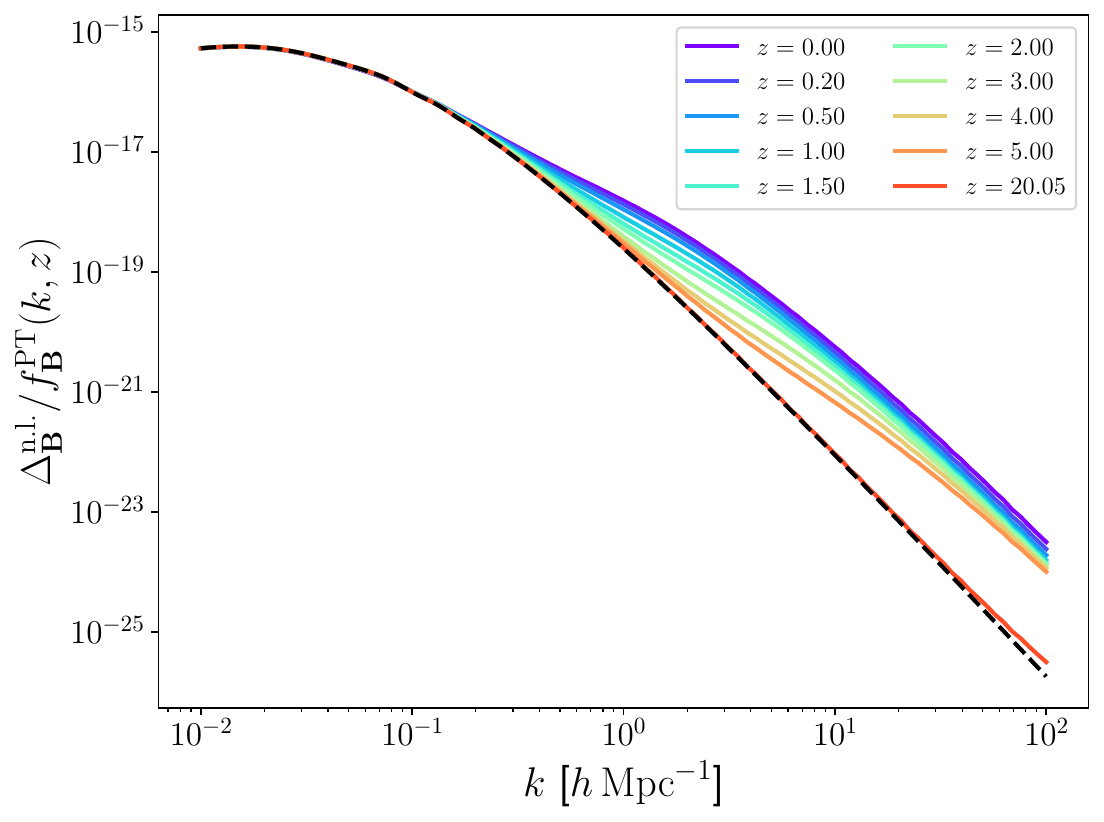}
    \hfill
    \includegraphics[width=0.49\textwidth]{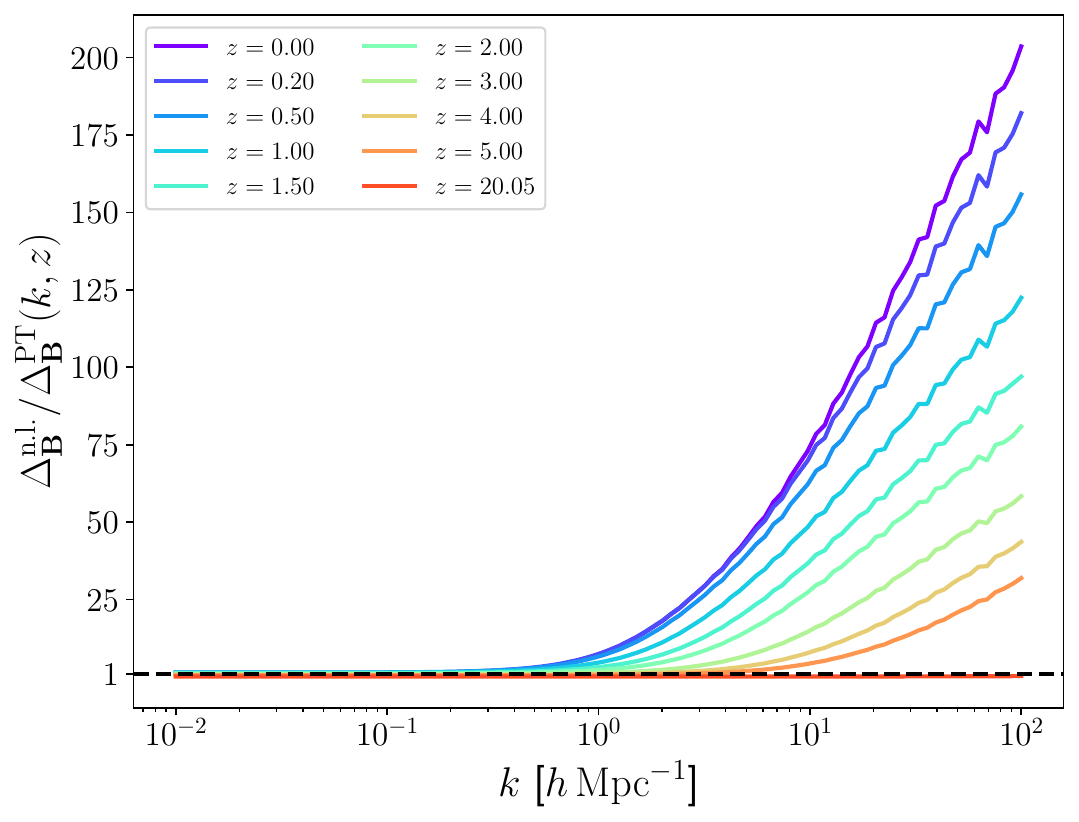}
    \caption{Evolution of the non-linear prediction for the dimensionless power spectrum of $\mathbf{B}$. In the left panel, this quantity is normalised by the analytic growth factor from perturbation theory, given by Equation~\eqref{eq:growth-factor-B-PT}. The black dashed line represents $\Delta_{\mathbf{B}}^\mathrm{PT}/f_{\mathbf{B}}^\mathrm{PT}$, which is only function of $k$. In the right panel, the ratio $\Delta_{\mathbf{B}}^\mathrm{n.l.}/\Delta_{\mathbf{B}}^\mathrm{PT}$ is shown to highlight the amplitude gain in the non-linear scenario.}
    \label{fig:Norm_B_evol}
\end{figure*}

Since the non-linear prediction of the vector potential power spectrum $\Delta_{\mathbf{B}}^\mathrm{n.l.}$ gives overall a good estimate for the fully non-linear vector potential, and its evolution, relevant information can already be retrieved from this quantity.
In order to highlight the redshift dependence of the power spectrum independently of its overall amplitude at a given redshift due to the linear growth of structure, we normalise the power spectrum $\Delta_{\mathbf{B}}^\mathrm{n.l.}$ by the analytic growth factor of $\Delta_{\mathbf{B}}^\mathrm{PT}$, which is \citep[see Equation B.4 of][]{adamekNbodyMethodsRelativistic2014}
\begin{equation}\label{eq:growth-factor-B-PT}
    f_{\mathbf{B}}^\mathrm{PT}(z) = \frac{g^2}{\mathcal{H}^2 \Omega_\mathrm{m}^2}\left[g-(1+z)g'\right]^2 \; ,
\end{equation}
where
\begin{equation}
    g(z) \equiv 2.5 \, g_{\infty} \Omega_\mathrm{m} \left[ \Omega_\mathrm{m}^{4/7} - (1 - \Omega_\mathrm{m}) + \left(1 + \frac{\Omega_\mathrm{m}}{2} \right) \left(1 + \frac{1 - \Omega_\mathrm{m}}{70} \right) \right]^{-1} 
\end{equation}
is the growth factor for the scalar potential and $g_\infty$ is chosen so that $g(0) = 1$ \citep[see Equation A.4 of ][]{luCosmologicalBackgroundVector2009}.
The left panel of Figure~\ref{fig:Norm_B_evol} shows $\Delta_{\mathbf{B}}^\mathrm{n.l.}/f_{\mathbf{B}}^\mathrm{PT}$ for different redshifts. 
Here, we can clearly see how the non-linear growth of vector modes is highly scale-dependent and how the power \WB{is} transferred over time. As expected, the non-linearity in the power spectrum of the gravito-magnetic vector potential is observed to increase with redshift from $z=20$ to $z=0$. 
From the right panel, one can quantify the increase of the amplitude of the non-linear approximation with respect to perturbation theory. In particular, at $z=0$, the non-linear amplitude is $\approx 175$ times larger than PT, for scales close to $k=100\; h$ Mpc$^{-1}$, while being $\approx75$ times larger for $k=10\; h$ Mpc$^{-1}$.

\begin{figure*}
    \centering
    \includegraphics[width=0.49\linewidth]{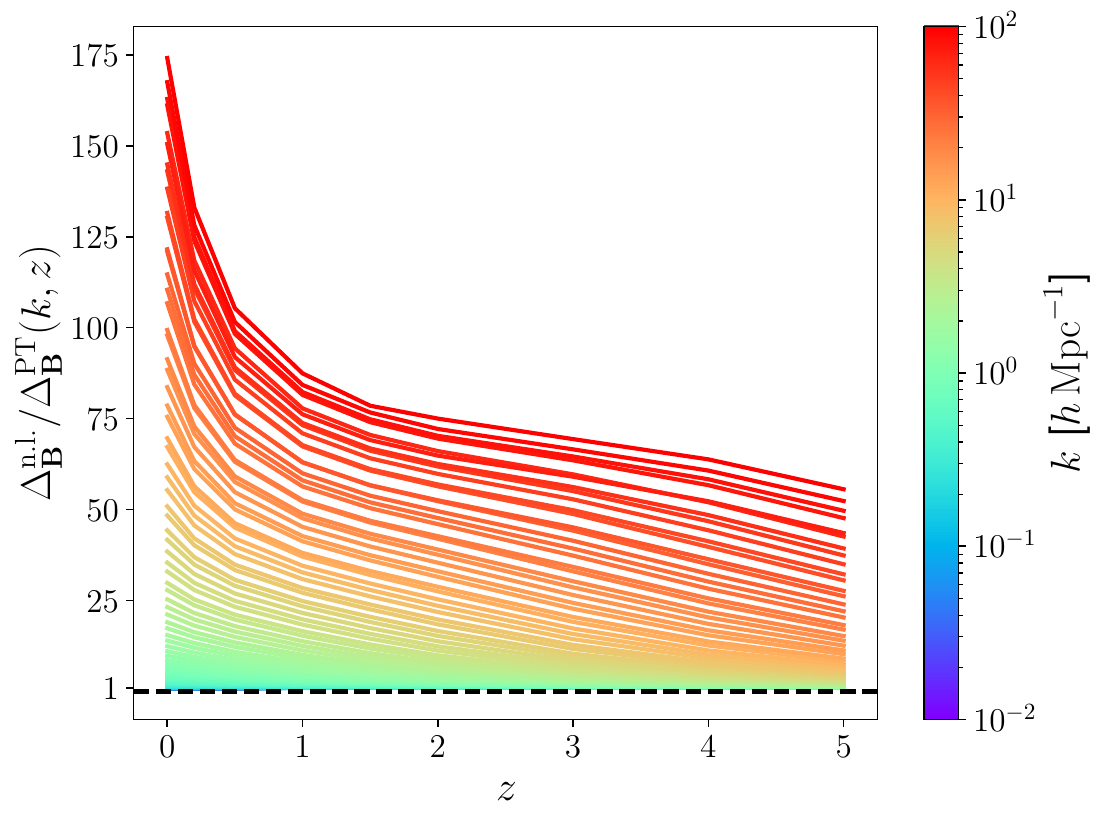}
    \hfill
    \includegraphics[width=0.49\linewidth]{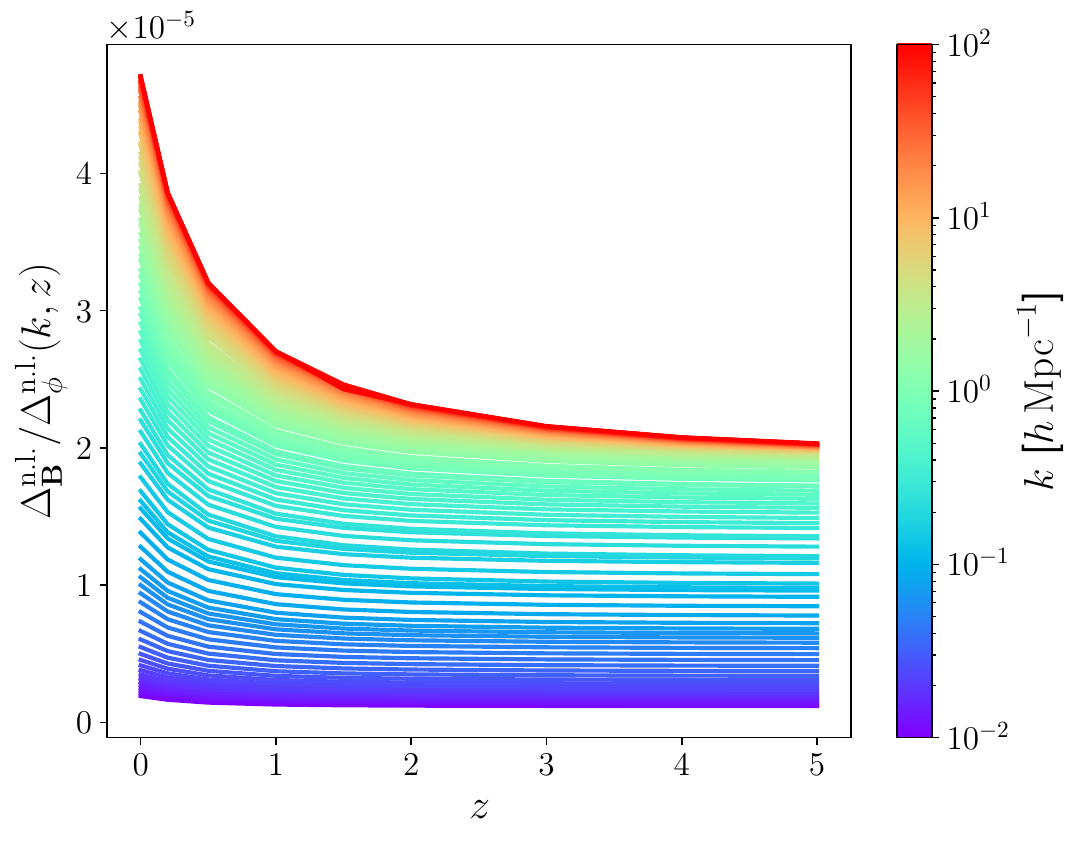}
    \caption{Left panel: ratio of the non-linear approximation with respect to PT for the dimensionless power spectrum of $\mathbf{B}$, as a function of the redshift and for different scales. The black dashed line represents the second-order PT. Right panel: ratio between the non-linear approximation $\Delta_{\mathbf{B}}^\mathrm{n.l.}$ and the non-linear power spectrum of the scalar potential $\Delta_{\phi}^\mathrm{n.l.}$, as a function of the redshift and for different scales.}
    \label{fig:B_z}
\end{figure*}

The amplitude gain is also represented in Figure~\ref{fig:B_z}, as a function of the redshift. 
For the smallest scales, in the range $10\text{--}100 \; h$ Mpc$^{-1}$, the pure non-linear evolution is intensively boosted from $z=1\text{--}2$ onwards.
However, as shown in the right panel and already stated at the beginning of this section, the ratio between the non-linear approximation for the power spectrum of the vector potential and the non-linear analogue for the scalar potential remains of the order of $10^{-5}$. Contrary to \citet{thomasFullyNonlinearPostFriedmann2015}, where this ratio is not monotonic over the redshift range (see Fig. 9 of the reference), here instead we find a monotonic increase over time. In particular, for the smallest scales, this ratio starts growing rapidly since $z=1\text{--}2$. 
Therefore, we can conclude that there is no indication pushing towards the direction of a relevant gravito-magnetic effect arising during a particular phase in the evolution of structures, other than the rapidly accelerated yet continuous growth of non-linearity at low redshift.



\section{Conclusions}\label{sec:conclusions}
On non-linear scales, cosmic structure formation is typically studied through Newtonian $N$-body simulations, while relativistic effects are only considered at large scales, of the order of the Hubble horizon, using perturbative methods, e.g.\ to study primordial non-gaussianity signatures in galaxy surveys. The post-Friedmann approach provides a unified framework for small and large scales that generalises the weak-field approximation to cosmology. In this framework, at the leading order in the $c^{-n}$ expansion, 
namely in the Newtonian limit of small peculiar velocities, a frame-dragging effect is expected to arise in terms of a gravito-magnetic vector potential. This potential is sourced by the transverse part of the momentum field and is thus expected to be non-zero if measured from Newtonian $N$-body simulations, as confirmed by previous works on cosmological scales \citep{bruniComputingGeneralrelativisticEffects2014a, thomasFullyNonlinearPostFriedmann2015}.

The IllustrisTNG project provides a suite of simulations that offer a unique opportunity to investigate this gravito-magnetic effect not only on cosmological scales but also in the highly non-linear regime, down to galactic scales, where structures have fully collapsed, have gone through the virialisation process and the vorticity that has been generated in the multi-stream regime becomes significant \citep{pueblasGenerationVorticityVelocity2009}. Our goal here has been the investigation of gravito-magnetic effects on these previosly unexplored smaller non-linear scales, using the \WB{post-Friedmann} framework on simulations of the IllustrisTNG project.

To this aim, three dark-matter-only simulations have been considered, i.e. TNG300-2-Dark, TNG100-2-Dark, and TNG50-2-Dark, which sample different non-linear regimes in the formation of structures.
\WB{Following the methodology developed by \citet{bruniComputingGeneralrelativisticEffects2014a} and \citet{thomasGravityNonlinearScales2015}}, the source fields of the gravito-magnetic potential have been extracted via the Delaunay Tessellation Field Estimator, that provides a detailed reconstruction of the density and velocity fields, from which power spectra were computed.

As expected, the matter power spectra from the TNG300-2-Dark, TNG100-2-Dark, and TNG50-2-Dark simulations show a good agreement with the non-linear HaloFit predictions, which is a commonly used semi-analytical model for non-linear growth of structures. However, for the smallest simulation, the suppression of power at large scales is evident, as a consequence of the finite size of the simulation box, which removes the contributions from long-wavelength perturbations. The momentum density field is more sensitive to these finite-volume effects, with a stronger suppression in smaller boxes. While this effect is well known for the matter density field, here we have shown for the first time that it is much stronger for the momentum density. This clearly indicates that the relevant non-linear terms are significantly affected by the lack of large-scale perturbations.

In our results the power spectra of the velocity gradients show the onset of non-linearity and the generation of vorticity. At around $k\approx1.5 \; h$ Mpc$^{-1}$ for the $205$ and $75$ $ h^{-1}$ Mpc simulations, and at around $k\approx3 \; h$ Mpc$^{-1}$ for the $35$ $ h^{-1}$ Mpc simulation, the velocity divergence shows a dip while the vorticity exhibits a peak and start to dominate at smaller scales as the system transitions into the non-linear regime, in agreement with previous studies \citep{pueblasGenerationVorticityVelocity2009, jelic-cizmekGenerationVorticityCosmological2018, barrera-hinojosaVectorModesLCDM2021}. 
This reveals that rotational (i.e. divergence-free) modes emerge naturally in cosmological simulations as a consequence of orbit crossing in collapsing regions, a process that becomes increasingly important as structures evolve beyond the single-stream approximation due to the collision-less nature of dark matter.

The analysis of the gravito-magnetic potential further highlights the effects of finite box size. Similarly to the momentum field, the power spectrum of the vector potential is systematically suppressed in smaller simulations, indicating that missing large-scale perturbations play a crucial role in sourcing the non-linearity of the gravito-magnetic field.

The simulation results are compared with both second-order perturbation theory and a non-linear approximation. This work shows for the first time that the non-linear prediction provides actually a good estimate for the non-linear power spectrum of the vector potential. 
A correction for the missing power in the gravito-magnetic vector potential power spectrum, inspired by a technique used in the Sunyaev-Zel'dovich effect literature~\citep{parkbaghyeonImpactBaryonicPhysics2018} and computed using the non-linear matter power spectrum from HaloFit, confirms that at large scales, the corrected spectra match perturbation theory predictions, while at small scales, the non-linear approximation remains a valid estimate.
This also confirms that the correction employed is actually accurate in reproducing consistent results across simulations of different size.

The ratio of the vector to scalar potential, corrected for the missing power, is found to be $\approx 2 \text{--}4 \times 10^{-5}$ across all simulations, consistent with previous studies \citep{bruniComputingGeneralrelativisticEffects2014a, thomasFullyNonlinearPostFriedmann2015, adamekGeneralRelativityCosmic2016, barrera-hinojosaVectorModesLCDM2021} and second-order perturbation theory \citep{luCosmologicalBackgroundVector2009}. Interestingly, the non-linear prediction for the vector power spectrum is found to reach an approximately constant value of $\approx 5 \times 10^{-5}$ at very small scales, which is only slightly larger than expected in 2nd-order PT, i.e. $\approx 3 \times 10^{-5}$.

Furthermore\WB{, differently from \citet{bruniComputingGeneralrelativisticEffects2014a} and \citet{thomasFullyNonlinearPostFriedmann2015}, the global solution for the gravito-magnetic potential over the entire simulation volume is also reconstructed from spectral analysis. Here,} a visual representation of its magnitude is presented for the first time for Newtonian $N$-body simulations. The gravito-magnetic potential, while weaker, is generally correlated with the scalar potential. It also shows clear non-local structures and organised vortical patterns that indicate a coupling between rotational flows and the dragging of spacetime.

As for the time evolution of the vector potential, 
the non-linear mode coupling amplifies the growth of vector perturbations beyond the perturbative expectation. At scales of the order of $k=100\; h$ Mpc$^{-1}$ the amplitude of the non-linear power spectrum is almost $200$ times larger than PT. The ratio between the vector and the scalar potentials increases monotonically over time, but remains of the order of $10^{-5}$.

These results suggest that frame-dragging remains a small effect overall, being of the order of $1$--$0.1$\% of the Newtonian potential, at least down to scales comparable to galaxies.
On the other hand, the small magnitude of the vector potential implies that its effect on the motion of particles in a simulation is negligible, \WB{given} the precision currently required in cosmology, indirectly confirming the validity of Newtonian $N$-body simulations in the context of $\Lambda$CDM on fully non-linear scales, as well as their use for computing standard cosmological observables.


However, the results obtained here for the gravito-magnetic potential have also been confirmed by cosmological studies performed with the fully relativistic code GRAMSES \citep{barrera-hinojosaVectorModesLCDM2021}. Contrary to the case of the post-Friedmann approach, which is restricted to weak-field regime approximations where relativistic terms are treated as small perturbations to the background FLRW metric, GRAMSES adopts the 3+1 formalism and the scalar and vector modes of the gravitational field are treated fully non-linearly. The code makes also use of the Adaptive Mesh Refinement in high-density regions so that it can reach spatial resolutions down to $2 \; h^{-1}$ kpc.
Here, indeed, a simulation of $256\;h^{-1}$ Mpc with mass resolution of $1.33 \times 10^9 \; h^{-1}\, \mathrm{M}_{\odot}$ was found to give a ratio between the gravito-magnetic and scalar potentials of the order of $10^{-3}$ inside dark matter halos of $10^{12.5}\text{--}10^{15} \; h^{-1}\, \mathrm{M}_{\odot}$.
However, while this contribution remains small on these scales, the actual computation of the frame-dragging for simulations of the size of TNG50-2-Dark, having a mass resolution of $2.9 \times 10^6 \; h^{-1}\, \mathrm{M}_{\odot}$, is yet to be performed using GRAMSES.

Importantly, as current and previous studies have focused exclusively on dark-matter-only simulations, the impact of baryons on such small scales still remains to be assessed. 
\CB{The presence of baryons will affect the source term for the gravito-magnetic field, which can be broken down into three components as in Equation~(\ref{eq:components}), from left to right: vorticity, density-weighted vorticity and $\mathbf{\nabla}\delta\times\mathbf{v}$. It has been shown that when baryonic physics is included in simulations, there is a suppression of the matter clustering on small scales due to processes such as AGN feedback~\citep{vanDaalen:2011,Saha:2024}. 
On the flip side, baryons contribute to the generation of vorticity via feedback-driven flows, shocks, and pressure gradients.
Hence, the net effect on the rotational component of the momentum field of matter, and hence on the gravito-magnetic potential, is not a priori clear. However, since Figure~\ref{fig:curlComponents_PS} shows that on small scales the $\mathbf{\nabla}\delta\times\mathbf{v}$ term is larger than the density-weighted vorticity term, it is likely that the net effect is a suppression of the gravito-magnetic field.}

\CB{In future, developing robust observational strategies to measure gravito-magnetic effects on cosmological scales may provide a novel way to test General Relativity beyond the linear regime. In particular, cross-correlation approaches combining weak-lensing surveys with CMB observations -- especially those sensitive to the kSZ effect signal -- may enhance the detectability of these otherwise suppressed effects~\citep{barrera-hinojosaLookingTwistProbing2022}. Since the gravito-magnetic field differs in cosmologies beyond $\Lambda$CDM, such measurements could offer a complementary probe to constrain dark matter, dark energy, and modified gravity models~\citep{thomasGravityNonlinearScales2015}.}


\section*{Funding}
WB and MC are indebted to the Italian Space Agency (ASI) for their continuing support through contract 2018-24-HH.0 and its addendum 201824-HH.1-2022 to the National Institute for Astrophysics (INAF).
CB-H is supported by the Chilean National Agency of Research and Development (ANID) through grant FONDECYT/Postdoctorado No. 3230512. MB has been supported by UK STFC Grants No. ST/S000550/1.

\section*{Acknowledgements}
We express our gratitude to the Reviewer for their time and insightful comments that contributed to improving the original manuscript.
We wish to thank Mario G. Lattanzi for useful comments and his continuous support. We also thank James Trayford, Christopher Lovell, Daniel B. Thomas, and Kazuya Koyama for helpful discussions. This work has made use of the SCIAMA supercomputer of the University of Portsmouth’s Institute of Cosmology and Gravitation. For the purpose of open access, we have applied a Creative Commons Attribution (CC BY) licence to any Author Accepted Manuscript version arising.

\section*{Data Availability}
 Supporting research data are available on reasonable request from the authors.



\bibliographystyle{mnras}
\bibliography{WBbiblio, other_refs} 




\appendix

\section{Convergence tests for power spectra}\label{app:main_gravito}
This appendix presents some convergence tests that have been performed to ensure the reliability of the result shown in Section~\ref{sec:results}. The effect due to the limited size of the simulation box is not reported here, as it is an integral part of the discussion addressed in the aforementioned section.

\subsection{Dependence on the DTFE grid resolution}\label{app:grid_res}
\begin{figure}
    \centering
    \includegraphics[width=\columnwidth]{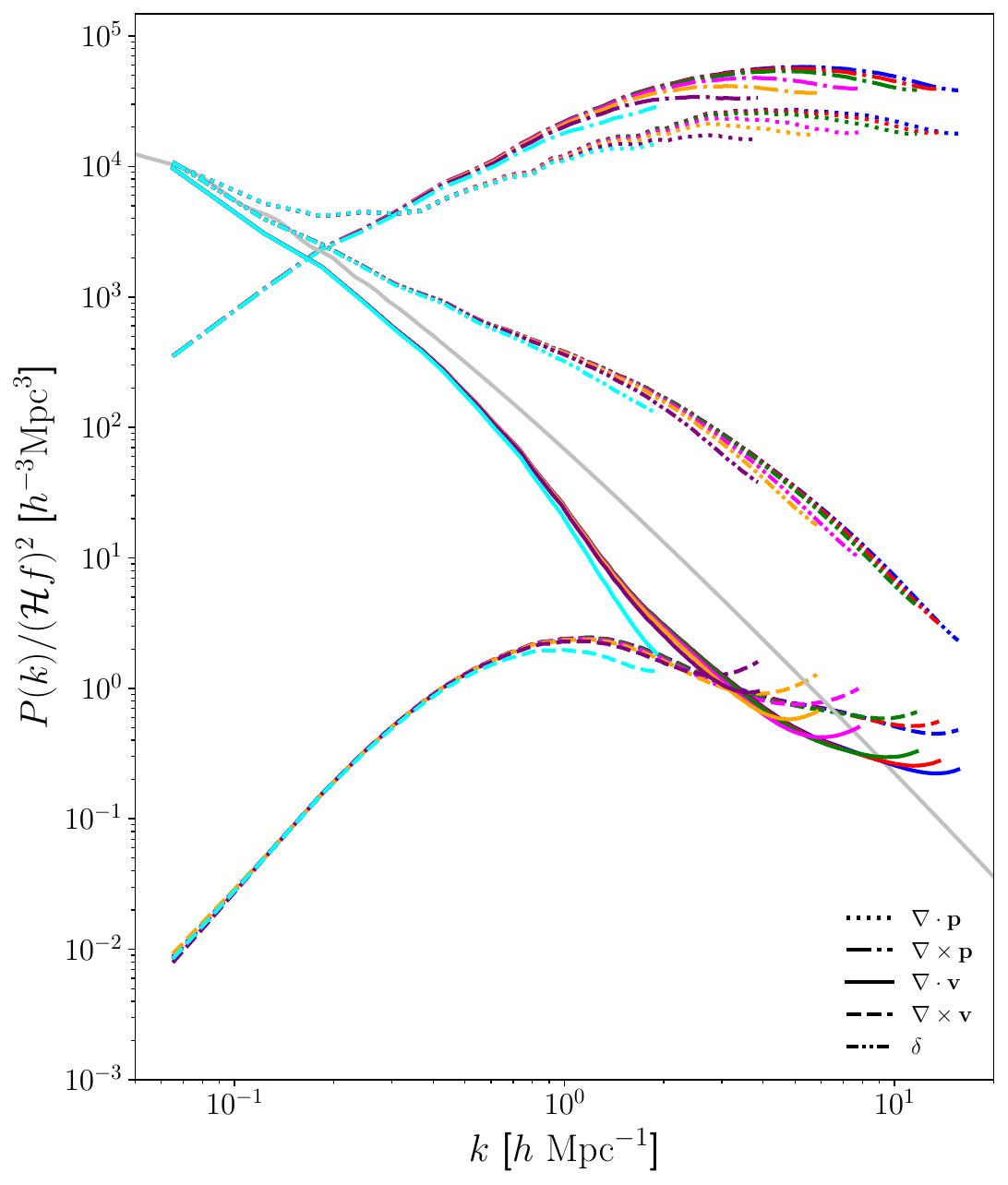}
    \caption{Power spectra of the velocity and momentum gradients, as detailed in Figure~\ref{fig:velMom_PS} for the simulation TNG300-3-Dark at $z=0$, for different grid resolutions of the DTFE: $N_\mathrm{grid}=128$ (cyan), $256$ (purple), $384$ (orange), $512$ (magenta), $768$ (green), $896$ (red), and $1024$ (blue). The divergence and vorticity of the velocity are represented as solid and dashed lines, respectively, while the divergence and vorticity of the momentum are shown with dotted and dash-dotted lines. The linear matter power spectrum is plotted as a reference (grey solid line), together with the non-linear matter power spectra \WB{(dot-dot-dashed line)} measured from the simulation for the corresponding $N_\mathrm{grid}$.}
    \label{fig:Ngrid_convergence}
\end{figure}

Here, the power spectra of the source terms of the gravito-magnetic potential are checked to have achieved convergence while varying the grid resolution of the DTFE. Figure~\ref{fig:Ngrid_convergence} shows that convergence is guaranteed for \WB{$N_\mathrm{grid} = 1024$} for all quantities, as \WB{the green ($N=768$), red ($N=896$), and blue ($N=1024$) lines are very close to each other}.

\subsection{Dependence on the mass resolution}\label{app:mass_res}
\begin{figure}
    \includegraphics[width=\columnwidth]{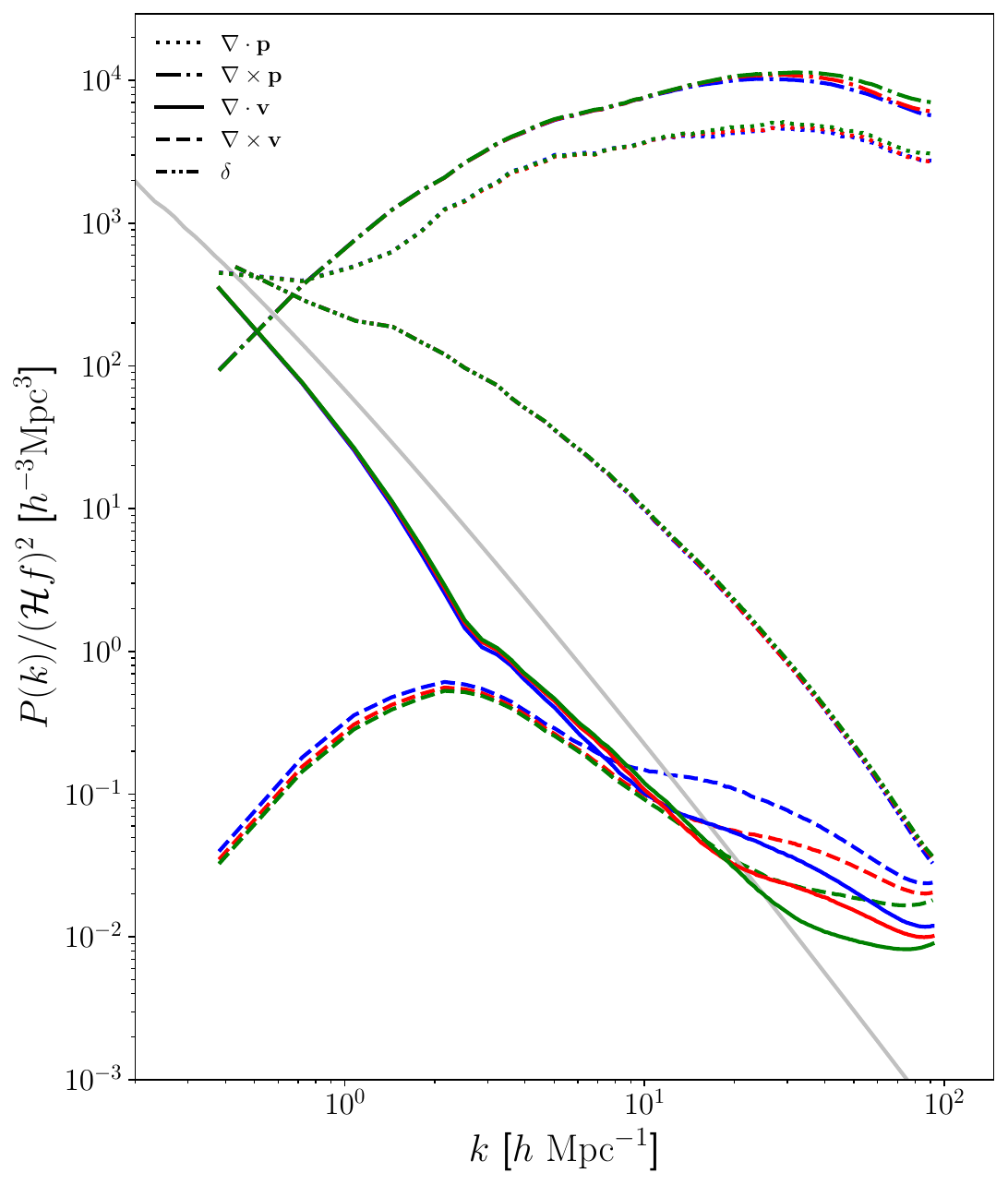}
    \caption{Sensitivity of the power spectra to the mass resolution of the 35 $ h^{-1}$ Mpc simulations, namely TNG50-4-Dark (blue), TNG50-3-Dark (red), and TNG50-2-Dark (green). As detailed in Figure~\ref{fig:velMom_PS}, the divergence and vorticity of the velocity are represented as solid and dashed lines, respectively, while the divergence and vorticity of the momentum are shown with dotted and dash-dotted lines. The linear matter power spectrum is plotted as a reference (grey solid line), together with the non-linear matter power spectra measured from the simulation.
    } 
    \label{fig:massRes}
\end{figure}

The dependence on the mass resolution of the simulation has also been investigated. For this test, the three simulations with $35$ $ h^{-1}$ Mpc were used: these have mass resolutions differing by a factor $8$, as TNG50-2-Dark consists of $1080^3$ particles of $2.9 \times 10^6 \; h^{-1}$ M$_\odot$, while TNG50-3-Dark consists of $540^3$ particles of $2.3 \times 10^7 \; h^{-1}$ M$_\odot$, and TNG50-4-Dark consists of $270^3$ particles of $1.9 \times 10^8 \; h^{-1}$ M$_\odot$. Figure~\ref{fig:massRes} shows how the mass resolution affects the power spectra of interest. It turns out that velocity is the quantity most significantly affected, while the matter power spectrum remains largely unchanged. The difference in the velocity divergence and vorticity power spectra is more pronounced at small scales due to the $k^2$ scaling factor. A similar behaviour is observed for the momentum density, although to a much lesser extent. Consequently, the ratio between the power spectra of the vector and scalar potentials is only slightly impacted at small scales, making the mass resolution dependence negligible for the final results presented in Section~\ref{sec:gravito-magneticPot}.  


\bsp	
\label{lastpage}
\end{document}